%% file: ms_arxiv.tex
\documentclass[twocolumn,tighten]{aastex631}
\usepackage{mathrsfs}
\usepackage{amsmath}

\newcommand{\hb}{\ifmmode {\rm H\beta} \else H$\beta$\fi}
\newcommand{\mgii}{\ifmmode {\rm Mg\ II} \else Mg {\sc ii}\fi}
\newcommand{\feii}{\ifmmode {\rm Fe\ II} \else Fe {\sc ii}\fi}
\newcommand{\heii}{\ifmmode {\rm He\ II} \else He {\sc ii}\fi}
\newcommand{\oiii}{\ifmmode {\rm [O\ III]} \else [O {\sc iii}]\fi}
\newcommand{\hei}{\ifmmode {\rm He\ I} \else He {\sc i}\fi}
\newcommand{\rfe}{\ifmmode {\rm \mathcal{R}_{Fe}} \else $\rm \mathcal{R}_{Fe}$\fi}
\newcommand{\dhb}{\ifmmode {\rm \mathcal{D}_{H\beta}} \else $\rm \mathcal{D}_{H\beta}$\fi}
\newcommand{\sersic}{$\rm S\acute{e}sic$}

\begin{document}

\title{DEPENDENCE OF VIRIAL FACTORS ON OPTICAL SPECTRAL PROPERTIES OF ACTIVE GALACTIC NUCLEI}

\author{Sen Yang}
\affiliation{Key Laboratory for Particle Astrophysics, Institute of High Energy Physics, Chinese Academy of Sciences, 19B Yuquan Road, Beijing 100049, People's Republic of China}
\affiliation{School of Physical Science, University of Chinese Academy of Sciences, 19A Yuquan Road, Beijing 100049, People's Republic of China}

\author[0000-0002-5830-3544]{Pu Du}
\affiliation{Key Laboratory for Particle Astrophysics, Institute of High Energy Physics, Chinese Academy of Sciences, 19B Yuquan Road, Beijing 100049, People's Republic of China}
\email{dupu@ihep.ac.cn, wangjm@ihep.ac.cn}

\author[0000-0001-9449-9268]{Jian-Min Wang}
\affiliation{Key Laboratory for Particle Astrophysics, Institute of High Energy Physics, Chinese Academy of Sciences, 19B Yuquan Road, Beijing 100049, People's Republic of China}
\affiliation{National Astronomical Observatories of China, Chinese Academy of Sciences, 20A Datun Road, Beijing 100020, People's Republic of China}
\affiliation{School of Astronomy and Space Science, University of Chinese Academy of Sciences, 19A Yuquan Road, Beijing 100049, People's Republic of China}

\begin{abstract}
Reverberation mapping (RM) has long been a powerful tool for measuring the masses of supermassive black holes (SMBHs) at the centers of active galactic nuclei (AGNs), but the precision of these mass measurements depends on the so-called virial factors. It has been demonstrated that the virial factors exhibit significant diversity, spanning approximately 1-2 orders of magnitude across different AGNs. However, the underlying physical drivers for the diversity have not yet been finalized. Here, adopting the SMBH mass -- spheroid luminosity relations of inactive galaxies with different bulge classifications, we calibrate the virial factors corresponding to the AGNs with pseudobulges (PB) and classical bulges (or elliptical hosts, CB) using the latest nearby RM sample. We investigate the correlations between virial factors and the AGN spectral properties, and find that for both PB and CB samples, the FWHM-based virial factors exhibit significant anti-correlations with the emission-line widths and profiles, while the $\sigma_{\rm line}$-based virial factors only show moderate anti-correlations with line widths for PB. We attribute these correlations mainly to the inclination angle or opening angle of the broad-line regions. Moreover, we establish new relations to give more precise virial factors and, in combination with the latest iron-corrected radius-luminosity relation, develop tentatively new single-epoch estimators of SMBH masses, which enable more accurate measurements of SMBH masses in large AGN samples.
\end{abstract}
    

\section{Introduction}
\label{sec:intro}
Reverberation mapping \citep[RM, e.g.,][]{Bahcall1972, Blandford1982, Peterson1993} is widely recognized as a powerful method for directly measuring the masses of supermassive black holes (SMBHs) located in the centers of active galactic nuclei (AGNs). Over the past 30 years, RM has been applied to over 200 AGNs, successfully determining the masses of their SMBHs \citep[e.g.,][]{Kaspi2000,Bentz2009b,Du2014,Hu2021,Bao2022,Yu2023,Shen2023}. In RM campaigns, the sizes of broad-line regions (BLRs) in AGNs can be determined by measuring the variations of the continuum fluxes and the delayed responses of the broad emission lines (BELs) through long-term spectroscopic monitoring. Combining with the velocity width ($V$) quantified by the FWHM or the dispersion ($\sigma_{\rm line}$) of the emission-line profile, the BH mass can be obtained by
\begin{equation}
    \label{eqn:rm}
	M_{\bullet} = f_{\rm BLR}\frac{V^2 R_{\rm BLR}}{G}
\end{equation}
where $R_{\rm BLR}=c\tau$ is the size of BLR, $\tau$ is the time delay of the
emission line with respect to the varying continuum, $c$ is the speed of light,
$G$ is the gravitational constant, and $f_{\rm BLR}$ is called ``virial
factor''. The factor $f_{\rm BLR}$ represents the unknown geometry
and kinematics of the BLR, and directly determines the accuracy of mass
measurements of SMBHs. The virial factor is different if using different
velocity-width measurement (FWHM or $\sigma_{\rm line}$) of the emission line.
In addition, the velocity width of the emission line can be measured both from
the mean and root-mean-square (rms) spectra in RM observations. Therefore, there
are virial factors for four velocity combinations -- $f_{\rm MF}$, $f_{\rm MS}$,
$f_{\rm RF}$, and $f_{\rm RS}$ (corresponding to the cases of FWHM or
$\sigma_{\rm line}$ from the mean or rms spectra). Here, ``M'' and ``R''
represent the mean and rms spectra, while ``F'' and ``S'' denote the FWHM and
$\sigma_{\rm line}$. 

In addition to the mass measurements obtained from the RM
observations, the values of virial factors have important implications for the
single-epoch mass estimates of SMBHs in AGNs, particularly for those in the
high-redshift universe during the era of the James Webb Space Telescope (JWST).
Recently, many SMBHs in the high-redshift universe have been found to exhibit
disproportionately large masses (relative to the dynamical masses of their host
galaxies) compared to their local counterparts \citep[e.g.,][]{Farina2022,
Goulding2023, Harikane2023, Pacucci2023, Maiolino2024}. However, there are also
reports of SMBH masses to host galaxy mass ratios that are consistent with the
local cases \citep[e.g.,][]{Izumi2019, Izumi2021, Ding2023}. These discoveries
are all based on single-epoch mass estimates, highlighting the importance of a
better understanding of the virial factors.

From an observational perspective, there are currently two approaches to determine the values of virial factors. The first approach is to calibrate the virial factor using the masses measured by methods independent from RM. Commonly used are the relations between the masses of the central SMBHs and the properties (e.g., the stellar velocity dispersion $\sigma_*$, the luminosity $L_{\rm bul}$, and the mass $M_{\rm *, bulge}$) of the spheroids (bulges or ellipticals) of inactive galaxies \citep[e.g.,][]{Onken2004, Park2012, Grier2013, Ho2014, Woo2015, Yu2019}. The underlying principle is based on the assumption that active galaxies and inactive galaxies have the same $M_{\bullet} - \sigma_*$, $M_{\bullet} - L_{\rm bul}$, or $M_{\bullet} - M_{\rm *, bul}$ relations. The average virial factor of an RM sample can be determined by comparing with the masses obtained by these relations. Although, due to the different RM samples used in the $f_{\rm BLR}$ calibrations and the evolving understanding on the $M_{\bullet}$-spheroid relations over time, the value of virial factor varies slightly across different studies. Specifically, the virial factors based on FWHM (from mean and rms spectra) are consistently around $1\sim1.5$, while those based on $\sigma_{\rm line}$ (from mean and rms spectra) are generally in the range of 4 to 5. The second approach is to obtain the virial factors by the dynamical modeling of BLRs, which yields the best model-dependent description of the BLR through fitting the RM data and gives the SMBH mass without assuming a virial factor \citep[e.g.,][]{Pancoast2011, Pancoast2014, Grier2017, Li2018, Villafana2023}. Therefore, it can derive the virial factors for individual AGNs. Additionally, the BLR dynamical modeling based on single-epoch spectra can also constrain the virial factors by fitting emission-line profiles \citep{kuhn2024}. Recently, the high-resolution interferometry provide a new approach to perform BLR dynamical modeling \citep[e.g.,][]{gravity2018, gravity2020, gravity2021, Abuter2024}. The joint analysis of RM and interferometry can give even stronger constraint to the virial factors \citep[e.g.,][]{wang2020, li2022}. The average values of virial factors obtained from dynamical modeling are generally consistent with those derived from the former approach \citep[e.g.,][]{Grier2017, Villafana2023}. However, the virial factors from both approaches exhibit wide range of values \citep[see, e.g.,][]{Woo2015, Villafana2023}, implying that the virial factors of individual objects may be highly diverse because of their different BLR geometry and kinematics.

Exploring the relationship between the virial factors and the AGN/BLR properties is an approach to understanding the physical origins of the diversity and to establish methods for improving the accuracy of SMBH mass measurements. \cite{Collin2006} separated the sample from \cite{Onken2004} into two populations based on the ${\rm FWHM}/\sigma_{\rm line}$ of their H$\beta$ emission lines, and found that the virial factors are slightly smaller in the population with larger ${\rm FWHM}/\sigma_{\rm line}$ (see Figures 9 and 10 in \citealt{Collin2006}). They speculated that the strength of a disk-wind component of BLRs contributes to the difference of the two populations, and the inclination effects play some roles in the population with smaller ${\rm FWHM}/\sigma_{\rm line}$. When dividing the sample into barred and non-barred galaxies, \cite{Graham2011} found that the virial factors of barred galaxies are smaller than those of the non-barred galaxies based on the $M_{\bullet} - \sigma_*$ at that time. Pseudobulges (PB) are a category of bulges characterized by disk-like density profiles and kinematics, believed to arise from the secular evolution of galaxies, in contrast to the formation processes of classical bulges (CB). Considering that the $M_{\bullet} - \sigma_*$ relation of PB is different from that of CB and ellipticals, \cite{Ho2014} evaluated the $f_{\rm BLR}$ factor by taking into account their bulge classifications, and found that the virial factors of PB are smaller by a factor of $\sim$2 compared with the CB and ellipticals. 

Recently, \cite{MejiaRestrepo2018} derived SMBH masses by fitting the accretion disk spectra in 37 AGNs at $z\sim1.5$ and obtained their virial factors. They claimed that the virial factors of H$\alpha$, H$\beta$, Mg {\sc ii}, and C {\sc iv} lines show strong anti-correlations their FWHM. However, as their sample lacked RM observations, they estimated the BLR sizes of their sample by adopting the radius-luminosity ($R_{\rm H\beta}$-$L_{5100}$) relation \citep[e.g.,][]{Kaspi2000, Bentz2013}. Based on almost the same sample as \cite{Ho2014}, \cite{Yu2019} investigated the virial factors and found that $f_{\rm MF}$ and $f_{\rm RF}$ show anti-correlations with FWHM, $\sigma_{\rm line}$, and ${\rm FWHM}/\sigma_{\rm line}$, whereas $f_{\rm MS}$ and $f_{\rm RS}$ display much weaker or no correlations. From dynamical modeling, \cite{Pancoast2014} found that the virial factors are correlated with inclination angles, but not with $M_{\bullet}$, $L_{5100}$, or ${\rm FWHM}/\sigma_{\rm line}$ of the emission-line profiles, where $L_{5100}$ is the monochromatic luminosity at 5100\AA. However, \cite{Grier2017} discovered that the virial factor is likely correlated with $M_{\bullet}$, inclination and opening angles of the BLR, but shows no correlation with $L_{5100}$. The latest compilation of dynamical modeling works \citep{Villafana2023} found evidence (marginal evidence) for a correlation between $f_{\rm MS}$ ($f_{\rm RS}$) and $M_{\bullet}$, marginal evidence for an anti-correlation of BLR thickness (inclination angle) with $f_{\rm MF}$ and $f_{\rm RF}$ ($f_{\rm MF}$, $f_{\rm RF}$, and $f_{\rm RS}$), and marginal evidence for a correlation between $f_{\rm RS}$ and ${\rm FWHM}/\sigma_{\rm line}$ from rms spectrum.

It is obvious that the dependency of virial factors on the BLR/AGN properties still remain controversial. Specifically, \cite{Collin2006}, \cite{MejiaRestrepo2018}, and \cite{Yu2019} found that the virial factors show anti-correlations with FWHM, $\sigma_{\rm line}$, or ${\rm FWHM}/\sigma_{\rm line}$, indicating that the AGNs with narrower BELs have larger virial factors. On the other hand, \cite{Graham2011} and \cite{Ho2014} suggested that the virial factors of AGNs with bars and PB are relatively smaller. It has been known that narrow-line type 1 AGNs (narrow-line Seyfert 1 galaxies, NLS1s) are more likely hosted in barred and in PB galaxies, compared to broad-line counterparts \citep[e.g.,][]{Xivry2011, Kim2017}. Therefore, on the contrary, it implies that the AGNs with narrower BELs have smaller virial factors. In addition, \cite{Villafana2023} discovered marginal evidence for a correlation between $f_{\rm RS}$ and ${\rm FWHM}/\sigma_{\rm line}$, further supporting this point. Therefore, it is crucial to clarify this ``paradox'' through the use of larger samples and more detailed analysis, as this will also advance our understanding of BLRs. 

In the present paper, we compile the latest RM sample with high-resolution observations from {\it Hubble Space Telescope (HST)}, calibrate the virial factors based on $M_{\bullet} - L_{\rm bul}$ relation, and investigate the correlation between the virial factors with the BLR/AGN properties. In Section \ref{sec:data_measurement}, we present the sample used in this paper and describe how we measure the host luminosities and the BLR/AGN properties. The calibrating procedure of the virial factors and their correlations with the BLR/AGN properties are provided in Section \ref{sec:analysis}. We provide some discussion in Section \ref{sec:discussion} and summarize our main conclusions in Section \ref{sec:summary}. We adopt the standard $\rm \Lambda CDM$ cosmology with the parameter of $H_0 = 67~{\rm km~s^{-1}~Mpc^{-1}}$, $\Omega_{\rm \Lambda} = 0.68$, and $\Omega_{\rm m} = 0.32$ \citep{Planck2014, Planck2020} in this paper.

\section{Data and Measurement}
\label{sec:data_measurement}

\subsection{Sample}
\label{sec:data_sample}
Our sample is mainly based on the compilation in \cite{Du2019},  which includes mostly low-redshift AGNs with $z \lesssim 0.2$. Additionally, we have incorporated some of the latest RM samples from \cite{Bao2022} and \cite{Hu2021}, as well as individual objects such as PG~2130+099 \citep{Hu2020}, PG~0026+129 \citep{Hu2020b}, PG~0923+201 \citep{Li2021}, IC~4329A \citep{Bentz2023}, Mrk~50 \citep{Barth2011, Barth2015}, Zw~229 \citep{Barth2011b}, Mrk~704 \citep{DeRosa2018}, I~Zw~I \citep{Huang2019}. In order to measure the luminosity of their host galaxies, we conducted a systematic search of this sample in the {\it HST} archive and identified 72 objects with high-quality images suitable for host and AGN decomposition. Our goal is to calibrate the virial factor as accurately as possible based on the luminosity of the host galaxies. Therefore, we focus on low-redshift AGNs and do not include the samples with higher redshifts from the Sloan Digital Sky Survey RM (SDSS-RM) project \citep{Shen2023} and the Australian Dark Energy Survey (OzDES) RM Program \citep{Malik2023, Yu2023}. The names, coordinates, redshifts, and Galactic extinctions are listed in Table \ref{tab:tar_info}.

\subsection{Decomposition of AGNs and host galaxies}
\label{sec:host_decomposition}

In order to extract the luminosities of the spheroids in the host galaxies of the RM AGNs, it is necessary to decompose the contribution of the bright point spread function (PSF) from the AGNs and the components (e.g., disks, bulges, and bars) of the host galaxies. The host decomposition of some objects has been presented in, e.g., \cite{Bentz2009a}, \cite{Bentz2013}, \cite{Du2014}, and \cite{Kim2017}. However, we perform the host composition for all of the objects in the sample using a same procedure in the present paper to ensure the consistency of the decomposition. 

\subsubsection{HST images}
\label{sec:HST_images}

The {\it HST} images of the objects were obtained using the instruments Advanced Camera for Surveys (ACS, \citealt{acs}), Wide Field Planetary Camera 2 (WFPC2, \citealt{wfpc2}), and Wide Field Camera 3 (WFC3, \citealt{wfc3}). ACS, WFPC2, and WFC3 have different channels and cameras. The detailed informations of the {\it HST} observations are provided in Table \ref{tab:tar_info}.

To calibrate the virial factors using the $M_{\bullet} - L_{\rm bul}$ relation in V band ($M_{\bullet} - L_{\rm V, bul}$), we opted to utilize the {\it HST} images taken with the filters close to the V band (e.g., F547M, F550M, F555W and F606W). If suitable images were not available, we chose to use the images taken with the filters at shorter or longer wavelengths (F438W, F675W, F791W etc.). Observations with higher signal-to-noise ratios (S/N) are consistently favored. The high contrast between the AGN (the central point source) and its host galaxy poses a challenge: a single exposure cannot ensure both a sufficient S/N ratio for the host galaxy and prevent saturation of the AGN. Fortunately, most observations have both short and long exposures \citep[e.g.,][]{Bentz2009a, Bentz2013}, which enables us to extract the host information even with very high contrast between the AGN and host components. 

We utilized the calibrated science images generated from the data processing pipeline of {\it HST} archive\footnote{\url{http://mast.stsci.edu}}. As the sharp peak (PSF) of science images generated by the pipeline is sometimes mis-classified and marked as cosmic rays, we used L.A.Cosmic package \citep{vanDokkum2001}\footnote{Which is provided in the \textsc{Astropy} image reduction package - \textsc{CCDPROC}} to identify and remove the cosmic rays. The removal of cosmic rays was performed for each individual science image. We then conducted a visual inspection of all processed images to ensure the preservation of the PSF. Subsequently, we aligned and combined the cosmic-ray-cleaned images, and corrected the distortion using DrizzlePac package \citep{DHbook}. It should be noted that we did not perform sky subtraction and cosmic-ray removal in the DrizzlePac operation.

\subsubsection{Generation of PSF}
\label{sec:psf}
Generating an accurate PSF is crucial in the host decomposition. The PSF varies slightly in different positions of the CCD, and the movements of the secondary mirror (SM) at different times also modulate the PSF profiles \citep{Kim2008, Makidon2006}. Additionally, the sampling of PSF and its sub-pixel displacement in the corresponding position of the AGN introduce uncertainties to some extent. Therefore, we follow these steps to generate the PSF model as accurately as possible.

We first followed the standard PSF simulation procedure by \texttt{TinyTim} \citep{Krist2011}. The SM movements were derived from the {\it HST} focus model\footnote{\url{https://www.stsci.edu/hst/instrumentation/focus-and-pointing/focus/hst-focus-model}}. During this step, an oversampling of 10 times is used when generating the PSF models by \texttt{TinyTim}. Secondly, we placed the PSF in the same position as the original AGN peak for each science frame, then undersampled the PSF to match the pixel size of the science frame. We then convolved the PSF with the charge diffusion kernel \citep{Krist2011}, and combined these PSF frames using DrizzlePac with the same parameters as the science frames. Finally, the final PSF model was extracted from the combined PSF image. We have checked that the generated PSF works well in the subsequent host decomposition.

\subsubsection{\texttt{GALFIT} fitting}
\label{sec:image_fitting}

We use \texttt{GALFIT} version 3.0 \citep{Peng2010} to model the surface brightness of the AGNs and their host galaxies. The following components are utilized in the image fitting: (1) a PSF (see Section \ref{sec:psf}) to model the AGN component, (2) several S\'ersic profiles to model the bulge (pseudobulge), disk and bar (if needed) components (only one S\'ersic for elliptical galaxy), and (3) a constant to model the sky background. The S\'ersic profile is expressed as
\begin{equation}
\Sigma(r) = \Sigma_{\rm e} \exp \left\{-\kappa \left[\left(\frac{r}{r_{\rm e}}\right)^{1/n}-1\right] \right\}
\end{equation}
where $\Sigma_{\rm e}$ is the surface brightness at the effective radius $r_{\rm e}$ (also known as half-light radius), $n$ is the S\'ersic index, and $\kappa$ is a variable coupled with $n$ (ensuring that half of the flux is within $r_{\rm e}$). The special case of the S\'ersic profile with $n=4$ corresponds to the de Vaucouleurs profile, and the case with $n=1$ is equivalent to an exponential disk. These two cases are widely adopted in the modeling of bulge (elliptical) and disk components of galaxies, respectively \citep{Bentz2009a, Du2014, Kim2017}. Fourier mode, coordinate rotation and truncation functions are also employed if the host galaxies show distortion or asymmetry.

The precision of the sky subtraction is crucial for accurate measurement of the host components. Following the method in \cite{Kim2017}, we initially determine the sky level using the \texttt{sky} task in IRAF and utilize it as the initial input for the \texttt{GALFIT} fitting. We include as large a region as possible in the fitting to encompass more sky area. After obtaining the best parameters in the fitting, we verify the sky background using the method provided in \cite{Kim2017} by checking if the cumulative flux reaches a constant at large radii.

After the fitting, the magnitudes of the host spheroids are converted to the V-band luminosities using the host template spectrum of the bulge from \cite{KC96}. For the images taken with red filters (e.g., F702W, F791W), their pass bands exceed the wavelength range of the bulge template \citep{KC96}. In such cases, we switch to using the template of S0 galaxies \citep{KC96} instead, which looks very similar to the bulge template at overlapping wavelengths. During the conversion, the template is redshifted and reddened by Galactic extinction of \cite{SFD98} (listed in Table \ref{tab:tar_info}). 

The uncertainties of the magnitudes of the host components decomposed by \texttt{GALFIT} are generally dominated by systematic errors \citep{Kim2008, Kim2017}. These systematic errors arise from (1) the uncertainties in the PSF model and (2) over- or under-estimation of the sky background. To evaluate the uncertainties caused by the PSF model, we collect the images of several (5-10) bright and unsaturated stars at similar positions of the CCD and taken with the same instruments and filters as each object. We generate a series of PSF models using these stars, and use these PSF models to perform the \texttt{GALFIT} fitting. The differences of the parameters obtained by these PSF models are regarded as the uncertainties caused by the PSF model. The uncertainties of the sky background determination may contribute to the uncertainties of the bulge magnitudes, we simply adopted 0.1 mag as extra uncertainties and added them to the error bars of the bulge magnitudes from the \texttt{GALFIT} fitting by quadratic summation in the following analysis \citep{Bentz2009a, Bentz2013}. For some objects (21 out of the 72 objects), the \sersic\ indexes of their bulge components cannot be well constrained due to their relatively poor S/N. In such cases, we fix their S\'ersic indexes as 4 and add additional uncertainties of 0.3 mag to their magnitudes \citep{Kim2008}. For the luminosity distance uncertainties, we follow \cite{Bentz2013} and include an error of 500 km/s in recession velocity. Furthermore, the template utilized in the k-correction (to V band) introduces additional uncertainties. In accordance with \cite{Kim2017}, we assess these uncertainties by examining various templates for the bulge spectra and consider the corresponding scatter as the errors. The final uncertainties of the magnitudes are derived by adding these uncertainties in quadrature. The best parameters of the \texttt{GALFIT} fitting are provided in Table \ref{tab:galfit}. The best fitting of the {\it HST} images and their residuals are provided in Figures \ref{fig:galfit_1}.

\subsection{Classification of pseudo and classical bulges}
\label{sec:pseudo_classical_bulges}

The principle behind calibrating the virial factors using the $M_{\bullet}$ -- spheroid relations ($M_{\bullet} - \sigma_*$, $M_{\bullet} - L_{\rm bul}$, or $M_{\bullet} - M_{\rm *, bul}$) is based on the assumption that both the active and inactive galaxies adhere to the same relations. Galaxies with PB are claimed to significantly displaced from the $M_{\bullet}$ -- spheroid relations of CB and ellipticals \citep[e.g.,][]{KH13}. Consequently, \cite{Ho2014} calibrated the virial factors of the AGNs with classical bulges (ellipticals) and pseudobulges separately, based on the $M_{\bullet} - \sigma_*$ relation. In the present paper, we follow \cite{Ho2014} but use the $M_{\bullet} - L_{\rm V,bul}$ to calibrate the virial factors of AGNs by separating their bulge classification. To classify the bulge type of the RM AGNs, we adopt the criterion used in \cite{Ho2014} and \cite{Kim2017}, where PB are defined as the objects with a S\'ersic index $n_{\rm b} < 2$, and also take into consideration that PB generally favor low bulge-to-total luminosity ratio of B/T $<0.2$ \citep[e.g.,][]{Fisher2008, Gadotti2009, Ho2014}. Our bulge classifications are generally in agreement with previous studies \citep{Guyon2006, Ho2014, Kim2017}. We found a PB percentage of 44\% (32/72), which is slightly higher than the 38.3\% reported by \cite{Ho2014} and \cite{Guyon2006}, but lower than the result of \cite{Fisher2008} (67\%). The elliptical and CB galaxies have the same $M_{\bullet} - \sigma_*$ relation \citep{KH13}, thus we do not distinguish between them sometimes in the latter text. Our bulge classifications are given in Table \ref{tab:galfit}. Some notes on individual objects are provided in Appendix \ref{app:notes}.

\subsection{Spectral properties}
\label{sec:agn_properties}

To investigate how the single-epoch spectral properties control the virial factors, we collect the spectral properties: (1) the flux ratio between \feii\ and \hb\ ($\rfe = F_{\rm Fe\,II}/F_{\hb}$), where the $F_{\rm FeII}$ represents the \feii\ flux between 4434 and 4684\AA\ and $F_{\hb}$ is the flux of broad \hb, (2-3) the ratios between FWHM and $\sigma_{\rm line}$ of broad \hb\ line (\dhb) from mean and rms spectra, which are considered as an indicator of the line profile, (4) the FWHM ratio between \feii\ and \hb, (5) the asymmetry ($\mathcal{A}$) of broad \hb, defined as $\mathcal{A} = [\lambda_{c}(3/4) - \lambda_{c}(1/4)] / \Delta\lambda(1/2)$ \citep{deRobertis1985}, where $\lambda_{c}(3/4)$ and $\lambda_{c}(1/4)$ are the central wavelengths at the 3/4 and 1/4 of the peak height, $\Delta\lambda(1/2)$ is FWHM, and positive (negative) $\mathcal{A}$ indicates stronger blue (red) wings of the line profile, (6-8) the equivalent width of \hb\ ($\rm EW_{H\beta}$), \oiii\ ($\rm EW_{[O\,III]}$), and \heii\ ($\rm EW_{He\,II}$), (9-12) the FWHM and $\sigma_{\rm line}$ of broad \hb\ emission lines from the mean and rms spectra. These spectral properties are the observables of BLRs (also referred to as ``BLR properties''). The majority of these spectral properties can be obtained from literatures, such as \cite{Du2019}. For the properties that cannot be found in such way, we conduct fitting of the spectra found in the public archive and measure these properties ourselves. When possible, we prioritize measuring them from the mean spectrum if corresponding RM data is available, as the mean spectra can effectively represent the properties over the corresponding period of time as RM. However, in the cases where RM data is not accessible, we resort to measuring these properties from single-epoch spectra found in literature or databases. In cases where multiple RM observations are available for a target, we average the values from the multiple measurements.

We perform spectral fitting using \texttt{DASpec}\footnote{The living code for \texttt{DASpec} is available at \url{https://github.com/PuDu-Astro/DASpec}, while the version used in the present paper is available at \url{https://doi.org/10.5281/zenodo.12578529}.}, a multi-component spectral fitting code with a graphical user interface. The core algorithm employed in \texttt{DASpec} is the Levenberg–Marquardt method \citep{Press1992}. We use the following components in the spectral fitting: (1) a power law to model the AGN continuum, (2) a spectral template of bulge provided in \cite{KC96} to model the host galaxy, (3) a \feii\ emission template constructed by \cite{Boroson1992}, (4) one or two Gaussians to model each of the broad emission lines (e.g., \hb, \heii\ $\lambda$4686, and \hei\ $\lambda$4471), (5) one or two Gaussians to model each of the narrow emission lines (e.g., \oiii\ $\lambda\lambda$4959,5007, \hb, and \heii\ $\lambda$4686, etc.). The fitting is performed in the windows 4170 - 4260\AA\ and 4430 - 5550\AA\ in the rest frame. All of the narrow emission lines are fixed to have the same line profile and velocity shift. The flux ratio between \oiii\ $\lambda$4959 and \oiii\ $\lambda$5007 is fixed to 1/3 \citep{Osterbrock2006}. In NLS1s, the narrow \hb\ lines are very weak and difficult to be deblended from the broad \hb. In such cases, we fix the flux of narrow \hb\ as one tenth of the \oiii $\lambda$5007 \citep{Du2014, Du2019}. In the spectral fitting, the host is fixed to have the flux extracted from the \texttt{GALFIT} fitting (see Section \ref{sec:image_fitting}). The flux of the host components decomposed by \texttt{GALFIT} in the aperture of the corresponding spectroscopic observation is adopted here. We examine the components of the host galaxies in the spectral fitting and they appear to be reasonable. We also test the \feii\ template from \cite{Park2022}, the fitting results are consistent. The line spread functions caused by the instrumental broadening are removed from the measurements of the line widths. The spectral properties (as well as the RM time lags and luminosities) are provided in Tables \ref{tab:rm_table} and \ref{tab:se_table}.

\section{Analysis}
\label{sec:analysis}

\subsection{Calibration of virial factors}
\label{sec:f_factor}

\begin{figure}
	\centering
	\includegraphics[width=0.45\textwidth]{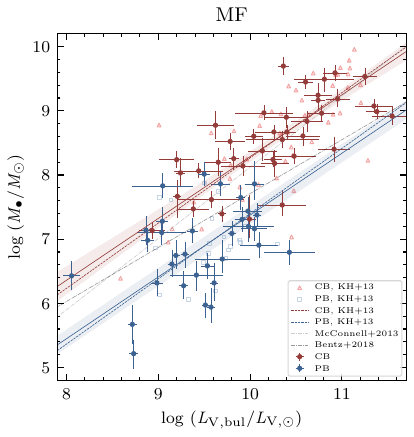}\includegraphics[width=0.45\textwidth]{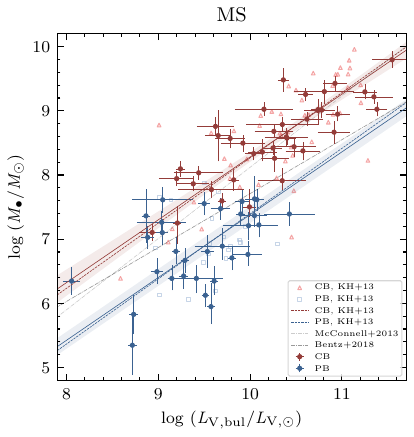}\
	\includegraphics[width=0.45\textwidth]{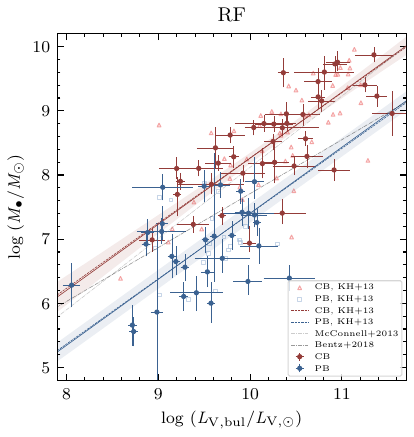}\includegraphics[width=0.45\textwidth]{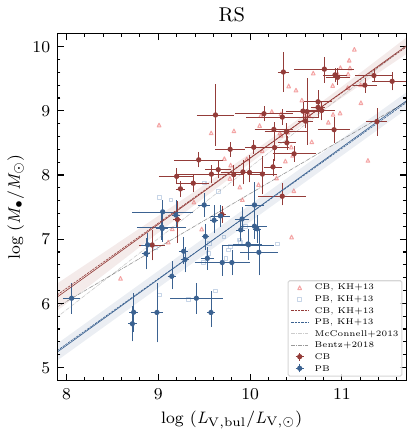}
	\caption{Calibration of virial factors based on the $M_{\bullet}-L_{\rm V,bul}$ relation. The four panels are the calibrations of $f_{\rm MF}$, $f_{\rm MS}$, $f_{\rm RF}$, and $f_{\rm RS}$, respectively. The red open triangles and blue open squares are the quiescent galaxies from \citet{KH13}, while the red and blue circles with error bars are the RM AGNs with CB and PB host galaxies, respectively. The red and blue solid lines and the regions are the results and confidence bands of the joint fitting (Approach I, see Section \ref{sec:f_factor}). The dashed and dash-dotted lines are the $M_{\bullet}-L_{\rm V,bul}$ relationships from previous works \citep{KH13, McConnell2013, Bentz2018}.}
	\label{fig:my_ml}
\end{figure}

To calibrate the virial factors, we perform a joint fitting for the combined sample of RM AGNs and inactive galaxies following the method as described in \cite{Woo2013} and \cite{Woo2015}. If we assume AGNs and inactive galaxies share the same importance in composing the $M_{\bullet} - L_{\rm V,bul}$ relation, we can obtain the slopes, intercepts, and the virial factors by minimizing (called Approach I)
\begin{equation}
\begin{split}
	\label{eqn:fitexy_joint}
	\chi^2=\sum_{i}^{M}\frac{(\log M_{\bullet,i} - \alpha \log L_{{\rm bul,}i} - \beta)^2}{\sigma_{\rm int}^{2} + (\alpha \sigma_{L_{\rm  bul},i})^2 + \sigma_{M_{\bullet},i}^2} + \\
	\sum_{j}^{N}\frac{(\log f + \log {\rm VP}_{\bullet,j} - \alpha \log L_{{\rm bul,}j} - \beta)^2}{\sigma_{\rm int}^{2} + (\alpha \sigma_{L_{\rm bul},j})^2 + \sigma_{{\rm VP}_{\bullet},j}^2}
\end{split}
\end{equation}
by Levenberg-Marquardt algorithm \citep{Press1992}, where the first summation reprsents the term of inactive galaxies, the second one is the term of RM AGNs,  $M,N$ are their numbers, ${\rm VP}_{\bullet} = V^2 R_{\rm BLR} / G$ is the virial product, $\sigma_{\rm int}$ is the intrinsic scatter of the relation, and $\sigma_{L_{\rm bul}}$, $\sigma_{M_{\bullet}}$, and $\sigma_{{\rm VP}_{\bullet}}$ are the uncertainties of the bulge luminosity, SMBH mass, and virial product, respectively. We adopt the sample of inactive galaxies in \citet{KH13}. However, the uncertainties of their V-band magnitudes are not provided in \cite{KH13}. For simplicity, we utilize a typical uncertainty of 0.2 dex in our analysis. As mentioned in previous sections, there are 4 combinations of velocity-width tracers (so thus 4 different VPs). We carry out the fitting separately for the four combinations. We perform the joint fitting for (1) the sample of the AGNs and inactive galaxies with PB (PB sample) and (2) the sample with classical bulges and ellipticals (CB sample), by using the FITEXY estimator \citep{Tremaine2002,Woo2013,Woo2015}. The calibrated virial factors in this case are called ``separate-sample'' virial factors. For comparison, we also perform the fitting using the total sample (PB $+$ CB), and the calibrated virial factors are called ``total-sample'' virial factors. The best parameters of the slope, intercepts, virial factors, and intrinsic scatters are provided in Table \ref{tab:lfrvf}. Because the dynamical range of the bulge luminosities is limited for the PB sample (see Figure \ref{fig:my_ml}), we fix the slope of the PB sample as that of the CB sample. 

For comparison, we also calibrate the virial factors using two additional approaches (called Approach II and Approach III). Approach II is similar to Approach I, but without including the intrinsic scatters. This approach can be expressed as 
\begin{equation}
\begin{split}
	\label{eqn:fitexy_joint_without_intrinsic}
	\chi^2=\sum_{i}^{M}\frac{(\log M_{\bullet,i} - \alpha \log L_{{\rm bul,}i} - \beta)^2}{(\alpha \sigma_{L_{\rm  bul},i})^2 + \sigma_{M_{\bullet},i}^2} +\\
	\sum_{j}^{N}\frac{(\log f_{\rm BLR} + \log {\rm VP}_{\bullet,j} - \alpha \log L_{{\rm bul,}j} - \beta)^2}{(\alpha \sigma_{L_{\rm bul},j})^2 + \sigma_{{\rm VP}_{\bullet},j}^2}.
\end{split}
\end{equation}
Approach III involves fixing the slope ($\alpha_0$) and intercept ($\beta_0$) to those of the $M_{\bullet} - L_{\rm V,bul}$ relation of inactive galaxies \citep{KH13} and only solving for the virial factors, which can be obtained by minimizing
\begin{equation}
	\label{eqn:fitexy_AGNs}
	\chi^2 = \sum_{j}^{N}\frac{(\log f_{\rm BLR} + \log {\rm VP}_{\bullet,j} - \alpha_0 \log L_{{\rm bul,}j} - \beta_0)^2}{(\alpha_0 \sigma_{L_{\rm bul},j})^2 + \sigma_{{\rm VP}_{\bullet},j}^2}.
\end{equation}
The virial factors obtained by Approach I are regarded as the primary results. The results from Approaches II and III are provided in Table \ref{tab:lfrvf} for comparison. 

Figure \ref{fig:my_ml} shows the $M_{\bullet}-L_{\rm V,bul}$ relation from Approach I, the blue and red points represent the PB and CB samples, respectively. It is evident that the correlations of $\sigma_{\rm line}$-based results are slightly tighter than the FWHM-based ones, both for total or PB/CB samples (see also the intrinsic error listed in Table \ref{tab:lfrvf}), indicating that $\sigma_{\rm line}$ is likely a better tracer for BLR velocities and for estimating $M_\bullet$ (support the claims in \citealt{Peterson2004}). Because inactive galaxies with PB exhibit smaller $M_{\bullet}$ compared to those with CB, our joint fitting results for the PB sample also deviate from those of the CB. The joint fitting of PB show significantly smaller intercepts than CB (see Table \ref{tab:lfrvf}). The offset of intercept between the PB and CB is about 1.0 dex for all of the cases (MF, MS, RF, and RS). As a result, the joint fitting (Approach I) of the total sample (PB $+$ CB) show steeper $M_{\bullet} - L_{\rm V,bul}$ relations. The virial factors calibrated by the PB sample in the present paper are $f_{\rm MF} = 0.67 \pm 0.28$, $f_{\rm RF} = 0.91 \pm 0.4$, $f_{\rm MS} = 2.22 \pm 0.86$ and $f_{\rm RS} = 2.4 \pm 0.93$, which are consistent with the values of $f_{\rm MF} = 0.9 \pm 1.9$, $f_{\rm RF} = 1.3 \pm 0.6$, $f_{\rm MS} = 3.2 \pm 1.6$, and $f_{\rm RS} = 4.8 \pm 2.0$ in \cite{Ho2014} within uncertainties. However, because of the larger sample, the uncertainties of our measurements are significantly smaller. In comparison, our virial factors calibrated by the CB sample are larger than the values in \cite{Ho2014} by factors of $2\sim3$ (see Table \ref{tab:lfrvf}). Additionally, when compared with the virial factors calibrated in \cite{Woo2015} without separating bulge types, our results calibrated using the total sample are also larger by factors of $\sim 2$. The rationale behind the larger CB virial factors is discussed in Section \ref{sec:comparison}.

\subsection{Correlations between virial factors and spectral properties}
\label{sec:scatters}

The virial factors calibrated by the $M_{\bullet}-L_{\rm V,bul}$ relations exhibit large scatters. The intrinsic scatters of the 4 cases (MF, MS, RF, and RS) are all around $0.5 \sim 0.6$ dex (see Table \ref{tab:lfrvf}). If we use the ``average'' virial factors from Table \ref{tab:lfrvf} and Eqn (\ref{eqn:rm}) to calculate the masses of SMBHs, the intrinsic scatters in the virial factors would contribute typical uncertainties of $0.5 \sim 0.6$ dex to the mass estimates. To investigate the correlations between the scatters of virial factors and the spectral properties obtained in Sections \ref{sec:agn_properties}, we define the deviation from the joint fitting (regarded as the virial factor of individual objects) as
\begin{equation}
\label{eqn:callogf}
	\log f = \alpha \times L_{\rm V,bul} + \beta - \log {\rm VP}_{\bullet},
\end{equation}
where $\alpha$ and $\beta$ are the slope and intercept obtained by Approach I in Section \ref{sec:f_factor}. Here, $f$ is $f_{\rm MF}$, $f_{\rm MS}$, $f_{\rm RF}$, or $f_{\rm RS}$, respectively. For objects that fall below the $M_{\bullet}-L_{\rm V,bul}$ relations, their $\log f$ values exceed the ``average'' virial factors of the sample obtained in Section \ref{sec:f_factor}. Essentially, the $\log f$ here can be considered as the individual virial factors if we assume that the masses predicted from the $M_{\bullet}-L_{\rm V,bul}$ relations are inherently accurate.

\begin{figure}
	\centering
	\includegraphics[width=\textwidth]{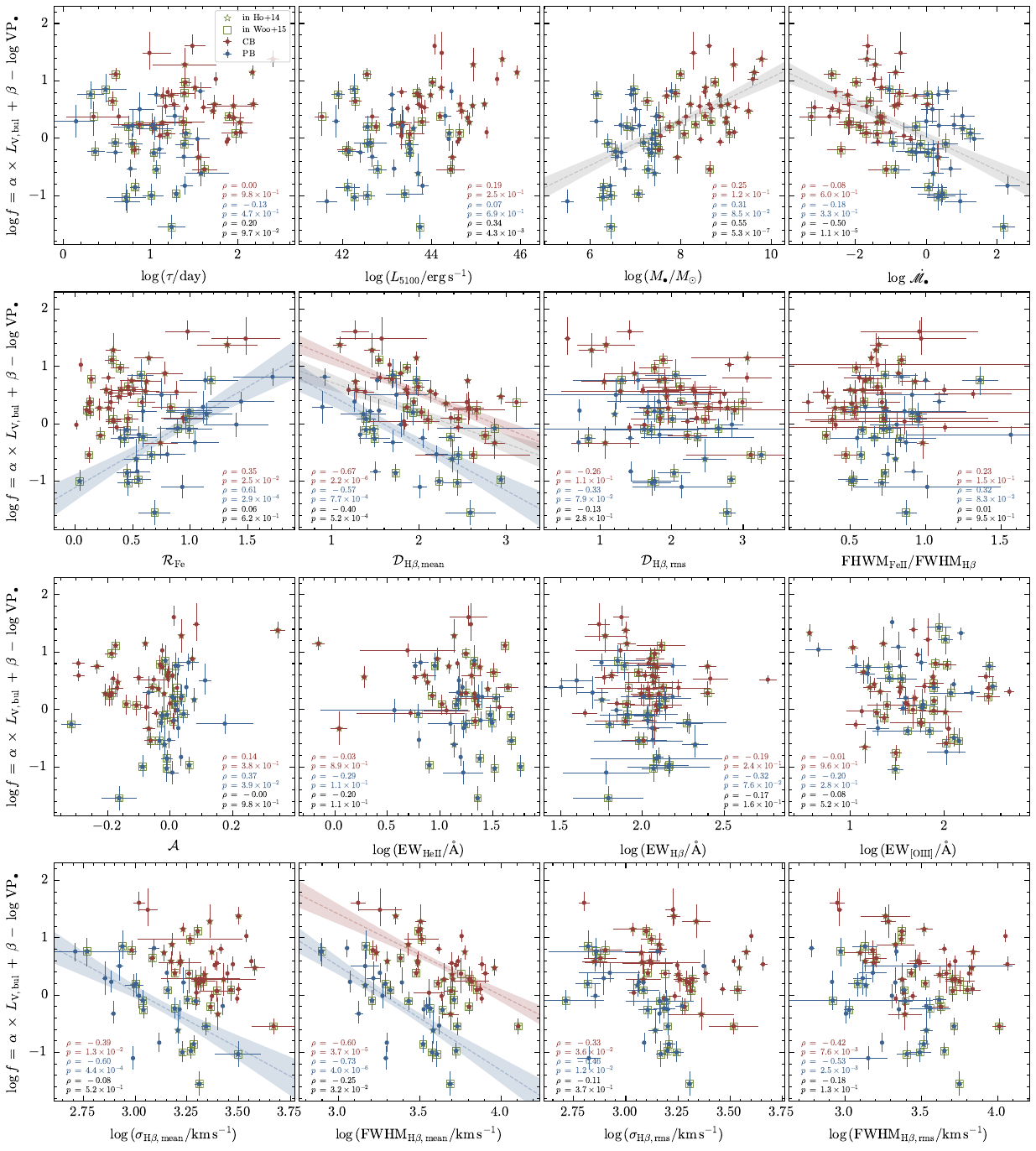}
	\caption{The correlations between $f_{\rm MF}$ and the spectral properties. The CB and PB objects are the red and blue circles, respectively. Spearman’s rank correlation coefficients ($\rho$) and the null probabilities ($p$) are provided in each panel (red for CBs, blue for PBs, and black for the total sample). The linear regressions and the corresponding confidence bands of the significant correlations with $\rm p<0.001$ are shown as red (CB), blue (PB), and gray (total) dashed lines and color regions. The circles overlapped with green squares and stars are the same objects as in \cite{Woo2015} and \cite{Ho2014}. }
	\label{fig:devi_f_sep_mf}
\end{figure}

\begin{figure}
	\centering
	\includegraphics[width=\textwidth]{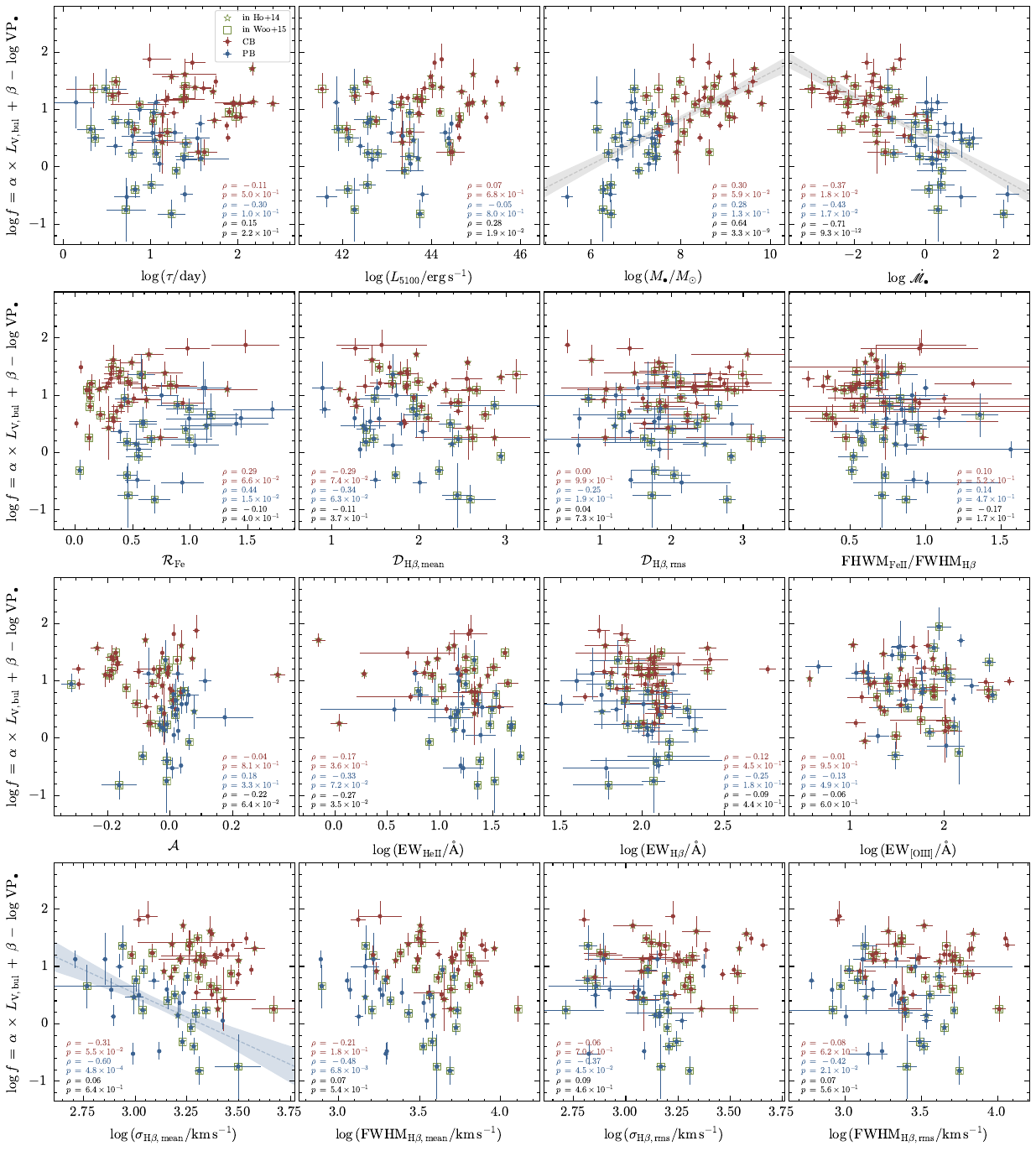}
	\caption{The correlations between $f_{\rm MS}$ and the spectral properties. The meanings of the symbols, lines, and regions are the same as in Figure \ref{fig:devi_f_sep_mf}.}
	\label{fig:devi_f_sep_ms}
\end{figure}

\begin{figure}
	\centering
	\includegraphics[width=\textwidth]{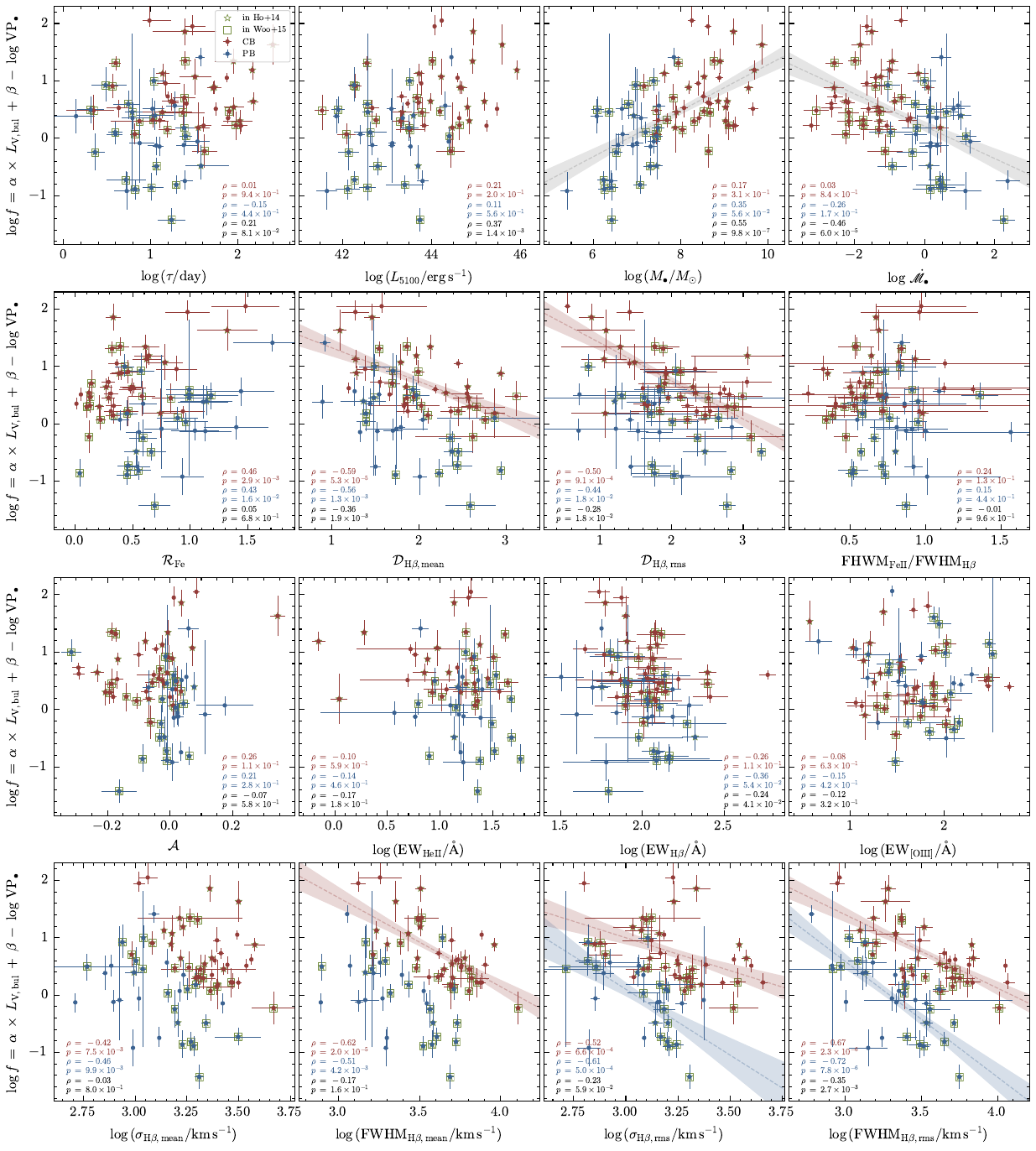}
	\caption{The correlations between $f_{\rm RF}$ and the spectral properties. The meanings of the symbols, lines, and regions are the same as in Figure \ref{fig:devi_f_sep_mf}. }
	\label{fig:devi_f_sep_rf}
\end{figure}

\begin{figure}
	\centering
	\includegraphics[width=\textwidth]{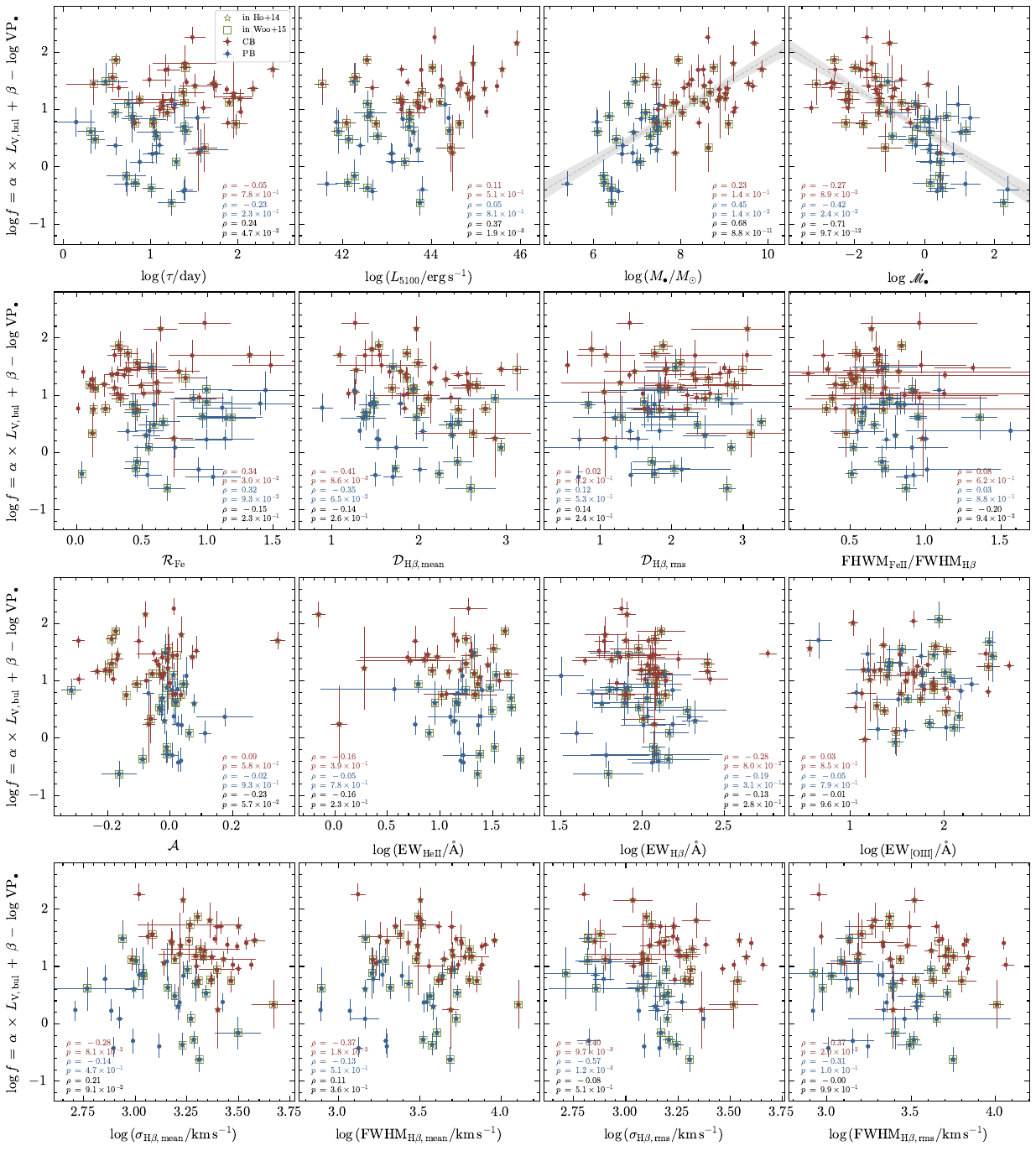}
	\caption{The correlations between $f_{\rm RS}$ and the spectral properties. The meanings of the symbols, lines, and regions are the same as in Figure \ref{fig:devi_f_sep_mf}. }
	\label{fig:devi_f_sep_rs}
\end{figure}

We plot the correlations between the deviations of the separate-sample virial factors and the spectral properties in Figures \ref{fig:devi_f_sep_mf}-\ref{fig:devi_f_sep_rs} (for MF, MS, RF, and RS, respectively). The case of the total-sample virial factors is provided in Appendix \ref{app:tot}. In addition, we also plot the correlations of the virial factors with the SMBH masses and dimensionless accretion rates ($\dot{\mathscr{M}}_{\bullet}$) in Figures \ref{fig:devi_f_sep_mf}-\ref{fig:devi_f_sep_rs}. We adopt the definition of accretion rate in \citep[e.g.,][]{Du2014, Du2016a} as 
\begin{equation}
\label{eqn:mdot}
\dot{\mathscr{M}}_{\bullet} = 20.1 \left(\frac{\ell_{44}}{\cos i}\right)^{3/2} m^{-2}_7,
\end{equation}
where $m_7=M_{\bullet}/10^{7}M_{\odot}$ is the SMBH mass, $\ell_{44}=L_{5100}/10^{44}~{\rm erg~s^{-1}}$ is the continuum luminosity at 5100\AA, and $i$ is the inclination angle of the accretion disk to the line of sight (LOS). For simplicity, an average of $\cos i = 0.75$ is adopted in this paper. To avoid the self-correlation, we calculate the $M_{\bullet}$ (and then $\dot{\mathscr{M}}_{\bullet}$) using their bulge luminosities (and the best parameters in Table \ref{tab:lfrvf}) rather than the virial products and the calibrated virial factors. The Spearman’s rank correlation $\rho$ and the corresponding null probabilities $p$ are marked in Figures \ref{fig:devi_f_sep_mf}-\ref{fig:devi_f_sep_rs}. 

Here we adopt $p<0.001$ as a criterion of the significant correlations, and provide the linear regressions for those significant correlations in Figures \ref{fig:devi_f_sep_mf}-\ref{fig:devi_f_sep_rs}. We use the linear regression defined as 
\begin{equation}
\label{eq:f_correction}
    \log f_{\rm BLR} = \alpha^{\prime} * {\cal{X}} + \beta^{\prime},
\end{equation}
where ${\cal{X}}$ is one of the BLR/AGN properties in Section \ref{sec:agn_properties}. In Figures \ref{fig:devi_f_sep_mf}-\ref{fig:devi_f_sep_rs}, the dashed lines and the regions in blue and red show the regressions and the confidence bands ($1\sigma$) for the PB and CB samples, respectively. The dashed lines and the regions in grey are those for the total samples. The slopes $\alpha^{\prime}$, intercepts $\beta^{\prime}$, and the intrinsic scatters ($\epsilon^{\prime}$) of the linear regressions are listed in Table \ref{tab:params_linear_regression}. 

Overall, for the PB sample, we find the following correlations:
\begin{itemize}
    \item $f_{\rm MF}$ shows anti-correlations with $\rm FWHM_{H\beta,mean}$, $\sigma_{\rm H\beta,mean}$, and ${\cal{D}}_{\rm H\beta,mean}$.
    
    \item $f_{\rm MF}$ shows a positive correlation with $\mathcal{R}_{\rm Fe}$.
    
    \item $f_{\rm MS}$ shows anti-correlations with $\sigma_{\rm H\beta,mean}$.
    
    \item $f_{\rm RF}$ shows anti-correlations with $\rm FWHM_{H\beta,rms}$ and $\sigma_{\rm H\beta,rms}$.
\end{itemize}
For the CB sample, we find the following correlations:
\begin{itemize}
    \item $f_{\rm MF}$ shows significant anti-correlations with $\rm FWHM_{H\beta,mean}$ and ${\cal{D}}_{\rm H\beta,mean}$.

    \item $f_{\rm RF}$ shows significant anti-correlations with $\rm FWHM_{H\beta,mean}$, $\rm FWHM_{H\beta,rms}$, $\sigma_{\rm H\beta,rms}$, ${\cal{D}}_{\rm H\beta,mean}$ and ${\cal{D}}_{\rm H\beta,rms}$.
\end{itemize}
For the total sample, we find:
\begin{itemize}
    \item $f_{\rm MF}$ shows anti-correlations with ${\cal{D}}_{\rm H\beta,mean}$.
    
    \item $f_{\rm MF}$, $f_{\rm MS}$, $f_{\rm RF}$, and $f_{\rm RS}$ all show positive correlations with $M_{\bullet}$ and anti-correlations with $\dot{\mathscr{M}}$ (not listed in Table \ref{tab:params_linear_regression}).
\end{itemize}
After establishing these correlations between the virial factors and the BLR properties, we can inversely utilize them to provide a better estimate of the virial factor in an AGN if the spectral properties can be measured from its spectrum. Finally, the correlations/anti-correlations between $f_{\rm MF}$, $f_{\rm MS}$, $f_{\rm RF}$, and $f_{\rm RS}$ with $M_{\bullet}$ and $\dot{\mathscr{M}}$ can not be simply used if we do not know the SMBH mass and accretion rate beforehand. However, these correlations do indicate that the virial factors are smaller in the AGNs with smaller SMBH masses and higher accretion rates.

\section{Discussion}
\label{sec:discussion}

\subsection{Physical driver of the correlations}
\label{sec:physical_driver}

The calibrated virial factors based on the FWHM ($f_{\rm MF}$ and $f_{\rm RF}$) tend to show significant anti-correlations with the line widths (FWHM or $\sigma_{\rm line}$), for either (or both) of the PB or CB samples. Additionally, the relationships between $f_{\rm MF}$ and ${\cal{D}}_{\rm H\beta,mean}$ (for PB and CB), and between $f_{\rm RF}$, ${\cal{D}}_{\rm H\beta,mean}$, and ${\cal{D}}_{\rm H\beta,rms}$ (only for CB) are also significant. From the perspective of eigenvector 1 sequence \citep[e.g.,][]{Marziani2001, Shen2014}, it has been demonstrated that accretion rate drives $\cal{R}_{\rm Fe}$, while the orientation dominates the dispersion of ${\rm FWHM}$ at fixed $\cal{R}_{\rm Fe}$. Some works about the line profiles \citep[e.g.,][]{Kollatschny2011} also suggest that the objects with narrower line widths (smaller ${\cal{D}}_{\rm H\beta}$) have thicker BLRs (can be considered to have larger opening angles). Furthermore, $f_{\rm MF}$ shows a positive correlation with $\cal{R}_{\rm Fe}$ for the PB sample. This correlation may stem from the tendency that the high-$\cal{R}_{\rm Fe}$ AGNs typically have smaller FWHM and ${\cal{D}}_{\rm H\beta}$ (see also the next paragraph). Considering that most of the correlations are between the virial factors and the line widths and profiles, the correlation between $f_{\rm MF}$ and $\cal{R}_{\rm Fe}$ could possibly be a secondary correlation induced by the correlations between the virial factors and the line widths/profiles. As a small test, we find that there is no residual correlation between $(f_{\rm MF} - f_{\rm MF}^{\rm fit})$ and ${\cal{R}}_{\rm Fe}$, where $f_{\rm MF}^{\rm fit}$ is the linear regression of $f_{\rm MF}$ vs. $\rm FWHM_{\rm H\beta,mean}$ (or ${\cal{D}}_{\rm H\beta,mean}$) obtained by Eqn (\ref{eq:f_correction}). Therefore, the inclination and opening angles are probably the primary physical driver of the strong anti-correlations between the virial factors and line widths and profiles.

\begin{figure}
    \centering
    \includegraphics[width=\textwidth]{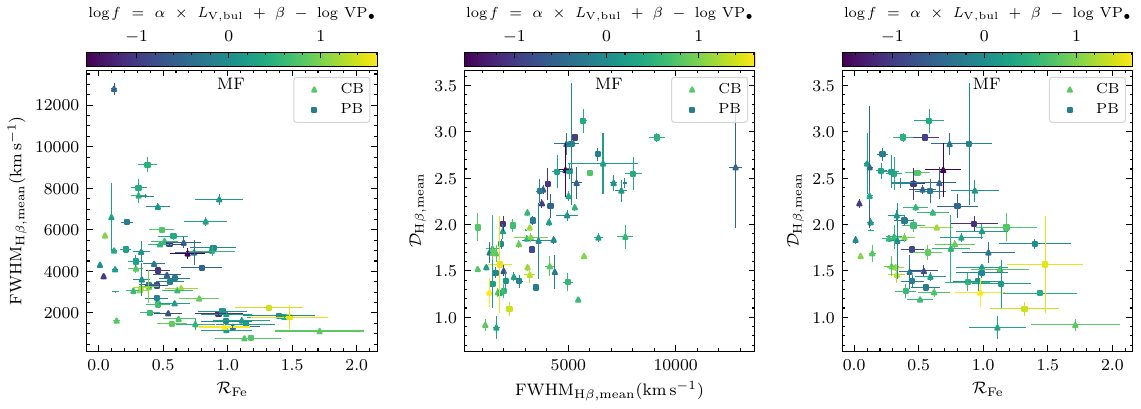}\
    \includegraphics[width=\textwidth]{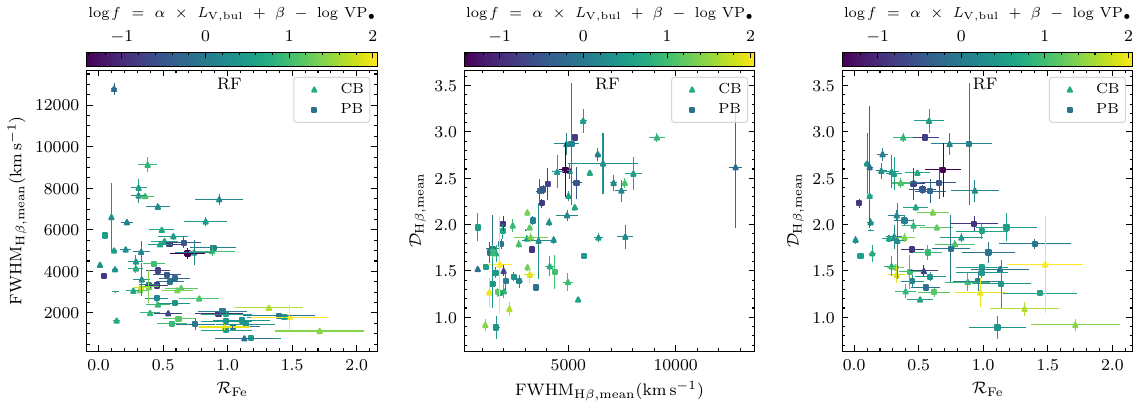}\
    \caption{Pairwise correlations between $\rm FWHM_{H\beta,mean}$, $\rm \mathcal{R}_{Fe}$ and $\rm \mathcal{D}_{H\beta,mean}$, color-coded by the virial factors (the upper panels for $f_{\rm MF}$ and the lower ones for $f_{\rm RF}$). The triangles and squares are the CB and PB objects, respectively.}
    \label{fig:binary_correlation}
\end{figure}

The $\rm FWHM_{H\beta,mean}$, ${\cal{D}}_{\rm H\beta,mean}$, and $\cal{R}_{\rm Fe}$ are mutually correlated \citep[e.g.,][]{Marziani2001, Shen2014, Du2016b, Kollatschny2011}. In Figure \ref{fig:binary_correlation}, we also plot the correlations between each pair of these three parameters, color-coded by the virial factors $f_{\rm MF}$ and $f_{\rm RF}$ as detailed in Section \ref{sec:scatters}. The tendencies of virial factors are more complicated than the cases in Figures \ref{fig:devi_f_sep_mf}-\ref{fig:devi_f_sep_rs}. In the correlations between $\rm FWHM_{H\beta,mean}$ and ${\cal{D}}_{\rm H\beta,mean}$, the smallest virial factors are located at larger ${\cal{D}}_{\rm H\beta,mean}$ at fixed $\rm FWHM_{H\beta,mean}$. However, the virial smallest factors only appear at largest  ${\cal{D}}_{\rm H\beta,mean}$ with lower $\rm FWHM_{H\beta,mean}$, while the largest virial factors are only located at the smallest ${\cal{D}}_{\rm H\beta,mean}$ and $\rm FWHM_{H\beta,mean}$, different from the simple anti-correlations between the virial factors and ${\cal{D}}_{\rm H\beta,mean}$ and $\rm FWHM_{H\beta,mean}$ in Figures \ref{fig:devi_f_sep_mf}-\ref{fig:devi_f_sep_rs}. Such complicated correlations may probably originate from the coordination of BLR thickness and (opening angle) and the inclination.

The $\sigma_{\rm line}$-based virial factors show less correlations with the spectral properties. The only relationships are the correlations between $f_{\rm MS}$ and $\sigma_{\rm H\beta,mean}$. This is likely due to the dominance of turbulent motion rather than Keplerian rotation in $\sigma_{\rm line}$ (e.g. \citealt{Kollatschny2011}, and its reduced susceptibility to the influence of the inclination angle.) 

The total-sample correlations between the virial factors and $M_{\bullet}$ are also significant. These correlations are evidently due to the deviation of the $M_{\bullet}-L_{\rm V,bul}$ relation of PB from that of CB. The galaxies with PB systematically exhibit smaller SMBH masses. Consequently, in Figures \ref{fig:devi_f_sep_mf}-\ref{fig:devi_f_sep_rs}, the PB (the blue points) are all situated in the lower left corners in the correlations of virial factors and $M_{\bullet}$. The situation is similar to the anti-correlations between the virial factors and the accretion rates. The smaller virial factors in AGNs with high accretion rates, coupled with their smaller BLR radii \citep[e.g.,][]{Du2015, Du2016a, Du2018a}, may help alleviate the conundrum of disproportionately over-massive SMBHs in AGNs in the high-redshift universe \citep[e.g.,][]{Farina2022, Goulding2023, Harikane2023, Pacucci2023, Maiolino2024}.

\subsection{Comparison with previous results}
\label{sec:comparison}

As mentioned in Section \ref{sec:analysis}, the average virial factors of the PB sample calibrated in this paper are consistent with the factors obtained in \cite{Ho2014}. However, the average virial factors of the CB sample calibrated here are larger than the values in \cite{Ho2014}. The large virial factors of the CB sample also cause the average virial factors calibrated from the total sample to be larger than those in the previous works \citep[e.g.,][]{Onken2004, Park2012, Grier2013, Woo2015}. Actually, the virial factors calibrated by the $M_{\bullet}$ -- spheroid relation or obtained in dynamical modeling of BLRs have very large scatters. The intrinsic scatters of the virial factors obtained in the present paper are roughly $0.5 \sim 0.6$ dex and the largest deviation of the virial factor of an individual object from the best fitting is $\rm \sim 1.4$ dex. Such large scatter means that the different samples used in different analysis may result in different virial factors. In Figures \ref{fig:devi_f_sep_mf}-\ref{fig:devi_f_sep_rs}, we have denoted the same objects as in \cite{Ho2014} and \cite{Woo2015} with stars and squares, respectively. It is evident that the CB sample in the present paper contains several more objects with very large virial factors, which cause the average virial factors calibrated by the CB sample (and the total sample) larger than those in the previous works. As a test, we select the same objects as \cite{Woo2015} and get the average virial factors (RF and RS) of 1.8 and 7.7 which are still larger than the results (1.12 and 4.47) in \cite{Woo2015}.

\begin{figure}
    \centering
    \includegraphics[width=0.5\textwidth]{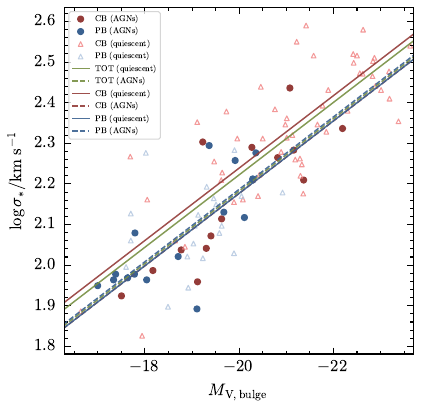}
    \caption{Comparison between the Faber-Jackson relationships of the AGNs and quiescent galaxies. The open triangles are the quiescent galaxies, and the circles are the AGNs. $M_{\rm V,bul}$ is the absolute V-band magnitude of the bulge. The solid and dashed lines are the linear regressions of the quiescent galaxies and AGNs with CB (red) and PB (blue). Those of the total sample are marked in green. The offsets between the AGNs and quiescent galaxies are obvious for the CB and total samples.}
    \label{fig:faber_jackson}
\end{figure}

Considering that the virial factors reported by \citet{Woo2015} and \citet{Ho2014} were based on the $M_{\bullet}-\sigma_{*}$ relationship, we present the correlation between stellar velocity dispersion ($\sigma_{*}$) and bulge magnitude (i.e. Faber-Jackson relationship) in Figure \ref{fig:faber_jackson}. The large circles represent the RM AGNs whose $\sigma_{*}$ can be found in the literature \citep{Ho2014}, and the open triangles represent the inactive galaxies. We found that the inactive galaxies with both PB and CB satisfy the same Faber-Jackson relationship; however, the CB AGNs seem biased towards smaller $\sigma_{*}$ and brighter bulge luminosity. Statistically, when we don't classify the bulge type, total AGN sample deviate from the Faber-Jackson relationship of inactive galaxies by about 0.04 dex. This offset propagates to a virial factor of about 0.2 dex, exactly the same deviation between our results and those of \citet{Woo2015}. For separate PB and CB samples, we found that AGN PB sample lie on the same line as their inactive counterparts, while the AGN CB sample have an offset of about 0.067 dex, equivalent to 0.3 dex for the virial factor. Both of them are consistent with the differences between our results and those of \citet{Ho2014}. 

The deviation of AGNs from the Faber-Jackson relationship of inactive galaxies could be attributed to three possible explanations: (a) the incompleteness of the AGN sample, (b) the intrinsic differences between AGNs and inactive galaxies, although this is contrary to the fundamental hypothesis of the calibration in the present paper (and all previous works), and (c) possible observational effects on the velocity dispersion measurements in AGNs, as they tend to reside at higher redshifts than the nearby quiescent galaxies used for the Faber-Jackson relationship, which could cause the apertures of spectra to cover larger areas of the host galaxies. Determining the main reason for this deviation is still an open question. If we consider the first explanation to be correct, we can make a correction to the virial factors calibrated in the present paper by systematically reducing the CB virial factors by 0.3 dex and the total-sample virial factors by 0.2 dex.

In \cite{Yu2019}, the correlations between the virial factors and several BLR/AGN properties (e.g., $\rm FWHM$, $\sigma_{\rm line}$, $\cal{R}_{\rm Fe}$, and $L_{5100}$) were investigated. Their study found correlations between $f_{\rm MF}$ and $\rm FWHM_{\rm H\beta,mean}$, and between $f_{\rm MF}$ and ${\cal{D}}_{\rm H\beta,mean}$ in the CB galaxies. Our results for the PB and CB samples also show similar correlations, but with higher significance. Although they did not show the calibrated virial factors of the PB, they did present the calibrations of the total sample (with PB scaled by $1/3.80$). Their calibrated virial factors ($f_{\rm MS}$ and $f_{\rm RS}$) from the total sample only show significant anti-correlations with Eddington ratios, which are similar to our finding in terms of  $\dot{\mathscr{M}}$ in the total sample.  

The significant anti-correlations between the virial factors and the line widths found in \cite{Collin2006}, \cite{MejiaRestrepo2018}, and \cite{Yu2019} appear contradictory to the results that AGNs with the bars and PB have relatively smaller virial factors in \citet{Graham2011} and \citet{Ho2014}, since NLS1s are more likely hosted in the barred and PB galaxies as compared to their broad-line counterparts \citep{Xivry2011,Kim2017}. Our results provide clarity on this matter: the virial factors of the PB and CB samples both show anti-correlations with the line widths (FWHM or $\sigma_{\rm line}$) and with the line profiles ($\cal{D}_{\rm H\beta}$), however, the average virial factors of the AGNs with PB are systematically smaller than those of the CB AGNs. In conclusion, accounting for the bulge type is crucial when investigating trends between the virial factor and spectral properties to avoid conflating the impact of galaxy morphology and that of structural BLR effects.

\subsection{Line-width- and line-profile-corrected virial factors}
\label{sec:new_estimators}

\begin{figure}
    \centering
    \includegraphics[width=0.45\textwidth]{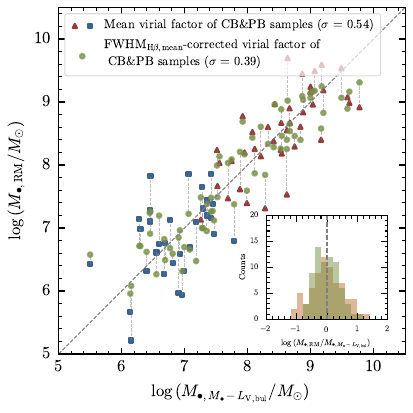}
    \includegraphics[width=0.45\textwidth]{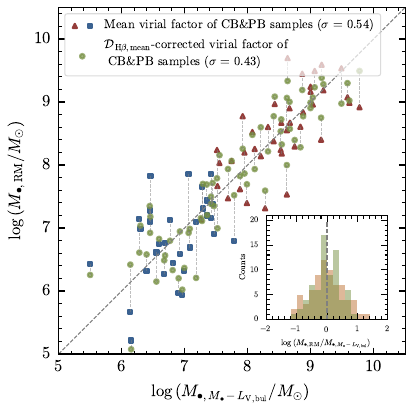}
    \caption{Comparisons between the SMBH masses derived from the $M_{\bullet}-L_{\rm V, bul}$ relations ($M_{\bullet,M_{\bullet}-L_{\rm V, bul}}$) and from the RM with the calibrated virial factors ($M_{\rm \bullet,RM}$). The red triangles (CB) and blue squares (PB) are the RM masses derived from the average virial factors from Table \ref{tab:lfrvf}, and the green circles are the masses obtained by using the $\rm FWHM_{\rm H\beta,mean}$- or ${\cal{D}}_{\rm H\beta,mean}$-corrected virial factors. The embedded panel is the distributions of $\log M_{\rm \bullet,RM}/M_{\bullet,M_{\bullet}-L_{\rm V, bul}}$ (green for $\rm FWHM_{\rm H\beta,mean}$- or ${\cal{D}}_{\rm H\beta,mean}$-corrected virial factors, and orange for the average virial factors). Here we only show the MF case as an example. It is clear that the circles from the corrected virial factors show significantly smaller scatters. The $\sigma$ in the panels denote the standard deviations of $\log (M_{\rm \bullet,RM}/M_{\bullet,M_{\bullet}-L_{\rm V, bul}})$.}
    \label{fig:M_M_RM_sep_mf}
\end{figure}

In Section \ref{sec:scatters}, we present the correlations between the virial factors and the spectral properties. The correlations of the PB and CB samples, respectively, are more significant than the correlations of the total sample (do not distinguish PB and CB). The significant separate-sample correlations are the relationship between the virial factors and $\rm FWHM$, $\sigma_{\rm H\beta}$, and ${\cal{D}}_{\rm H\beta}$ (representing the line profile) from the mean or rms spectra. Therefore, utilizing these correlations, we can provide more accurate estimates of the SMBH masses, which are superior to the SMBH masses based on the average virial factors. If we know the bulge types of the objects in advance, we can use the line-width-corrected or line-profile-corrected separate-sample FWHM-based virial factors from Eqn (\ref{eq:f_correction}) and the parameters in Table \ref{tab:params_linear_regression} (or also considering the offset caused by the difference between the $M_{\bullet}-\sigma_{*}$ and $M_{\bullet}-L_{\rm bul}$ relations in Section \ref{sec:comparison}). We show the comparisons between the $M_{\bullet}$ calculated from the luminosities of the host spheroids ($M_{\bullet,M_{\bullet}-L_{\rm V,bul}}$) and those obtained from the average and the corrected virial factors ($M_{\rm \bullet,RM}$) in Figure \ref{fig:M_M_RM_sep_mf}. As examples, we only provide here the $M_{\rm \bullet,RM}$ calculated by using the $\rm FWHM_{H\beta,mean}$- and ${\cal{D}}_{\rm H\beta,mean}$-corrected $f_{\rm MF}$. For clarity, we do not plot the error bars in order to show clearly the differences between the SMBH masses using the average virial factors and the line-width- and line-profile-corrected virial factors. We use the standard deviation of $M_{\rm \bullet,RM} / M_{\bullet,M_{\bullet}-L_{\rm V,bul}}$ to evaluate the scatters. It is obvious that the $M_{\bullet}$ from the line-width- and line-profile-corrected virial factors shows significantly smaller scatters than the cases with the average virial factors. This suggests that the line-width and line-profile-corrected virial factors indeed provide better estimates of the SMBH masses in RM AGNs. If we do not know the bulge types of the AGNs in advance, we can use the ${\cal{D}}_{\rm H\beta,mean}$-corrected $f_{\rm MF}$ or $f_{\rm RF}$ for the total sample to get improved estimates of the SMBH masses. 

\subsection{New single-epoch SMBH mass estimators}

The RM campaign of super-Eddington accreting massive black holes (SEAMBHs) discovered that the SEAMBHs deviate significantly from the traditional $R_{\rm H\beta}-L_{5100}$ relation \citep[e.g.,][]{Du2015, Du2016a, Du2018a}. Recently, \cite{Du2019} introduced a new scaling relation ($R_{\rm H\beta}-L_{5100}-{\cal{R}_{\rm Fe}}$) that provides more accurate estimates of the BLR sizes for the AGNs with both of low and high accretion rates. Their $R_{\rm H\beta}-L_{5100}-{\cal{R}_{\rm Fe}}$ relation is given by $\log\, (R_{\rm H\beta}/{\rm lt-days}) = (1.65\pm0.06) + (0.45\pm0.03) \log \ell_{44} - (0.35\pm0.08) {\cal{R}}_{\rm Fe}$. Therefore, by combining this with the line-width- and line-profile-corrected virial factors provided in Section \ref{sec:scatters}, we can tentatively establish new single-epoch estimators of the SMBH masses, based on the current very limited sample in the present paper.

From the line-width- and line-profile-corrected $f_{\rm MF}$ and the $R_{\rm H\beta}-L_{5100}-{\cal{R}_{\rm Fe}}$ relation, the new estimators for the SMBH masses of the PB AGNs are
\begin{equation}
\label{eq:MSE_DHb_PB}
\begin{split}
    \log (M_{\bullet}/M_{\odot}) = (0.45\pm0.03) \log \ell_{44} - (0.35\pm0.08) {\cal{R}}_{\rm Fe}\\
     + 2\log {\rm (FWHM_{\rm H\beta}/km~s^{-1})} - (0.82\pm0.19){\cal{D}_{\rm H\beta,mean}}\\
     + (2.25\pm0.35),
\end{split}
\end{equation}
\begin{equation}
\label{eq:MSE_Rfe_pb}
\begin{split}
    \log (M_{\bullet}/M_{\odot}) = (0.45\pm0.03) \log \ell_{44} + (0.82\pm0.29) {\cal{R}}_{\rm Fe}\\
     + 2\log {\rm (FWHM_{\rm H\beta}/km~s^{-1})} -(0.17\pm0.24),
\end{split}
\end{equation}
\begin{equation}
\label{eq:MSE_FWHM_PB}
\begin{split}
    \log (M_{\bullet}/M_{\odot}) = (0.45\pm0.03) \log \ell_{44} - (0.35\pm0.08) {\cal{R}}_{\rm Fe}\\
       + (0.18\pm0.33)\log {\rm (FWHM_{\rm H\beta}/km~s^{-1})} \\
       + (6.89\pm1.12)
       , 
\end{split}
\end{equation}

\begin{equation}
\label{eq:MSE_sigma_PB}
\begin{split}
    \log (M_{\bullet}/M_{\odot}) = (0.45\pm0.03) \log \ell_{44} - (0.35\pm0.08) {\cal{R}}_{\rm Fe}\\
    +  2\log {\rm (FWHM_{\rm H\beta}/km~s^{-1})}\\
    - (1.96\pm0.51)\log {\rm (\sigma_{\rm H\beta}/km~s^{-1})}\\
    + (6.88\pm1.59),
\end{split}
\end{equation}
and the new estimators for the CB AGNs are
\begin{equation}
\label{eq:MSE_FWHM_CB}
\begin{split}
    \log (M_{\bullet}/M_{\odot}) = (0.45\pm0.03) \log \ell_{44} - (0.35\pm0.08) {\cal{R}}_{\rm Fe}\\
       + (0.58\pm0.26)\log {\rm (FWHM_{\rm H\beta}/km~s^{-1})} + (6.6\pm0.94)
       , 
\end{split}
\end{equation}
and
\begin{equation}
\label{eq:MSE_DHb_mean_CB}
\begin{split}
    \log (M_{\bullet}/M_{\odot}) = (0.45\pm0.03) \log \ell_{44} - (0.35\pm0.08) {\cal{R}}_{\rm Fe}\\
     + 2\log {\rm (FWHM_{\rm H\beta}/km~s^{-1})} - (0.61\pm0.11){\cal{D}_{\rm H\beta,mean}}\\
     + (2.7\pm0.24).
\end{split}
\end{equation}
It is worth pointing out that Eqn (10) employs the $\mathcal{R}_{\rm Fe}$-correct $f_{\rm MF}$ and Eqn (11) employs the $\rm FWHM_{H\beta}$-corrected $f_{\rm MF}$. If the bulge types are not known in advance, the new estimator is
\begin{equation}
\label{eq:MSE_DHb_tot}
\begin{split}
    \log (M_{\bullet}/M_{\odot}) = (0.45\pm0.03) \log \ell_{44} - (0.35\pm0.08) {\cal{R}}_{\rm Fe}\\
     + 2\log {\rm (FWHM_{\rm H\beta}/km~s^{-1})} - (0.53\pm0.13){\cal{D}_{\rm H\beta,mean}}\\
     + (2.19\pm0.28),
\end{split}
\end{equation}

From the line-width-corrected $f_{\rm MS}$ and the $R_{\rm H\beta}-L_{5100}-{\cal{R}_{\rm Fe}}$ relation, the new estimator for the SMBH masses of the PB AGNs is

\begin{equation}
\label{eq:MSES_sigma_PB}
\begin{split}
    \log (M_{\bullet}/M_{\odot}) = (0.45\pm0.03) \log \ell_{44} - (0.35\pm0.08) {\cal{R}}_{\rm Fe}\\
       +(0.35\pm0.48)\log {\rm (\sigma_{\rm H\beta}/km~s^{-1})} + (6.42\pm1.51),
\end{split}
\end{equation}

\begin{figure}
    \centering
    \includegraphics[width=0.45\textwidth]{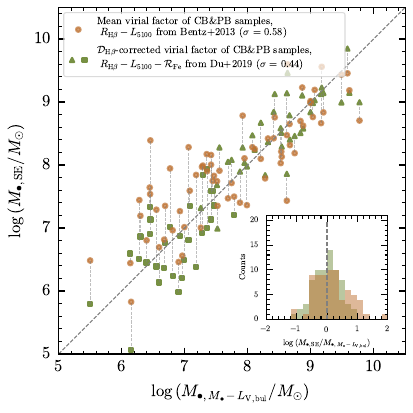}
    \includegraphics[width=0.45\textwidth]{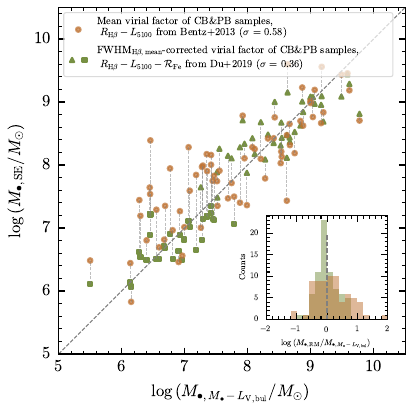}
    \caption{Comparisons between the $M_{\bullet,M_{\bullet}-L_{\rm V, bul}}$, the SMBH masses from the new single-epoch estimators, and those from the traditional single-epoch estimator. The orange circles are the single-epoch masses ($M_{\rm \bullet,SE}$) from the traditional estimator (the average virial factor and the $R_{\rm H\beta}-L_{5100}$ relation in \citealt{Bentz2013}), and the green triangles (CB)/ squares (PB) are the those from the new estimators in Section \ref{sec:new_estimators}. The embedded panel is the distributions of $\log M_{\rm \bullet,SE}/M_{\bullet,M_{\bullet}-L_{\rm V, bul}}$. The $\sigma$ in the panels denote the standard deviations of $\log (M_{\rm \bullet,SE}/M_{\bullet,M_{\bullet}-L_{\rm V, bul}})$.}
	\label{fig:M_M_se}
\end{figure}

As examples, we present the comparisons of the SMBH masses from the $M_{\bullet} - L_{\rm V,bul}$ relation, and the single-epoch SMBH masses ($M_{\rm \bullet,SE}$) from the ${\cal{D}_{\rm H\beta,mean}}$-corrected and the $\rm FWHM_{\rm H\beta,mean}$-corrected estimators and from the traditional estimator (the $R_{\rm H\beta}-L_{5100}$ from \citealt{Bentz2013} with the average total-sample virial factor) in Figure \ref{fig:M_M_se}. It is obvious that the new SMBH mass estimators can all yield SMBH masses with significantly smaller scatters than those from the traditional method. Moreover, it is surprising to note that the Eqn (\ref{eq:MSE_FWHM_PB}) exhibits weak dependence on $\rm FWHM_{\rm H\beta,mean}$. It should be noted that these new single-epoch estimators are derived under the assumption of the existence of an intrinsic scatter-free $M_{\bullet} - L_{\rm V,bul}$ relation, and the premise that all AGNs adhere to this relationship. On the one hand, the validity of this assumption requires further investigation. On the other hand, these estimators will propagate the scatter of the $M_{\bullet} - L_{\rm V,bul}$ relationship to the measurement errors of SMBH masses of the AGNs. This is also the reason why the scatters in Figures \ref{fig:M_M_RM_sep_mf} and \ref{fig:M_M_se} are larger than those in Figure 6 of \cite{Du2019}. Furthermore, it is important to emphasize that these estimators are only based on very limited sample in the present paper. It could be possibly necessary to apply the correction to the average virial factors (0.3 dex for CB and 0.2 dex for total) as discussed in Section \ref{sec:comparison}. A more detailed BLR model is also required to fully comprehend these correlations in the future.

\section{Summary}
\label{sec:summary}

We employed the latest nearby RM sample and the $M_{\bullet}-L_{\rm V,bul}$ relations of inactive galaxies in \cite{KH13} to calibrate the virial factors. We classified the RM objects into the PB and CB samples based on their high-resolution {\it HST} images, and calibrated the virial factors for the two individual samples and for the total PB$+$CB sample of the RM AGNs. We found that the virial factors have large intrinsic scatters ($0.5 \sim 0.6$ dex). Then we systematically investigated the scatters of the virial factors and found strong anti-correlations between the virial factors and the line widths or line profiles from the mean or rms spectra for both of the PB and CB samples. The physical driver of these relationships is probably the inclination or opening angle of BLR. Based on the correlations between the virial factors and the spectral properties, we established relationships to get the corrected virial factors which can provide more accurate RM masses. At the end, we tentatively established new single-epoch mass estimators for large AGN samples by taking into account the dependencies of the virial factors on the line widths and line profiles.

\begin{acknowledgments}

We thank C. Hu, D.-W. Bao, and S.-S. Li for their assistance in organizing the mean and rms spectra used for measuring the spectral properties, and all the members of the IHEP AGN Group for their helpful discussions. We acknowledge the support by National Key R\&D Program of China (grants 2021YFA1600404, 2023YFA1607903), by the National Science Foundation of China through grants {NSFC-12333003, -12022301, -11991051, -11991054, -11873048, -11833008}, by Grant No. QYZDJ-SSW-SLH007 from the Key Research Program of Frontier Sciences, Chinese Academy of Sciences (CAS), and by the China Manned Space Project CMS-CSST-2021-A06. 

Some of the data presented in this article were obtained from the Mikulski Archive for Space Telescopes (MAST) at the Space Telescope Science Institute. The specific observations analyzed can be accessed via \dataset[doi: 10.17909/bf5m-v928]{https://doi.org/10.17909/bf5m-v928}
\end{acknowledgments}

\software{\texttt{GALFIT} \citep{Peng2010}, \texttt{TinyTim} \citep{Krist2011}, DrizzlePac \citep{DHbook}, \texttt{DASpec} \citep{Du2024}}

\clearpage
\appendix 

\section{Notes for Individual Objects}
\label{app:notes}

Here, we give some comments on the host decomposition and bulge classification of each individual object.

{\it Mrk 335.} We obtained a B/T of 0.26, which falls between the values reported by \citet{Kim2017} (0.21) and \citet{Bentz2009a} (0.39).

{\it PG 0007+106.} We can see spiral arms (or tidal structures) in the UVIS1/F547M image. We fitted the host galaxy with hybrid of spiral arms and a disk, and obtained B/T = 0.29 and $n = 1.8$. \citet{Bentz2018} reported B/T of $\sim$ 0.16, 0.28, 0.49 (for their inner bulge, bulge, bulge+inner bulge, respectively) and $n = 1.1$. Considering the relatively high B/T, we classified this target as a CB galaxy.

{\it I Zw1.} Compared with the F105W images, its F438M images contain less signal from the host galaxy. If we released the \sersic\ index as a free parameter in the fitting, the \texttt{GALFIT} model tended towards a nonphysical model. To address this, we fixed $n = 1.73$ as provided by \citet{Huang2019}. This results in a B/T of 0.31, which is smaller than the value reported by \citet{Huang2019} (B/T = 0.52). In addition, the F105W and F814W images revealed clear spiral structures that are barely visible in F438W images. As a result, both B/T values (0.31 or 0.52) should be considered as upper limits. Finally, we classified this galaxy as a PB one.

{\it PG 0052+251.} The HRC/F550M image reveals a faint disk component, while WFPC2/F675W image shows a flat disk and a ring-like structure. \citet{Kim2017} fitted this target with a fixed $n = 4$ bulge and a flat disk ($n \sim 0.2$). We obtained $n \sim 2.1$, however our B/T $\sim$ 0.75 is consistent with \citet{Kim2017}. Following \citet{Ho2014}, we classified this object as a CB galaxy. 

{\it Fairall 9.} There is a circumnuclear ring in the central region of this object \citep{Bentz2009a, Kim2017}. We found it is necessary to add an extra component with a radius of $\sim 0.1^{\prime\prime}$ (3 pixels) to compensate the mismatch of the PSF. We calculated the PSF magnitude from its 5100\AA\ flux in the spectroscopic observation, and obtained a consistent value ($\sim$ 15) as the PSF magnitude from our \texttt{GALFIT} fitting ($\sim$ 14.8). The parameters of our bulge and disk components are similar to those of \citet{Bentz2009a}.

{\it Mrk 590.} We obtained ${\rm B/T} = 0.34$ and $n = 1.33$. B/T falls within the range between \citet{Kim2017} (0.26) and \citet{Vika2011} (0.48), and is consistent with the result in \citet{Bennert2010}. Additionally, we note that the size of galaxy exceeds the FoV, thus B/T should be thought as an upper limit. Therefore, we classified this target as a PB galaxy.

{\it Mrk 1044.} It is an S-type galaxy (HyperLeda and \citealt{Xivry2011}). We can see faint spiral structures from its {\it HST} image and an obvious ring in the \texttt{GALFIT} residuals. We obtained a PB (${\rm B/T} = 0.24$ and $n = 1$) similar to the result in \citet{Kim2017}

{\it 3C 120.} It is an S0 galaxy with a tidal arm (HyperLeda and \citealt{de1991third}). We adopted a bulge + disk model in our fitting, and obtained B/T = 0.16 and $n = 0.7$, which are consistent with the results in \citet{Bentz2009a} (B/T = 0.2 and $n = 0.67$) and \citet{Kim2017} (B/T = 0.2). As mentioned in \citet{Ho2014}, a model with a single host component can not be excluded. In accordance with them, we masked the tidal arm, adopted the single-component model, and classified this target as an elliptical.

{\it Ark 120.} The HRC image only covers the central region of the host galaxy. Its host galaxy can extends over $\rm \sim 1'$ (the HRC FoV is only $\rm \sim 25''$). The inner ring/spiral is masked during the fitting. We obtained a B/T of 0.35, falling between the results reported by \citet{Bentz2009a} (B/T = 0.49) and \citet{Kim2017} (B/T = 0.25). Despite its \sersic\ index ($n=1.73$) slightly smaller than the criteria, considering the large B/T, we still classified this target as a CB galaxy \citep{Ho2014}.

{\it UGC 3374.} This target is a bar-dominated spiral galaxy with relatively faint spiral arm and disk (HyperLeda, \citealt{de1991third}, and \citealt{Veron2006}). We classified this target as a CB galaxy since the steep gradient of the surface brightness of the bulge component.

{\it Mrk 6.} We masked the dust lane in the core region during the fitting.

{\it Mrk 374.} It is identified as a barred spiral galaxy according to \cite{Contini1998} and \cite{Robinson2019}. The bar component appears to be more compact with a higher S\'{e}rsic parameter ($n \sim 1$) compared to the regular bar structures ($n \sim 0.5$).

{\it Mrk79.} The host galaxy exceeds the FoV of the HRC image, and only the central region is covered. By fixing the bar/disk S\'{e}rsic index to 0.5, we obtained ${\rm B/T} = 0.27$ and $n = 3$. Our B/T is smaller than that reported in \citet{Kim2017} (0.37) and \citet{Bennert2010} (0.41), but larger than the result in \citet{Bentz2009a} (${\rm B/T} = 0.19$).

{\it Mrk 382.} It is classified as a barred spiral galaxy. Its nuclear region in the {\it HST} image is saturated. Though we masked out the saturated pixels, \texttt{GALFIT} still tends to fit the central region with a bright and compact S\'{e}rsic component. Therefore, we fixed the PSF magnitude to 16.3, a value simply calculated from 5100\AA\ flux. The small FoV also suggests that the ${\rm B/T} = 0.18$ obtained here may be an upper limit. Consistent with \citet{Kim2017}, we classified this target as a PB galaxy.

{\it PG 0804+761.} \citet{Bentz2009a}, \cite{Kim2017}, and \citet{Bennert2010} reported an exponential bulge profile. We obtained a slightly steeper profile ($n = 1.2$). \citet{Guyon2006} reported this target as bulge-dominated, and the high contrast between the AGN and the host made it difficult to add an additional disk profile in the fitting. Thus, we also classified this target as an elliptical \citep{Ho2014}.

{\it NGC 2617.} HyperLeda\footnote{\url{http://atlas.obs-hp.fr/hyperleda/}} suggests that this is a barred galaxy. However, in contrast, \citet{ned95galcls} classified its morphology as a type of Sc, indicating no bar exists. After visually inspecting the data, we prefer to conclude that there is no obvious bar structure. We also tested the fitting with a bar-added model, and the roundness ($b/a = 0.75$) of the bar component suggests that it is more likely a bulge. Finally, we employed a bulge+disk model and obtained ${\rm B/T} = 0.25$ and $n = 6.67$.

{\it PG 0844+349.} Due to the low SN ratio of the ACS/HRC/F550M image, we utilized ACS/WFC1/F625W image instead. The host galaxy exhibits complex spiral structures. \citet{Kim2017} obtained ${\rm B/T} = 0.64$. We fixed the bulge S\'{e}rsic index as 4, and obtained a ${\rm B/T}$ (0.64) very close to that reported by \citet{Kim2017}.

{\it Mrk 704} is more likely to be a barred spiral galaxy according to \cite{Malkan1998}. The presence of the bar is necessary in the \texttt{GALFIT} fitting in the present paper. \cite{Xivry2011} reports ${\rm B/T} = 0.43$ and $n = 2.88$, while our result gives a flat bulge component with ${\rm B/T} = 0.35$ and $n = 1.6$. This difference may be primarily due to different {\it HST} data used in the two analyses. The WFC3/UVIS/F547M image used in our analysis isn't saturated. Taking the presence of the bar into consideration, we classified the bulge as a PB.

{\it Mrk 110.} The low S/N ratio makes it difficult to discern any visible host structure at the outer region of the HRC image. Moreover, the descent gradient indicates that the host galaxy exceeds the field of view (FoV) of the HRC image. We obtained ${\rm B/T} = 0.28$ and $n = 2$, which is a much brighter and steeper result compared to that of \citet{Kim2017} (${\rm B/T} = 0.2$ and $n = 1.64$). For consistency, we classified this target as a CB galaxy.

{\it PG 0923+201.} This target appears compact in the image, and \citet{Hamilton2008} reports it as an elliptical galaxy. \citet{Kim2017} also fitted an elliptical model to its host image by fixing $n = 4$. Following them, we fixed $n = 4$ in out fitting process.

{\it PG 0923+129.} HyperLeda reports it as an S0 galaxy with a clear circumnuclear ring. We obtained a B/T of 0.3 and $n = 0.83$. Although the B/T does not meet the PB criteria \citep{Fisher2008}, the exponential-like profile led us to classify the this object as a PB galaxy.

{\it SDSS J093302.68+385228.0.} HyperLeda classifies this target as an elliptical, however, we can see clear structure in the WFC1/F615W image - a spiral arm extending to the central region and forming a circumnuclear ring, which is different from the traditional elliptical characteristic \citep{Fisher2008}. The central region of the image is saturated, and we masked this region in fitting process. Our fitting reports ${\rm B/T} = 0.14$ and $n = 0.92$, leading us to classify the bulge as a PB.

{\it PG 0953+414.} The host galaxy is faint and compact \citep{Guyon2006}. As discussed in \citet{Ho2014}, it is challenging to robustly constrain the \sersic\ index for this object. Therefore, following \citet{Bentz2009a} and \citet{Kim2017},  we fixed the \sersic\ index as 1.5 during the \texttt{GALFIT} fitting.

{\it NGC 3227.} \citet{KH13} reported this target as a PB galxy with ${\rm B/T} = 0.1$. The central region is heavily affected by dusk lanes. We masked the dust lanes in the central region and obtained ${\rm B/T} = 0.34$ and $n = 1.42$. Except for the S\'{e}rsic index, our bulge model is in good agreement with \citet{Kim2017}. Our B/T should also be considered as an upper limit due to the small FoV of the HRC. 

{\it Mrk 142.} \citet{Bentz2013} suggested this object does not have a detectable bulge, however the existence of a bar is clear. On the other hand, \citet{Du2014} suggested that there is a bulge component in the host. When employing three components in the fitting (bulge, bar, and disk), we found evidence of a PB (${\rm B/T} = 0.2$ and $n = 0.85$) in this object. Additionally, the CCF time lag of the RM campaign in \citet{Bentz2009b} is reported to be 2.87 days with respect to the V-band continuum, which deviates significantly from the re-analysis in \citet{Li2013} based on a Bayesian approach ($\tau_{\rm H\beta} = 15.3 \pm 2.7$ days). We employed the result from \citet{Li2013} in the present paper.

{\it PG 1100+772.} The central region of the F814W image is saturated.

{\it NGC 3516.} Consistent with \citet{Kim2017}, we also found a PB component in the fitting ($n = 1.7$). However, we obtained a large ${\rm B/T}$ of 0.6, which falls within the B/T range (0.52-0.86) reported by \citet{Bentz2009a}. Due to the small FoV of the HRC image, the B/T here should be an upper limit.

{\it SBS 1116+583A.} Our fitting reports B/T = 0.14 and $n = 1.6$, leading us to classify this target as a PB \citep{Ho2014}. The features of secular evolution in the image also supports our classification.

{\it Arp 151.} Also known as Mrk 40. It is an S0 galaxy (HyperLeda and \citealt{de1991third}) with no obvious spiral-like structures, and is located in an interacting system. \citet{Kim2017} obtained ${\rm B/T} = 0.25$ and $n = 4.5$ for this object. Our analysis yielded a similar $n = 4.4$, but a notably larger ${\rm B/T} = 0.5$. The discrepancy may be attributed to the PSF saturation in the WFPC2 image used in \cite{Kim2017}.

{\it NGC 3783.} It is a barred spiral galaxy with a ring-like structure surrounding the bulge. Our result indicates the presence of a PB ($n = 1.6$), which is consistent with \citet{Kim2017} ($n = 1.87$).

{\it MCG+06-26-012.} This is a barred spiral galaxy. \citet{Schade2000} decomposed the host components and reported a R-band ${\rm B/T} = 0.27$, consistent with the results of \citet{Kim2017}. However, their fitting region cannot cover the entire disk, so their B/T should be considered as an upper limit. We obtained a flatter and fainter bulge component ($n = 1.03$ and ${\rm B/T} = 0.074$). Considering the existence of a bar, we classified the bulge as a PB.

{\it UGC 06728.} This is a barred lenticular galaxy \citep{Bentz2021}, and our fitting gives a result similar to that of \citet{Bentz2021}. We obtained a slightly larger bulge ($R_{\rm e} = 2.96^{\prime\prime}$ and $B/T = 0.16$) than their result ($R_{\rm e} = 2.11^{\prime\prime}$ and B/T = 0.11).

{\it Mrk 1310.} HyperLeda and the SDSS-GZ2 catalogue \citep{sdssgz2} classify this target as an elliptical, however, a stellar ring is evident in its \texttt{GALFIT} residual from our fitting. \citet{Bentz2013} suggested this target as a ringed spiral galaxy. After masking the stellar ring, we obtained a ${\rm B/T} = 0.11$, consistent with that in \citet{Kim2017} and smaller than the result in \citet{Schade2000} (${\rm B/T} = 0.23$). The bulge can be modeled by a standard de Vaucauleurs profile ($n = 4$). Following \citet{Ho2014}, we classified this objects as a CB galaxy.

{\it NGC 4051.} We obtained $n = 1.2$ and ${\rm B/T} = 0.38$. Considering the small FoV, the B/T is likely overestimated. Our result is also in good agreement with that of \citet{Kim2017}.

{\it NGC 4151.} The host galaxy extends beyond the FoV of the HRC. Following \citet{Kim2017}, we employed the standard bulge/disk model ($n = 4/1$). The residual image shows a faint ring, which may result from the small mismatch between model and the image. We obtained a B/T of 0.53, which falls within the range between \citet{Kim2017} (0.32) and \citet{Bentz2009a} (0.72).

{\it PG 1211+143.} Due to the low SN ratio of the host galaxy in the HRC/F550M image, we employed WFC1/F625W image instead. We found a bar-like structure near the PSF, which was masked during the fitting. We obtained a B/T of 0.37, similar to the results of \citet{Bentz2009a} and \citet{Bennert2010}, and less than the result in \citet{Kim2017}. The bulge parameter still falls within the CB region \citep{Ho2014}.

{\it Mrk 202.} The central region of the host galaxy has a circumnuclear ring, which we masked out during the \texttt{GALFIT} fitting. We obtained ${\rm B/T} = 0.44$ and $n = 3.6$, indicating the presence of a CB in the host galaxy. \citet{Bentz2013} employed an extra component to fit the bulge ($+$ ring) in their fitting. Additionally, \citet{Ho2014} classified this target as PB galaxy mainly due to the existence of the circumnuclear ring. Despite our fitting parameter falling within the CB range, we still classified this target as a PB galaxy. 

{\it NGC 4253.} Different from \citet{Kim2017}, we employed an unsaturated WFC3/UVIS image in the \texttt{GALFIT} fitting. We found that an extra component is necessary to compensate the inaccuracy of the PSF model \citep{Bentz2013}. Our result is similar to those reported in \citet{Bentz2013}, \cite{Kim2017}, and \cite{Xivry2011}.

{\it Mrk 50} has been classified as an S0 galaxy by \citet{Malkan1998} and \citet{Veron2006}, which differs from the classification by \citet{Kim2017}. In our analysis, an unsaturated WFC3/UVIS/F547M image was employed in the \texttt{GALFIT} fitting. We found no obvious features related to secular evolution in the \texttt{GALFIT} residual, and a CB component worked well ($n = 4$ and ${\rm B/T} = 0.26$). Therefore, classifying this object as a CB galaxy is reasonable. Single epoch properties were measured from the scanned figures in \citet{Barth2011}.

{\it PG 1226+023.} There are some ring-like structures in the residuals in both our result and that of \citet{Kim2017}.

{\it PG 1229+204.} In comparison with \citet{Kim2017}, we included an additional bar component in the fitting \citep{Hamilton2008, Ho2014}. We obtained B/T = 0.21 and $n = 1.16$, consistent with the previous works in \cite{Bentz2009a} and \cite{Bennert2010}.

{\it NGC 4593.} There is an apparent circular dust lane around the nucleus, and we masked it in the \texttt{GALFIT} fitting. The object is a barred spiral galaxy, and the WFC3/UVIS FoV can covers most regions of this target. Our fitting result gives that the bulge is flat and faint (${\rm B/T} = 0.16$ and $n = 0.96$), which is consistent with \citet{Kormendy2006, Kim2017, Ho2014}. We also classified it as a PB galaxy.

{\it IRAS F12397+3333.} The central region of its {\it HST} image is saturated (around the central $\sim$ 5 pixels), and is excluded during the fitting. Although HyperLeda reports that this object has an S-type morphology, its spiral structures (or other features related to secular evolution) are not obvious. Our bulge parameters (${\rm B/T} = 0.65$ and $n = 4.1$) are consistent with those of \citet{Mathur2012} (${\rm B/T} = 0.59$ and $n = 3.45$). Therefore, we classified it as a CB galaxy.

{\it NGC 4748.} An S0 galaxy (2MASX J12521292-1324388) overlays on its host, with a distance between them of $\rm \sim$ 0.7 Mpc. However, no interaction signature can be seen on the host morphology. Similar to \citet{Xivry2011}, we obtained ${\rm B/T} = 0.14$ and $n = 1.6$. Considering the existence of a bar and stellar ring, we also classified this target as a PB, which is consistent with \cite{Ho2014}.

{\it PG 1307+085.} The central region of the image suffers from saturation. HyperLeda and \citet{Ho2014} classified it as an elliptical galaxy.

{\it PG 1310-108.} The central region of the image is saturated, and we masked this area during the fitting. To minimize the systematic error, we fixed the PSF at 15.78 mag (derived by converting the continuum flux from \citealt{Bao2022}). Subsequently, we obtained B/T = 0.2 and $n = 2.6$. Considering the steep surface brightness gradient around the nucleus,we classified this target as a CB. 

{\it MCG -6-30-15.} The UVIS image reveals a central dust lane, which could potentially influence the fitting results. We employed a mask for the dust lane following the strategy of \citet{Hu2016} and obtained similar bulge properties: $n = 0.83$ and B/T = 0.04 (\citealt{Hu2016} provided B/T = 0.06 and $n = 1.29$). \citet{Hu2016} and our fitting both classified this galaxy as a PB.

{\it NGC 5273.} The WFPC2 FoV is insufficient to fully cover this target. Our fitting result indicates the presence of a CB (${\rm B/T} \sim 0.34$ and $n \sim 4.3$).

{\it IC 4329A.} It is an edge-on S0-a galaxy (see HyperLeda, \citealt{de1991third}, and \citealt{Khorunzhev2012}). Its host galaxy is heavily affected by the dust lanes, thus we masked out those dust lanes in our fitting. We obtained a flatter and fainter bulge component (B/T = 0.19 and $n = 0.49$) than that in \citet{Kim2017} (B/T = 0.28 and $n = 1$). Considering the flat morphology, we classified this object as a PB galaxy. It should be  noted that the dust lanes heavily influence the measurements of the bulge morphology and luminosity, thus we didn't employed this target for the analysis in Section \ref{sec:analysis}. 

{\it PG 1351+695.} This is an S0 galaxy (HyperLeda, \citealt{de1991third}, and \citealt{Bentz2009a}). Some dust lanes overlay the central region \citep{Kim2017}. We masked out the dust lanes in the \texttt{GALFIT} fitting and obtained a similar bulge component (B/T = 0.21 and $n = 1$) to \citet{Kim2017} (B/T = 0.21, $n = 1.3$). 

{\it PG 1351+640.} In contrast to the bulge $+$ disk models used in \citet{Kim2017}, we fitted this target with a single $n = 2.7$ \sersic\ profile. \citet{Kim2017} reported B/T $\sim$ 0.92, which is also bulge-dominated. 

{\it NGC 5548.} The bulge S\'{e}rsic index is found to be $n = 3.9$, with a corresponding B/T ratio of 0.36. These results are in good agreement with the results reported by \citet{Kim2017} (${\rm B/T} = 0.39$ and $n = 4.13$). Moreover, our measured B/T ratio is significantly smaller than the ground-based R-band results obtained by \cite{Virani2000} (${\rm B/T} = 0.57$).

{\it PG 1426+015.} \citet{Kim2017} and \citet{Guyon2006} classified this target as an elliptical. Our fitting employed bulge and disk components and yielded B/T $\sim$ 0.7, slightly larger than the value reported in \citet{Schade2000} (B/T $\sim$ 0.54).

{\it Mrk 817.} There are clear bar and spiral components in the image \citep{Ho2014,Deo2006}. \citet{Kim2017} employed a bar component in their fitting, however, \citet{Bentz2009a} and \citet{Bennert2010} did not. We found that the bar component did not significantly affect the bulge profile, and thus still included it in our fitting. Additionally, the host galaxy exceeded the FoV of the HRC image. 

{\it PG 1440+356.} \citet{Kim2017} suggested that this object is a PB galaxy. Using the UVIS/F547M image, we obtained a slightly smaller and flatter PB (B/T = 0.15 and $n = 0.31$) compared to the result in \citet{Kim2017} (B/T = 0.3 and $n = 0.64$). The UVIS/F547M image in our fitting shows a more significant disk component than the HRC/F625W image used by \cite{Kim2017}. Both \citet{Kim2017} and our fittings indicate the bulge is a PB.

{\it PG 1448+273.} It is a spiral galaxy (HyperLeda and \citealt{Petrosian2008}). Our fitting gives $n = 4$ and B/T = 0.35. Therefore, we classified this object as a CB galaxy.

{\it Mrk 1511.} It is classified as a barred spiral galaxy according to HyperLeda and \cite{Xivry2011}. The WFC3/UVIS/F547M image is not saturated, allowing for a more accurate fitting of the central region compared to \citet{Kim2017}. Our result suggests a small PB with ${\rm B/T} = 0.02$ and $n = 1.75$.

{\it PG 1534+580.} Although HyperLeda and \cite{de1991third} both report that this object is an elliptical, the UVIS1/F547M image shows clear faint spiral/disk structures. \citet{Kim2017, Bentz2009a, Xivry2011} also suggest that using two host components in the fitting is better. We obtained B/T = 0.34, which is consistent with the result in \citet{Kim2017} (B/T $\sim$ 0.42).

{\it PG 1535+547.} This object has a barred spiral galaxy, however the spiral structures are not bright. By modeling the bar/disk component with a $n = 1$ \sersic\ profile, we obtained B/T = 0.2 and $n = 0.6$. Therefore, we classified this object as PB galaxy.

{\it Mrk 493.} The nuclear region is saturated, and we obtained similar results compared to \citet{Kim2017} and \citet{Wang2014a}, confirming the presence of a PB in this object. 

{\it PG 1613+658.} This target is an elliptical galaxy according to HyperLeda and the previous studies \citep{Bentz2009a, Guyon2006}. Following \citet{Kim2017}, we fitted the spheroid with the Fourier modes in \texttt{GALFIT}.

{\it PG 1617+175.} It is an elliptical galaxy \citep{Guyon2006}. We employed the WFC1/F625W image which has higher SN ratio than that of the PC/F547M image. There is some ring-like structure in the residual image. 

{\it PG 1626+554.} We used the UVIS1/F475M image in our \texttt{GALFIT} fitting because of its optical central wavelength. The WFC3/F105W image suggests that two components are needed for a good fitting to its host galaxy. We fixed the structure parameters (e.g., $n$, $R_{\rm e}$) as those derived from the WFC3/F105W image. The resulting B/T = 0.25 and $n = 4$ lead us to classify the bulge as a CB.

{\it PG 1700+518.} In contrast to \citet{Kim2017}, we set the \sersic\ index of the bulge as a free parameter, and obtained a single and flat component for the host galaxy. Some structure located at the north side of the nucleus seems weird. We masked this structure in our fitting. Similar to \citet{Ho2014, Guyon2006, Bentz2009a}, we classified this target as an elliptical galaxy.

{\it 3C 382.} It is a compact object, and was classified as an elliptical in HyperLeda. \citet{Kim2017} also classified this object as an elliptical galaxy.

{\it 3C 390.3.} Its compact galaxy makes it challenging to decompose the host components. Our 1D and 2D results suggest that it is an elliptical, which is in agreement with the conclusions in \citet{Kim2017} and \citet{Bennert2010}. The F550M total magnitude from our fitting is 15.2, almost the same as the V-band magnitude of 15.3 reported in \cite{3c390vband} and \cite{3c390ana}.

{\it Zw 229-015.} Also known as Z~229-15. HyperLeda reports this target as a barred spiral galaxy. \citet{Bentz2018} reported B/T = 0.143 and $n = 1.1$. We obtained a slightly flatter and fainter bulge (B/T = 0.09, $n = 0.77$). Both \cite{Bentz2018} and our fitting result classified this object as a PB galaxy.

{\it NGC 6814.} It is a barred spiral galaxy (HyperLeda and \citealt{de1991third}). Our fitting reports a PB with $n = 1.5$ and ${\rm B/T} = 0.03$. 

{\it Mrk 509.} As discussed in \citet{Kim2017}, the presence of non-elliptical features suggests that Mrk 509 may not to be an elliptical galaxy. Our fitting with two host components resulted in a B/T of 0.58 and $n$ of 4, indicating that the bulge is dominated  (consistent with the ground-based observation in \citealt{Ho2014}).

{\it PG 2130+099.} This is a spiral galaxy \citep{Guyon2006, de1991third}. We employed the WF2/F555W image which has a much higher SN ratio than the HRC/F550M image, though the PSF is saturated. We fixed the PSF flux as 14.5 mag derived from the HRC/F550M (unsaturated), since F555W and F550M image have the similar central wavelengths. We obtained B/T = 0.31 and $n = 0.43$, which are consistent with the results in \citet{Kim2017} and \citet{Bentz2009a}. Therefore, we classified this object as a PB galaxy. 

{\it NGC 7469.} The inner spiral/ring structures  were masked out during the fitting, resulting in a lower ${\rm B/T} = 0.18$ and larger $n$ of approximately $1.7$ for the bulge component, in comparison with the results in \citet{Bentz2009a} (${\rm B/T} \sim 0.7$, $n \sim 1.3$). Both the parameters and structures suggest the presence of a PB in this object.

\section{Comparison of \texttt{GALFIT} results with previous works\label{app:gal_comp}}

\begin{figure}
    \centering
    \includegraphics[width=0.96\textwidth]{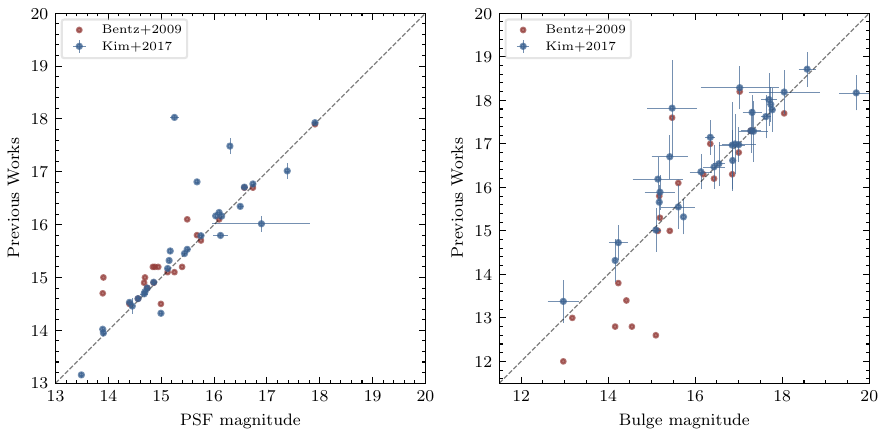}
    \caption{Comparisons of the PSF \& bulge magnitudes between our \texttt{GALFIT} fitting and those from the previous works in \cite{Bentz2009a} and \cite{Kim2017}.}
    \label{fig:comp_galfit_results}
\end{figure}

Here we show some comparisons between our \texttt{GALFIT} results and those from the previous works in \cite{Bentz2009a} and \cite{Kim2017}. The comparisons of the magnitudes of the PSF and bulge components are provided in Figure \ref{fig:comp_galfit_results}. In general, our fitting results are consistent with those from \cite{Bentz2009a} and \cite{Kim2017}. Most of the points are located around the diagonal lines, indicating the reliability of our \texttt{GALFIT} fitting and the host decomposition. Some outliers are due to the PSF saturation or different images used in the fitting.

\section{Correlations analysis of the virial factors based on total sample\label{app:tot}}

\begin{figure}
    \centering
    \includegraphics[width=\textwidth]{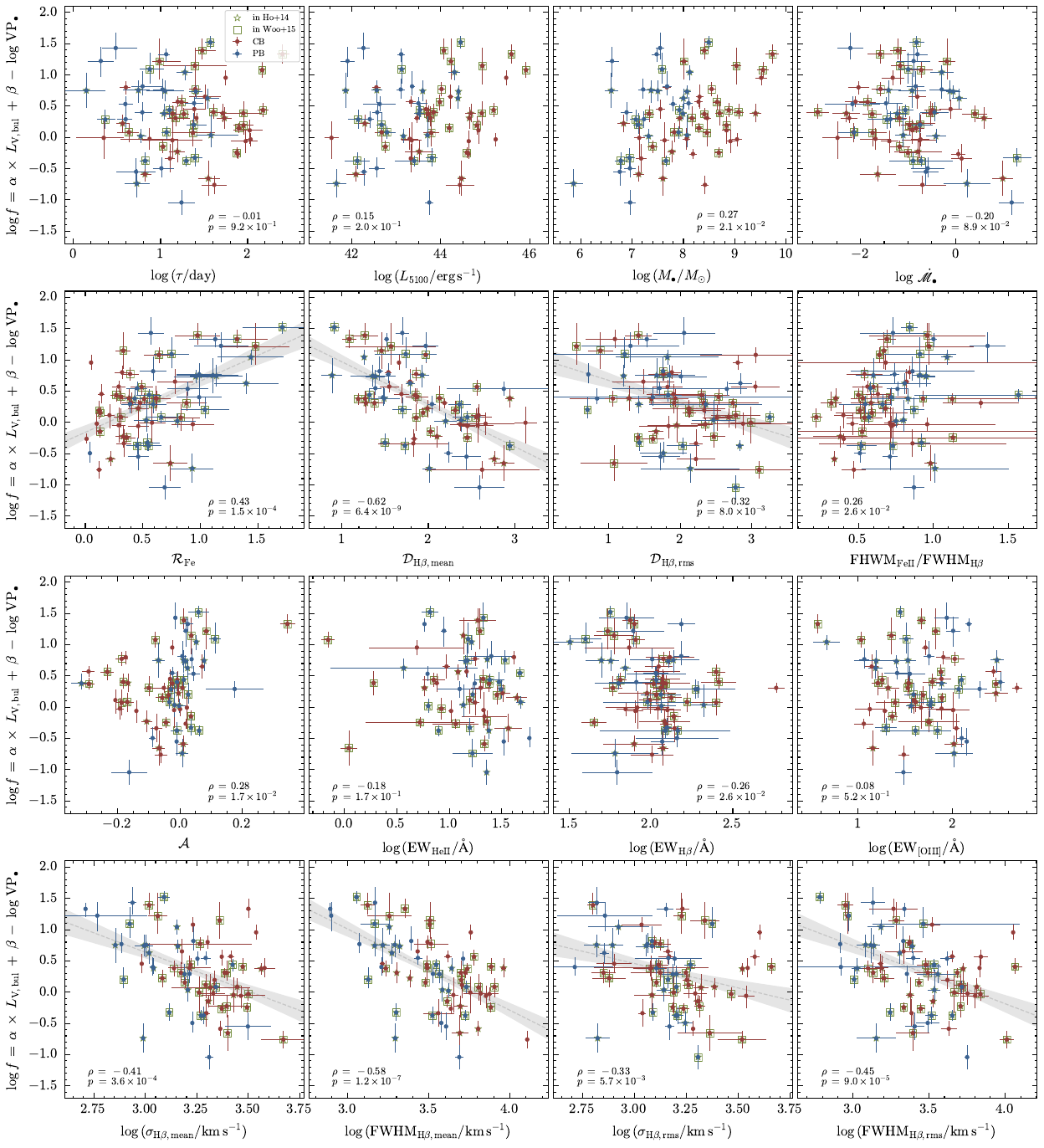}
    \caption{The correlations between $f_{\rm MF}$ and the spectral properties for the total-sample calibration (do not distinguish PB and CB). The meanings of the symbols, lines, and regions are the same as in Figure \ref{fig:devi_f_sep_mf}. }
    \label{fig:devi_f_tot_mf}
\end{figure}

\begin{figure}
    \centering
    \includegraphics[width=\textwidth]{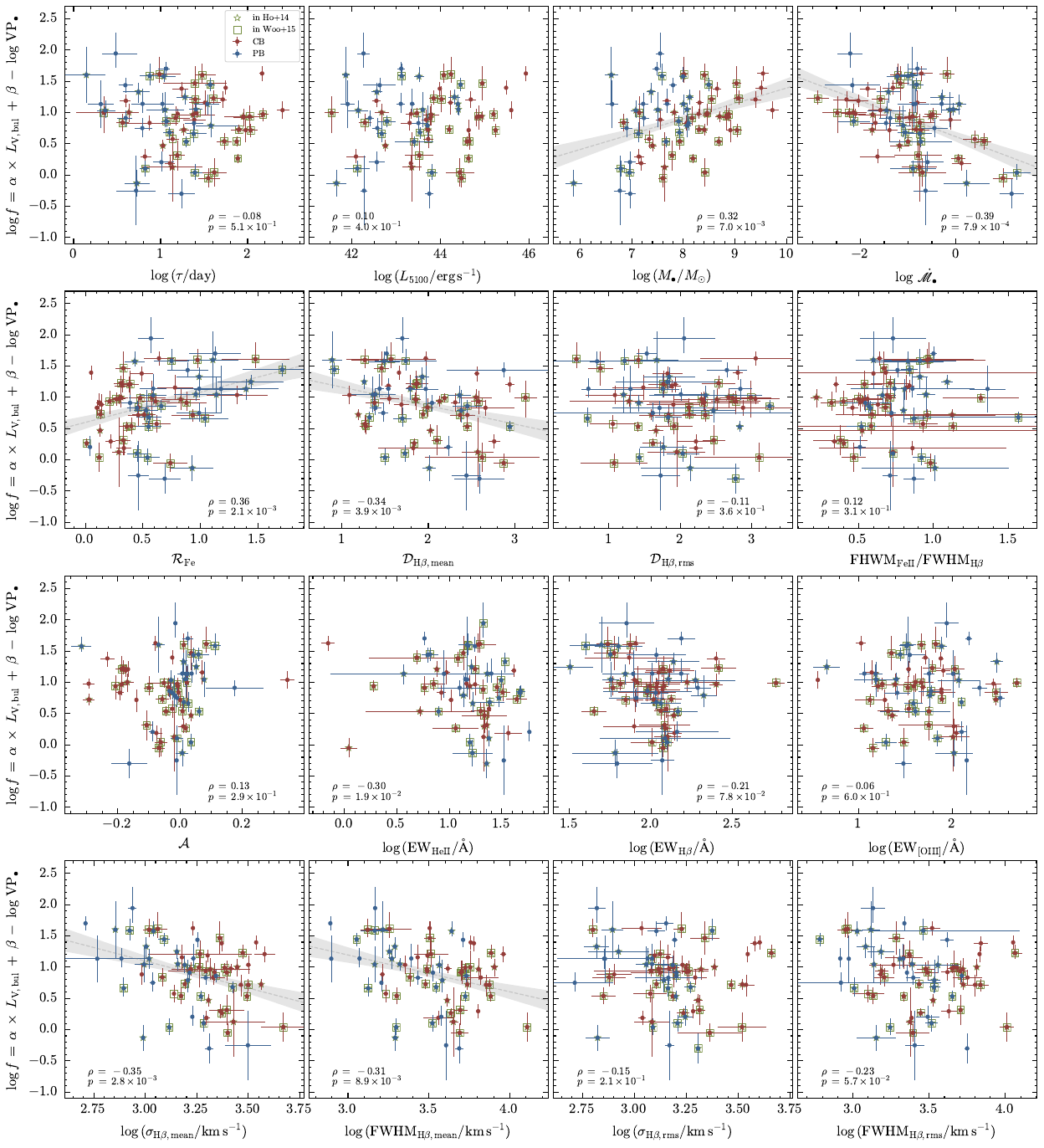}
    \caption{The correlations between $f_{\rm MS}$ and the spectral properties for the total-sample calibration (do not distinguish PB and CB). The meanings of the symbols, lines, and regions are the same as in Figure \ref{fig:devi_f_sep_mf}.}
    \label{fig:devi_f_tot_ms}
\end{figure}

\begin{figure}
    \centering
    \includegraphics[width=\textwidth]{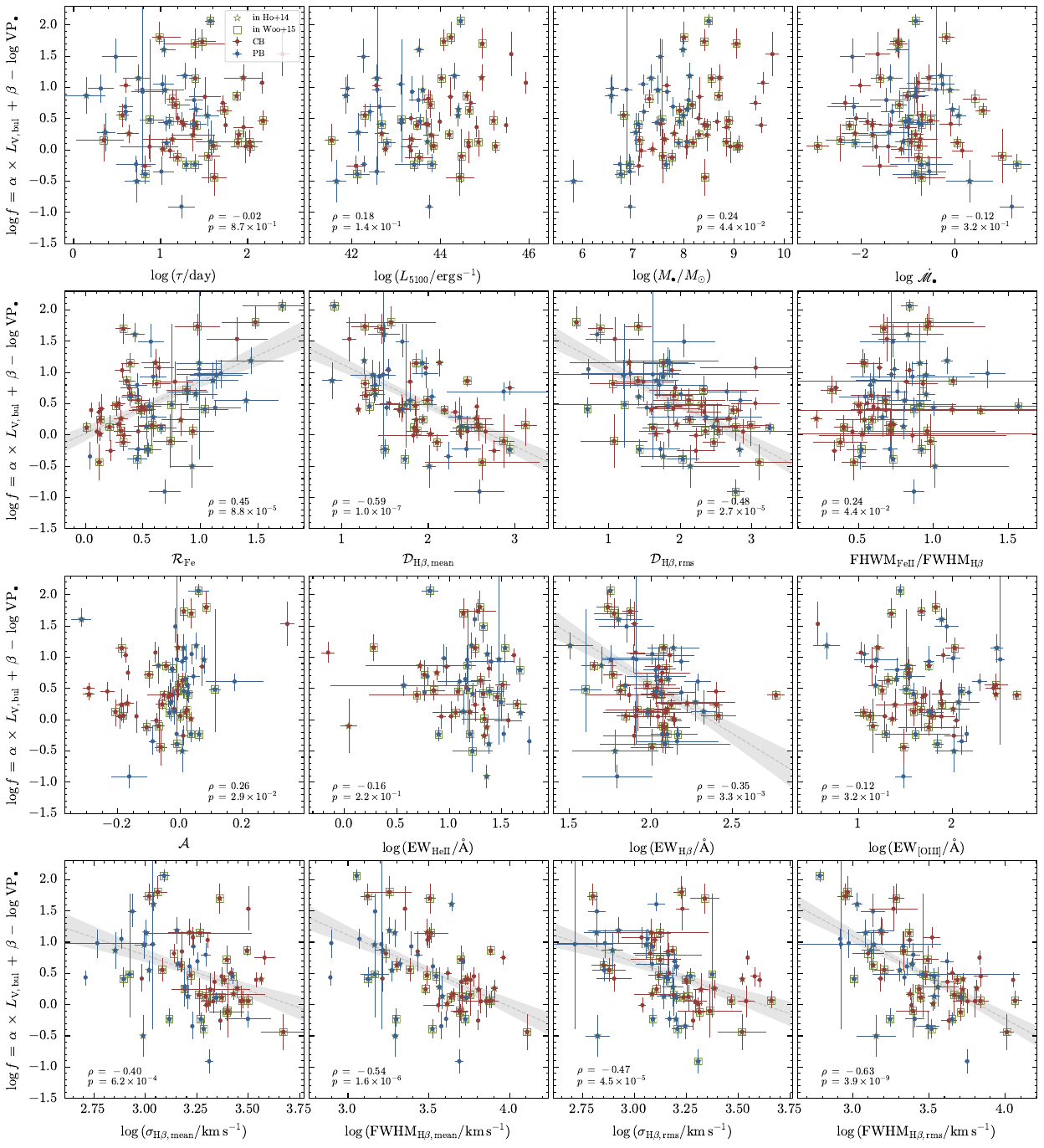}
    \caption{The correlations between $f_{\rm RF}$ and the spectral properties for the total-sample calibration (do not distinguish PB and CB). The meanings of the symbols, lines, and regions are the same as in Figure \ref{fig:devi_f_sep_mf}.}
    \label{fig:devi_f_tot_rf}
\end{figure}

\begin{figure}
    \centering
    \includegraphics[width=\textwidth]{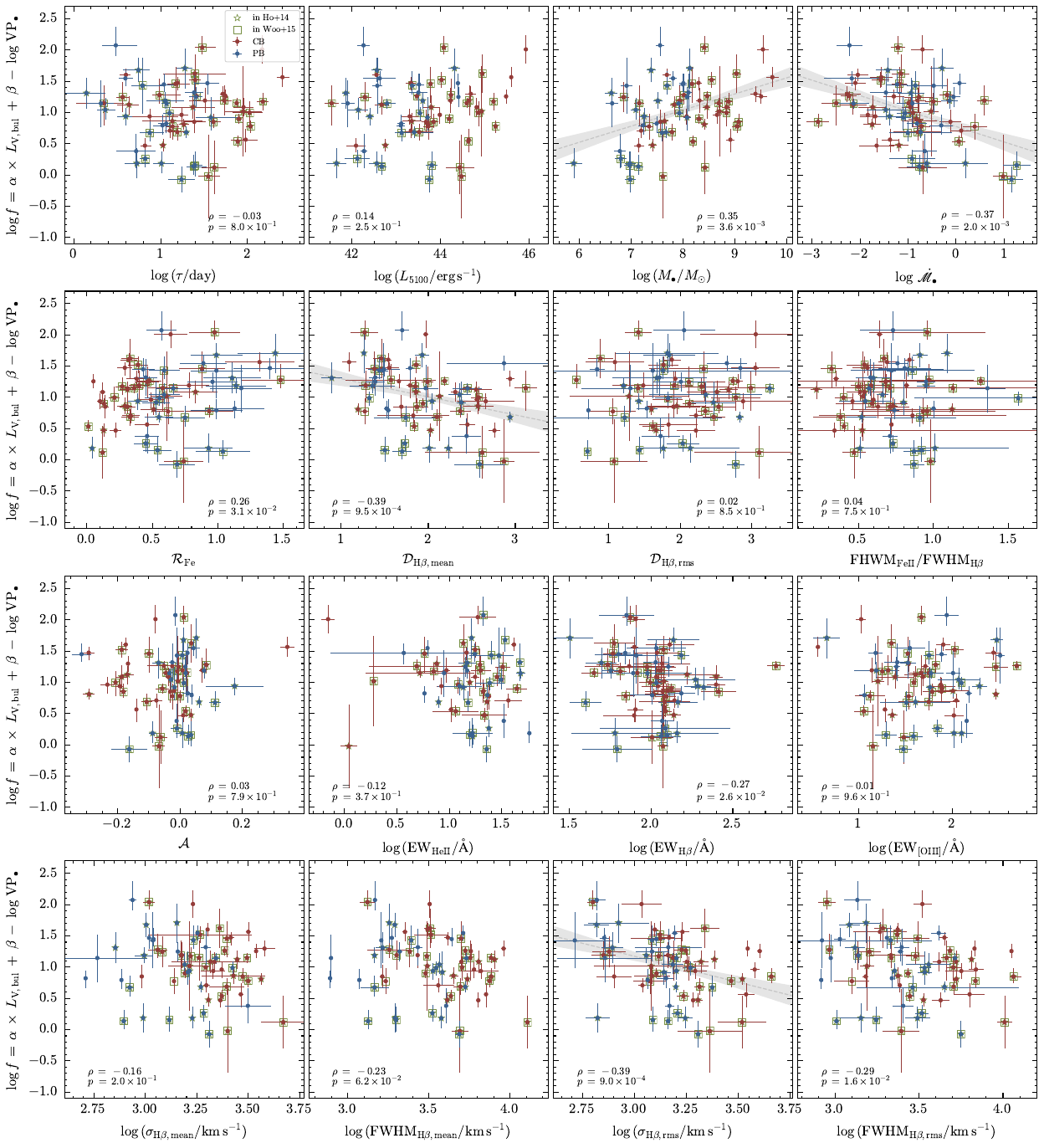}
    \caption{The correlations between $f_{\rm RS}$ and the spectral properties for the total-sample calibration (do not distinguish PB and CB). The meanings of the symbols, lines, and regions are the same as in Figure \ref{fig:devi_f_sep_mf}.}
    \label{fig:devi_f_tot_rs}
\end{figure}

For comparison, we plot the correlations between the deviations of the total-sample virial factors and the spectral properties of in Figures \ref{fig:devi_f_tot_mf}-\ref{fig:devi_f_tot_rs}.

\clearpage

\input{gal_rst.tex}

\clearpage

\startlongtable
\begin{deluxetable}{lcccccccc}
\tablecolumns{7}
\setlength{\tabcolsep}{3pt}
\tablecaption{Target Informations\label{tab:tar_info}}
\tabletypesize{\scriptsize}
\tablehead{
    \colhead{Name}  &
    \colhead{RA}    &
    \colhead{DEC}   &
    \colhead{$A_{\rm V}$}    &
    \colhead{$z$}     &
    \colhead{Inst./Aper./Filt.} &
    \colhead{Exp. Time} &
    \colhead{Dataset}
}
\startdata
Mrk 335&00:06:19.5&+20:12:10.5&0.11&0.026&ACS/HRC/F550M&2040&J9MU010\\
PG 0007+106&00:10:31.0&+10:58:29.5&0.313&0.087&WFC3/UVIS1/F547M&2236&ICNO010\\
PG 0026+129&00:29:13.7&+13:16:04.0&0.221&0.145&WFPC2/PC1/F547M&1445&U9MU520\\
I Zw1&00:53:34.9&+12:41:36.2&0.198&0.061&WFC3/UVIS2-M1K1C-SUB/F438W&340&IBY8070\\
PG 0052+251&00:54:52.1&+25:25:39.0&0.146&0.154&WFPC2/WF2/F675W&1991&U402220\\
Fairall 9&01:23:45.8&-58:48:20.8&0.081&0.046&ACS/HRC/F550M&1020&J8SC040\\
Mrk 590&02:14:33.6&-00:46:00.1&0.113&0.026&ACS/HRC/F550M&1020&J8SC050\\
Mrk 1044&02:30:05.5&-08:59:53.3&0.107&0.016&WFC3/UVIS1/F547M&725&IBGU100\\
\enddata
\tablecomments{$A_{\rm V}$ is the Galactic extinction at V-band \citep{SFD98}. ``Inst./Aper./Filt.'' is the instrument/aperture/filter employed in the {\it HST} observation used in the present paper. ``Exp. Time'' is the exposure time. The redshifts ($z$) are derived from NASA\/IPAC Extragalactic Database (\url{https://ned.ipac.caltech.edu/}).}
\end{deluxetable}


\startlongtable
\begin{deluxetable*}{lccccccc}
\tablecaption{Results of \texttt{GALFIT} Fitting\label{tab:galfit}}
\tablewidth{0pt}
\tabletypesize{\scriptsize}
\setlength{\tabcolsep}{3pt}
\tablehead{
\colhead{Name}				&
\colhead{Bulge Type}        &
\colhead{$\rm m_{\rm st}$}	&
\colhead{$R_{\rm e}$}		&
\colhead{$n$}				&
\colhead{$b/a$}			&
\colhead{$\rm P.A.$}		&
\colhead{Note}				
\\
\colhead{}					&
\colhead{$\rm (B/T)$}		&
\colhead{}					&
\colhead{$\rm (arcsec)$}	&
\colhead{}					&
\colhead{}					&
\colhead{(deg)}				&
\colhead{}
}

\startdata
Mrk 335&C&14.68$\pm$0.083&&&&&PSF\\
&0.26$\pm$0.045&16.85$\pm$0.304&0.42$\pm$0.31&4&0.74$\pm$0.178&12.69$\pm$37.113&Bulge\\
&&15.73$\pm$0.32&2.57$\pm$0.668&1&0.91$\pm$0.026&69.13$\pm$24.551&Disk\\
PG 0007+106&C&15.99$\pm$0.046&&&&&PSF\\
&0.29$\pm$0.022&17.42$\pm$0.12&0.34$\pm$0.088&1.85$\pm$0.59&0.67$\pm$0.085&34.81$\pm$5.219&Bulge\\
&&16.47$\pm$0.317&5.11$\pm$0.632&1&0.72$\pm$0.022&49.13$\pm$22.481&Disk, f\\
PG 0026+129&E&15.49$\pm$0.031&&&&&PSF\\
&1&17.31$\pm$0.218&3.79$\pm$1.764&4&0.75$\pm$0.044&161.74$\pm$3.33&Bulge, f\\
\enddata
\tablecomments{``B/T'' is the bulge-to-total luminosity ratio. ``$\rm m_{st}$'' is the apparent magnitude in ST magnitude system. ``$R_{\rm e}$'' is the effective radius. ``$n$'' is the \sersic\ indexes. ``$b/a$'' is the axial ratio. ``P.A.'' is the position angle. ``f'' in the last column denotes that the fourier mode is used in the fitting. ``C'', ``P'', and ``E'' represents the classifications of CB, PB and elliptical. The parameters without errors are fixed in the fitting. We note here that the uncertainties here include the error bars from the \texttt{GALFIT} and the uncertainties from the PSF mismatch (see Section \ref{sec:image_fitting}). The systematic errors resulting from the over- or under-estimation of the sky background (0.1 mag) and those caused by fixing the \sersic\ index (0.3 mag for some objects) are not included here, but are added in quadrature to the total error bars in the analysis in Section \ref{sec:analysis}. }
\end{deluxetable*}


\startlongtable
\begin{deluxetable*}{lccccccccc}
\tablecolumns{10}
\tablecaption{Measurements of RM campaigns\label{tab:rm_table}}
\tabletypesize{\scriptsize}
\setlength{\tabcolsep}{1pt}
\tablehead{
    \colhead{Name}                          &
    \colhead{$\rm FWHM_{H\beta, mean}$}     &
    \colhead{$\rm \sigma_{H\beta, mean}$}   &
    \colhead{$\rm FWHM_{H\beta, rms}$}      &
    \colhead{$\rm \sigma_{H\beta, rms}$}    &
    \colhead{$\rm \tau_{H\beta}$}           &
    \colhead{$\rm log\,\mathit{L}_{5100}$}  &
    \colhead{$\rm EW(H\beta)$}              &
    \colhead{$\rm EW([O\, III])$}              &
    \colhead{References}
    \\
    \colhead{}                              &
    \colhead{$\rm (km\,s^{-1})$}            &
    \colhead{$\rm (km\,s^{-1})$}            &
    \colhead{$\rm (km\,s^{-1})$}            &
    \colhead{$\rm (km\,s^{-1})$}            &
    \colhead{$\rm (days)$}                  &
    \colhead{$\rm (erg\, s^{-1})$}          &
    \colhead{(\AA)}                   &
    \colhead{$\rm (erg\, s^{-1})$}          &
    \colhead{}
}
\startdata
Mrk 335&$1748_{-120}^{+120}$&$1259_{-60}^{+60}$&$836_{-348}^{+348}$&$589_{-98}^{+98}$&$17_{-3.2}^{+2.5}$&$43.51_{-0.07}^{+0.07}$&$133.37\pm8.57$&$39.42\pm3.24$&8\\
&$2096_{-170}^{+170}$&$1470_{-50}^{+50}$&$1561_{-22}^{+61}$&$1287_{-33}^{+18}$&$8.7_{-1.9}^{+1.6}$&$43.69_{-0.06}^{+0.06}$&$110.5\pm22.3$&$31.8\pm2.4$&1,5\\
&$1792_{-3}^{+3}$&$1380_{-6}^{+6}$&$1629_{-145}^{+145}$&$917_{-52}^{+52}$&$16.8_{-4.2}^{+4.8}$&$43.76_{-0.06}^{+0.06}$&$119.7\pm23.3$&$36.9\pm1.9$&1,3,5\\
&$1679_{-2}^{+2}$&$1371_{-8}^{+8}$&$1375_{-357}^{+357}$&$948_{-113}^{+113}$&$12.5_{-5.5}^{+6.6}$&$43.84_{-0.06}^{+0.06}$&$111.2\pm21.1$&$31.1\pm1.6$&1,3,5\\
&$1724_{-236}^{+236}$&$1542_{-66}^{+66}$&$1025_{-35}^{+35}$&$1293_{-64}^{+64}$&$14.3_{-0.7}^{+0.7}$&$43.74_{-0.06}^{+0.06}$&$89.5\pm19.5$&$38.6\pm2.5$&1,5,7\\
&$\bf{1714_{-58}^{+58}}$&$\bf{1378_{-21}^{+21}}$&$\bf{1257_{-301}^{+302}}$&$\bf{1188_{-216}^{+215}}$&$\bf{13.75_{-2.37}^{+2.37}}$&$\bf{43.72_{-0.12}^{+0.12}}$&$\bf{122.8\pm18.11}$&$\bf{34.49\pm3.88}$&\ldots\\
PG 0007+106&$\bf{5301_{-28}^{+33}}$&$\bf{2424_{-41}^{+39}}$&$\bf{4832_{-11}^{+10}}$&$\bf{1766_{-9}^{+11}}$&$\bf{25.8_{-1.1}^{+1.2}}$&$\bf{43.46_{-0.06}^{+0.06}}$&$\bf{582.73\pm70.85}$&$\bf{491.28\pm58.56}$&2\\
PG 0026+129&$3374_{-27}^{+27}$&$2274_{-4}^{+4}$&$2735_{-578}^{+578}$&$2446_{-88}^{+88}$&$11.7_{-7.8}^{+7.4}$&$44.9_{-0.01}^{+0.01}$&$70.96\pm8.26$&$26.68\pm2.47$&11\\
\enddata
\tablecomments{References: (1) \cite{Du2019}, (2) \cite{Bao2022}, (3) \cite{Collin2006}, (4) \cite{Santos2001}, (5) \cite{Yu2020}, (6) \cite{Kaspi2005}, (7) \cite{Grier2012}, (8) \cite{Hu2021}, (9) \cite{Hu2015}, (10) \cite{Bentz2013}, (11) \cite{Hu2020b}, (12) \cite{Hu2020}, (13) \cite{Li2021}, (14) \cite{Du2018b}, (15) \cite{Peterson2004}, (16) \cite{Santos1997}, (17) \cite{Du2015}, (18) \cite{Fausnaugh2017}, (19) \cite{Kaspi2000}, (20) \cite{Bentz2009b}, (21) \cite{Denney2010}, (22) \cite{Bentz2016a}, (23) \cite{Zhang2019}, (24) \cite{Denney2006}, (25) \cite{Barth2013}, (26) \cite{Bentz2014}, (27) \cite{Bentz2007}, (28) \cite{Lu2016}, (29) \cite{Pei2017}, (30) \cite{Dietrich2012}, (31) \cite{Peterson2014}, (32) \cite{Bentz2023}, (33) \cite{Barth2015}, (34) \cite{Barth2011}, (35) \cite{Barth2011b}, (36) \cite{DeRosa2018}, (37) \cite{Huang2019}, (38) \cite{Hu2016}, (39) \cite{Brotherton2020}, (40) \cite{Du2018a}}
\end{deluxetable*}


\startlongtable
\begin{deluxetable*}{lcccccc}
\tablecolumns{7}
\tablecaption{Spectral Properties\label{tab:se_table}}
\tabletypesize{\scriptsize}
\tablehead{
    \colhead{Name}                          &
    \colhead{$\mathcal{R}_\mathrm{Fe}$}     &
    \colhead{$\rm FHWM_{Fe\,II}/FWHM_{H\beta}$}      &
    \colhead{$\rm \mathcal{A}$}           &
    \colhead{$\rm EW(He\,{II})$}              
    \\
    \colhead{}                              &
    \colhead{}            &
    \colhead{}            &
    \colhead{}                  &
    \colhead{(\AA)}                   
}
\startdata
PG 0007+106&0.48&$\rm 1.32\pm0.25$&$\rm -0.052\pm0.054$&too weak\\
I Zw1&1.71&$\rm 0.84\pm0.06$&$\rm 0.059\pm0.034$&$\rm 6.53\pm1.2$\\
Mrk 6&0.51&$\rm 1.12\pm0.67$&$\rm -0.292\pm0.019$&too weak\\
Mrk 704&0.55&$\rm 1.56\pm0.7$&$\rm 0.012\pm0.007$&$\rm 24.26\pm7.92$\\
PG 0923+201&0.94&$\rm 0.96\pm0.46$&$\rm 0\pm0.001$&too weak\\
PG 0923+129&0.59&$\rm 0.86\pm0.41$&$\rm 0.007\pm0.015$&$\rm 25.1\pm12.19$\\
PG 1100+772&0.05&$\rm 0.68\pm5.74$&$\rm -0.025\pm0.006$&$\rm 4.91\pm5.15$\\
Mrk 50&0.13&$\rm 0.52\pm1.23$&$\rm 0.035\pm0.003$&$\rm 21.6\pm10.6$\\
\enddata
\tablecomments{Most of the spectral properties utilized in this paper are from \cite{Du2019}. Here, we only list the additional objects included in this study. }
\end{deluxetable*}


\begin{deluxetable*}{ccccccccccccc}
\tablecaption{Joint Fitting and Calibrations of Virial Factors\label{tab:lfrvf}}
\tablewidth{0pt}
\tabletypesize{\scriptsize}
\setlength{\tabcolsep}{2pt}
\tablehead{
\colhead{VP} &
\colhead{Param.} &
\colhead{} &
\colhead{CB} &
\colhead{} &
\colhead{} &
\colhead{} &
\colhead{PB} &
\colhead{} &
\colhead{} &
\colhead{} &
\colhead{PB$+$CB} &
\colhead{} 
\\ \cline{3-5} \cline{7-9} \cline{11-13}
\colhead{} &
\colhead{} &
\colhead{Approach I} &
\colhead{Approach II} &
\colhead{Approach III} &
\colhead{} &
\colhead{Approach I} &
\colhead{Approach II} &
\colhead{Approach III} &
\colhead{} &
\colhead{Approach I} &
\colhead{Approach II} &
\colhead{Approach III} 
}
\startdata
MF&$\rm \alpha$&0.96$\pm$0.09&1.46$\pm$0.05&[1.02]&&[0.96]&[1.46]&[1.02]&&1.11$\pm$0.08&1.48$\pm$0.03&[1.27]\\
MF&$\rm \beta$&-1.35$\pm$0.9&-6.49$\pm$0.53&[-1.9]&&-2.26$\pm$0.14&-7$\pm$0.07&[-2.76]&&-3.05$\pm$0.78&-6.83$\pm$0.33&[-4.67]\\
MF&$f$&3.31$\pm$0.85&2.79$\pm$0.36&3.24$\pm$0.2&&0.67$\pm$0.28&0.58$\pm$0.11&0.62$\pm$0.03&&2.16$\pm$0.53&1.58$\pm$0.17&1.93$\pm$0.09\\
MF&$\rm \epsilon$&0.48&$\ldots$&$\ldots$&&0.58&$\ldots$&$\ldots$&&0.58&$\ldots$&$\ldots$\\
&&&&&&&&&&&&\\
MS&$\rm \alpha$&0.98$\pm$0.08&1.24$\pm$0.04&[1.02]&&[0.98]&[1.24]&[1.02]&&1.11$\pm$0.07&1.35$\pm$0.03&[1.27]\\
MS&$\rm \beta$&-1.5$\pm$0.84&-4.26$\pm$0.43&[-1.9]&&-2.4$\pm$0.13&-4.95$\pm$0.06&[-2.76]&&-3.03$\pm$0.73&-5.52$\pm$0.31&[-4.67]\\
MS&$f$&12.73$\pm$3.05&11.26$\pm$1.27&12.63$\pm$0.75&&2.22$\pm$0.86&2.08$\pm$0.33&2.14$\pm$0.13&&7.81$\pm$1.82&6.45$\pm$0.63&6.99$\pm$0.33\\
\enddata
\tablecomments{
The slopes of the joint fitting for PB samples are fixed to those of the CB results, and marked with square brackets. $\rm \alpha$ and $\rm \beta$ are the slope and intercept of the joint fitting of the $M_{\bullet}-L_{\rm V,bul}$ relation. $f$ is the virial factor. $\epsilon$ is the intrinsic error. The slope and intercept of Approach III are obtained by the quiescent galaxies in \cite{KH13}.
}
\end{deluxetable*}

\begin{deluxetable*}{lcccccccccccc}
\tablecolumns{13}
\tablecaption{Linear Regression of Correlations between Virial factors and Spectral properties\label{tab:params_linear_regression}}
\tabletypesize{\scriptsize}
\setlength{\tabcolsep}{2pt}
\tablehead{
    \colhead{} &
    \colhead{} &
    \colhead{} &
    \colhead{CB} &
    \colhead{} &
    \colhead{} &
    \colhead{} &
    \colhead{PB} &
    \colhead{} &
    \colhead{} &
    \colhead{} &
    \colhead{PB+CB} &
    \colhead{} 
        \\ 
    \cline{3-5} \cline{7-9} \cline{11-13}
    \colhead{VP} &
    \colhead{Param.} &
    \colhead{$\alpha^{\prime}$} &
    \colhead{$\beta^{\prime}$} &
    \colhead{$\epsilon^{\prime}$} &
    \colhead{} &
    \colhead{$\alpha^{\prime}$} &
    \colhead{$\beta^{\prime}$} &
    \colhead{$\epsilon^{\prime}$} &
    \colhead{} &
    \colhead{$\alpha^{\prime}$} &
    \colhead{$\beta^{\prime}$} &
    \colhead{$\epsilon^{\prime}$} 
}
\startdata
MF&$\mathcal{D}_\mathrm{H\beta,mean}$&$-0.61\pm0.11$&$1.76\pm0.23$&0.31&&$-0.82\pm0.19$&$1.31\pm0.34$&0.45&&$-0.53\pm0.13$&$1.25\pm0.27$&0.56\\
&$\mathcal{R}_\mathrm{Fe}$&\ldots&\ldots&\ldots&&$1.17\pm0.28$&$-1.11\pm0.23$&0.45&&\ldots&\ldots&\ldots\\
&$\rm log\,(FWHM_{H\beta,mean}/km\,s^{-1})$&$-1.42\pm0.26$&$5.66\pm0.94$&0.32&&$-1.82\pm0.33$&$5.95\pm1.12$&0.4&&\ldots&\ldots&\ldots\\
&$\rm log\,(\sigma_{H\beta,mean}/km\,s^{-1})$&\ldots&\ldots&\ldots&&$-1.96\pm0.51$&$5.94\pm1.59$&0.48&&\ldots&\ldots&\ldots\\ \\
MS&$\rm log\,(\sigma_{H\beta,mean}/km\,s^{-1})$&\ldots&\ldots&\ldots&&$-1.65\pm0.48$&$5.48\pm1.51$&0.42&&\ldots&\ldots&\ldots\\ \\
RF&$\mathcal{D}_\mathrm{H\beta,mean}$&$-0.59\pm0.13$&$1.91\pm0.27$&0.4&&\ldots&\ldots&\ldots&&\ldots&\ldots&\ldots\\
\enddata
\end{deluxetable*}

\end{document}

%% file: gal_rst.tex
\begin{figure}
	\centering
	\includegraphics[width=0.9\textwidth]{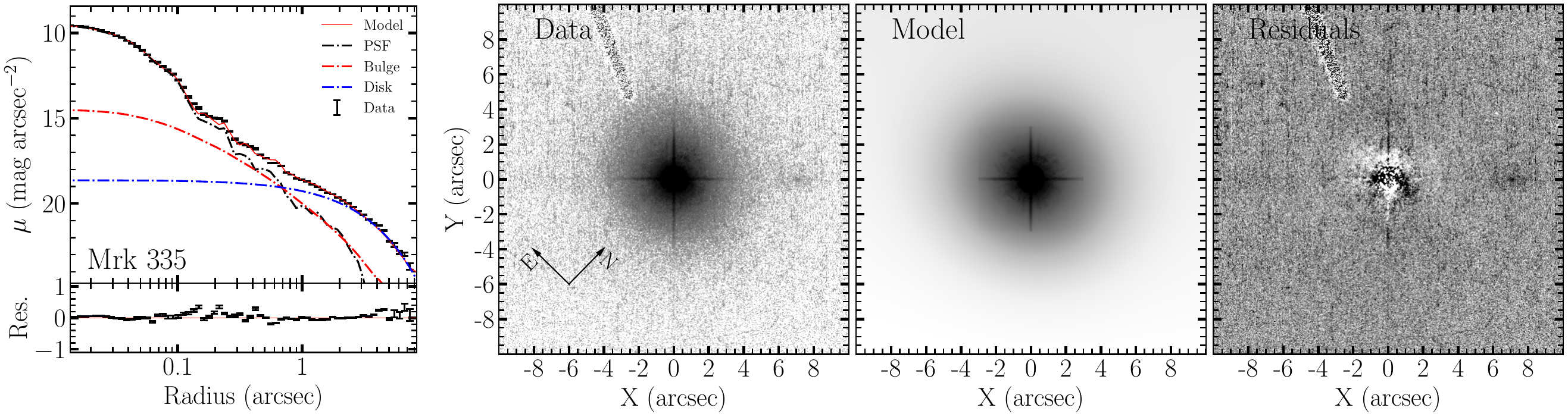}
	\includegraphics[width=0.9\textwidth]{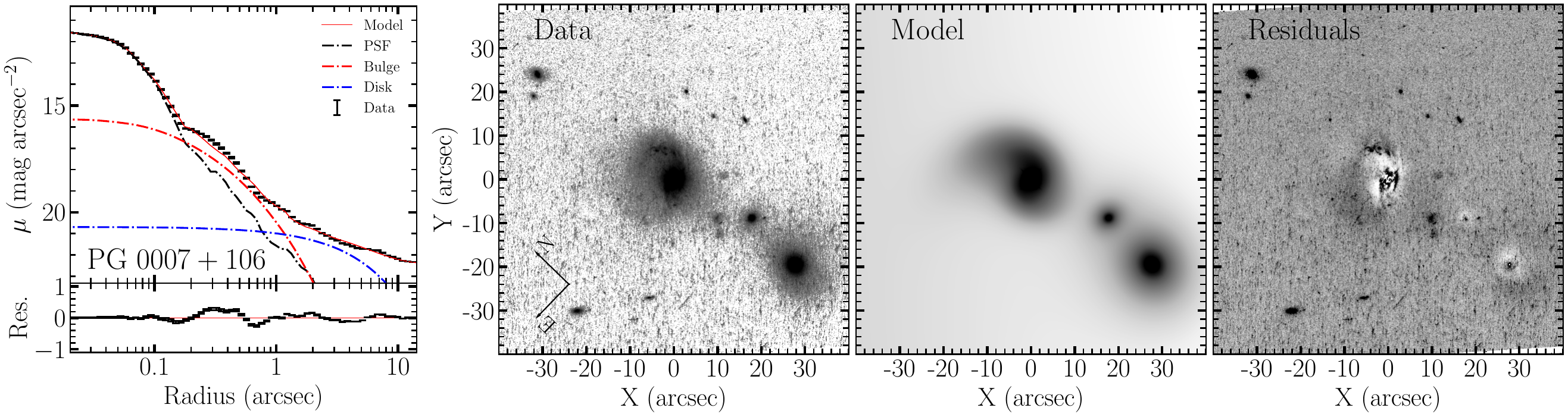}
	\includegraphics[width=0.9\textwidth]{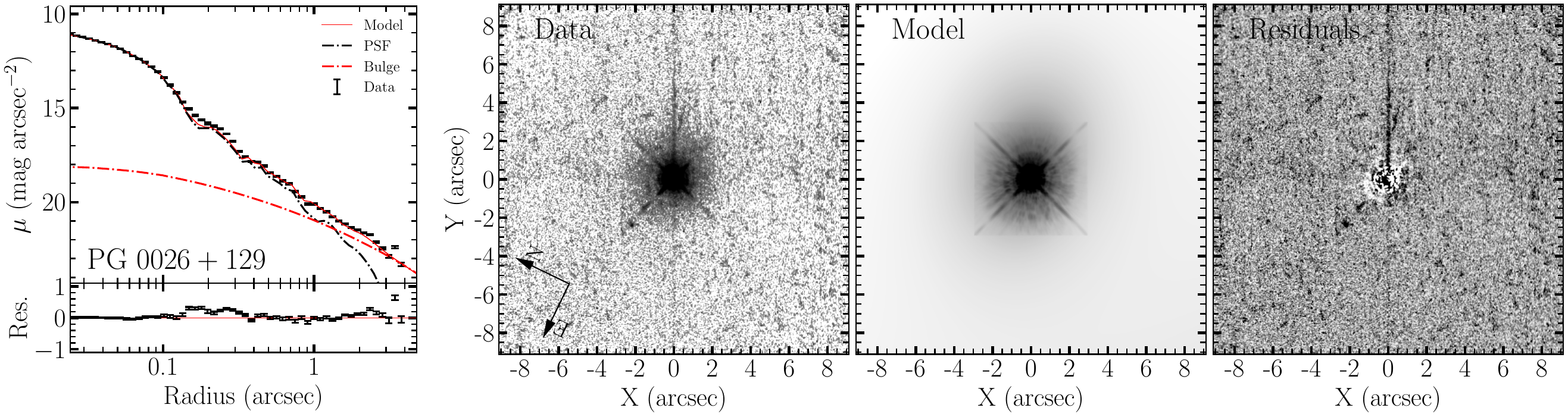}
	\includegraphics[width=0.9\textwidth]{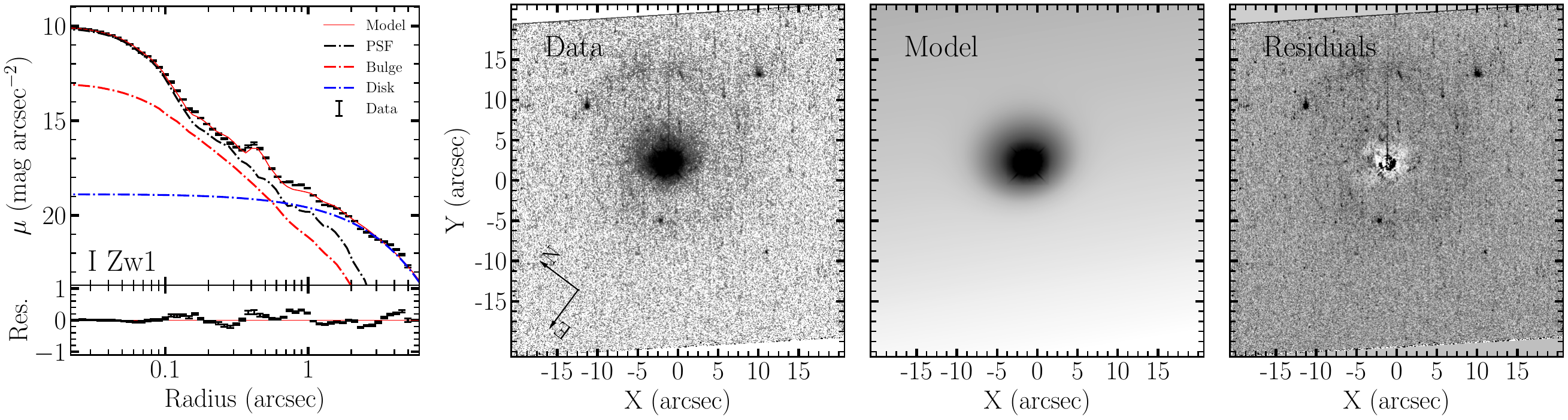}
	\includegraphics[width=0.9\textwidth]{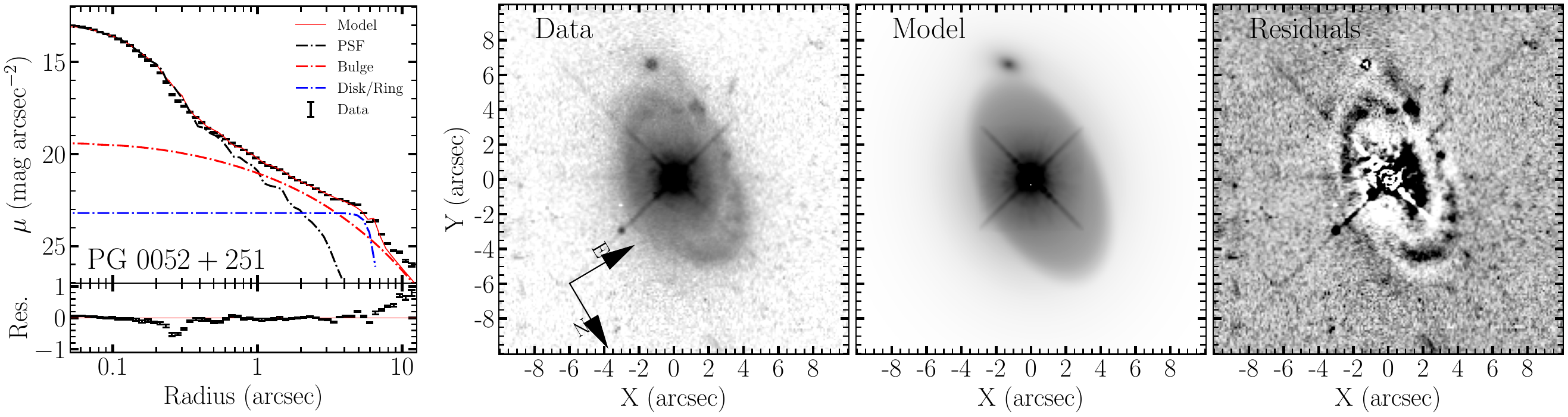}
	\caption{\texttt{GALFIT} fitting results. From left to right, the first columns are the one-dimensional surface brightness of their host components. The points with error bars are the {\it HST} data, the red solid lines are the best-fit models, the dash-dotted lines with different colors are the surface-brightness profiles of the PSFs, bulges, bars, and etc. The gray regions are masked out in the fitting. The second columns show the {\it HST} images. The third columns are the best-fit models. The last columns are the residual images. The names of the objects are marked in the first columns.}
	\label{fig:galfit_1}
\end{figure}

\begin{figure}
	\centering
	\includegraphics[width=0.9\textwidth]{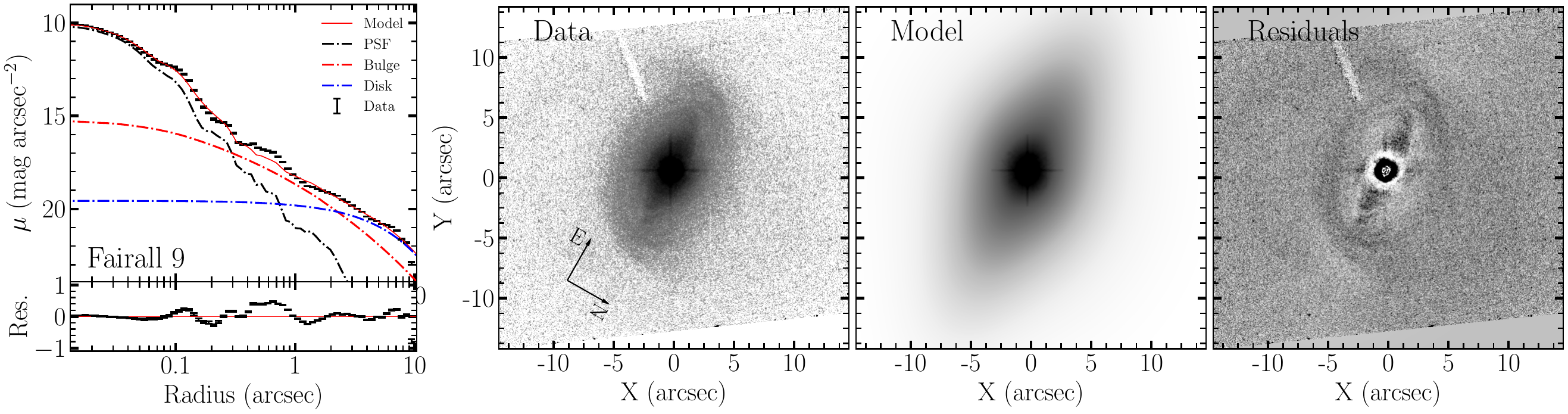}
	\includegraphics[width=0.9\textwidth]{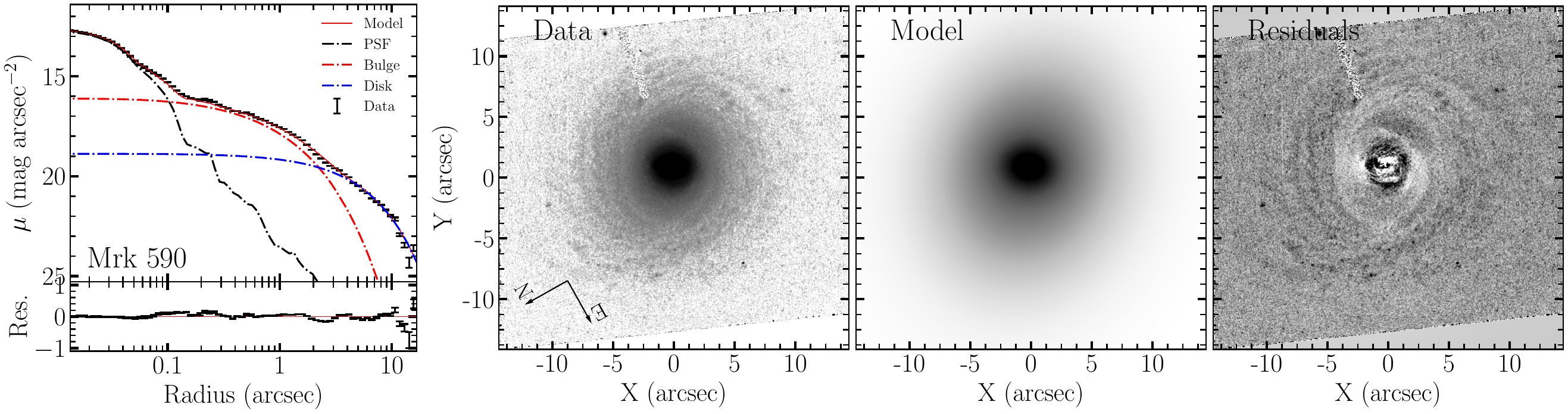}
	\includegraphics[width=0.9\textwidth]{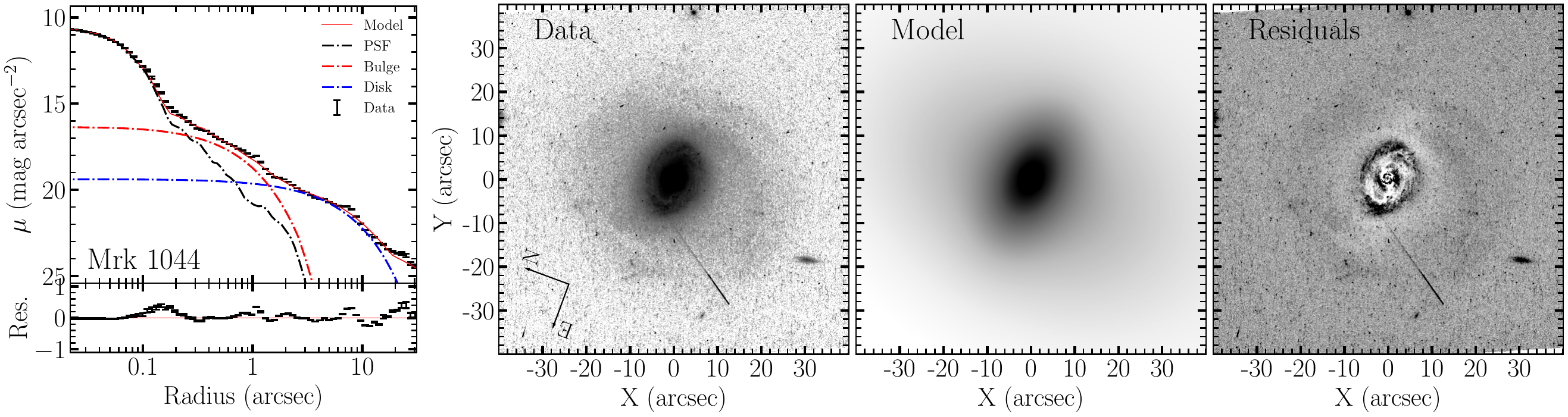}
	\includegraphics[width=0.9\textwidth]{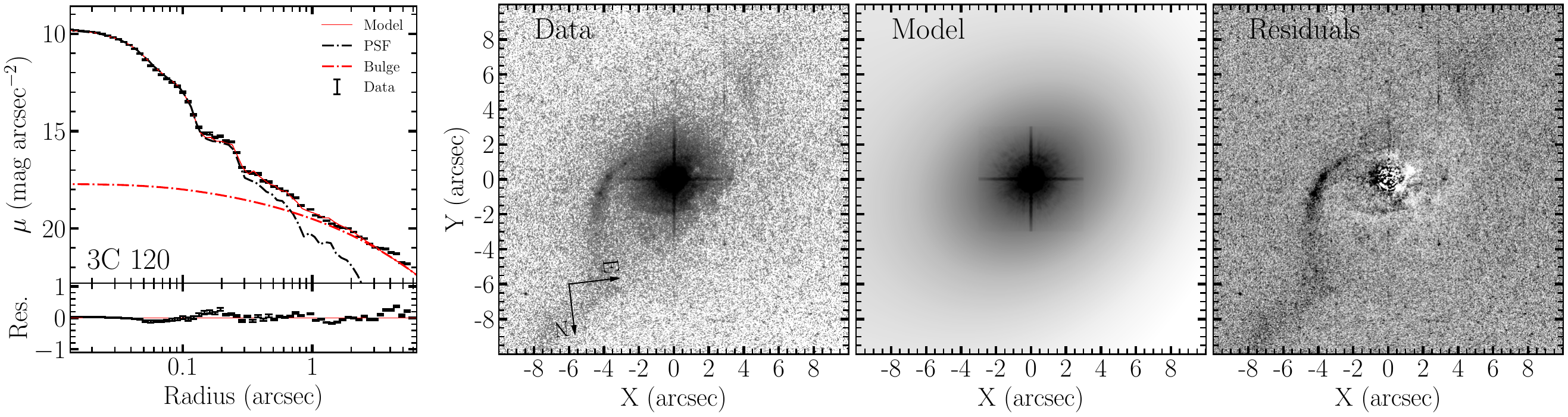}
	\includegraphics[width=0.9\textwidth]{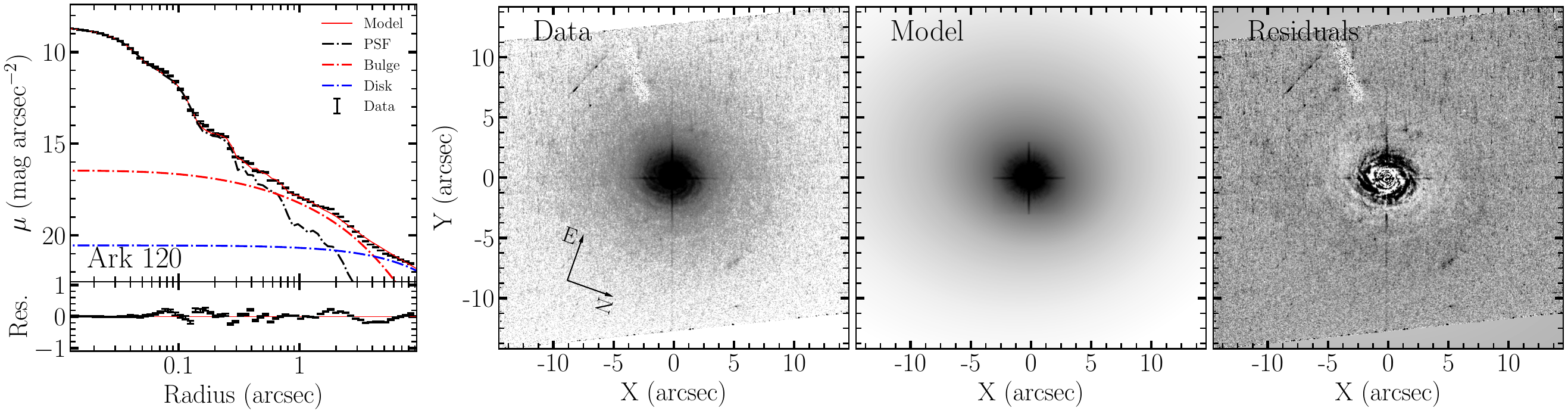}
	\figurenum{\ref{fig:galfit_1}}
	\caption{(Continued.)}
\end{figure}

\begin{figure}
	\centering
	\includegraphics[width=0.9\textwidth]{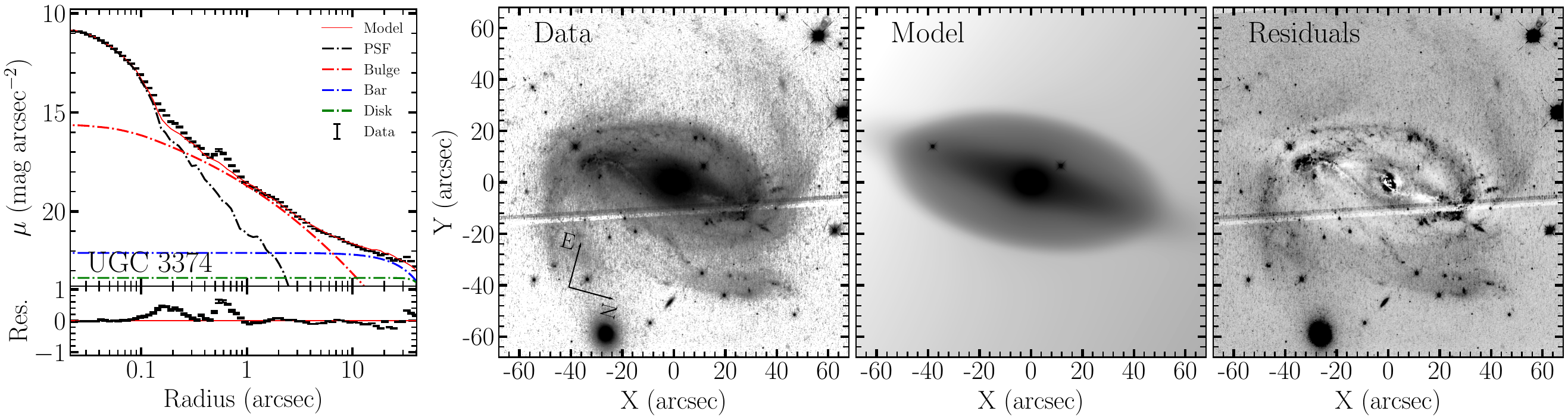}
	\includegraphics[width=0.9\textwidth]{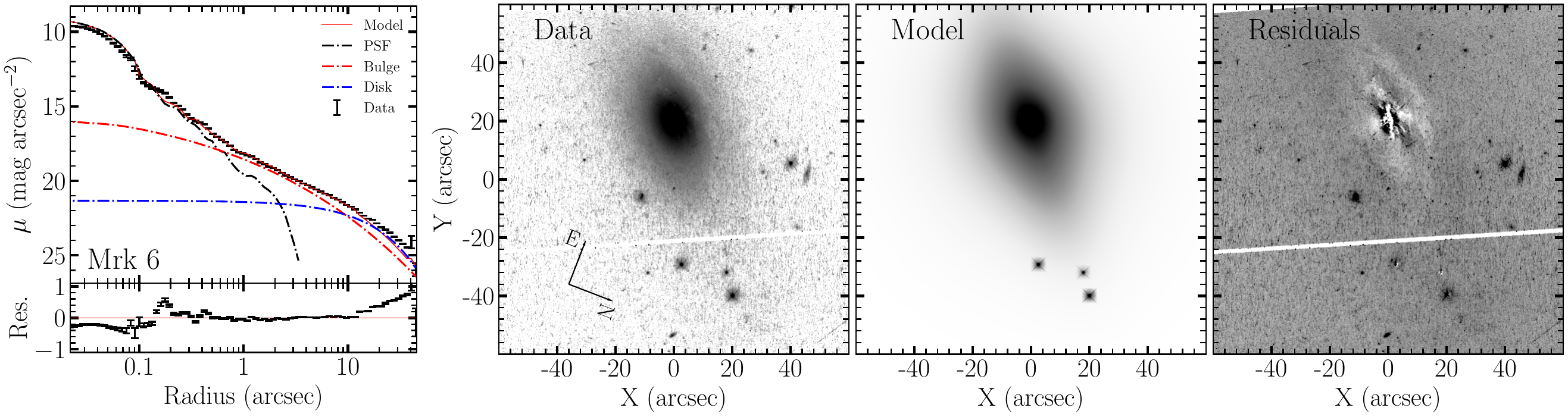}
	\includegraphics[width=0.9\textwidth]{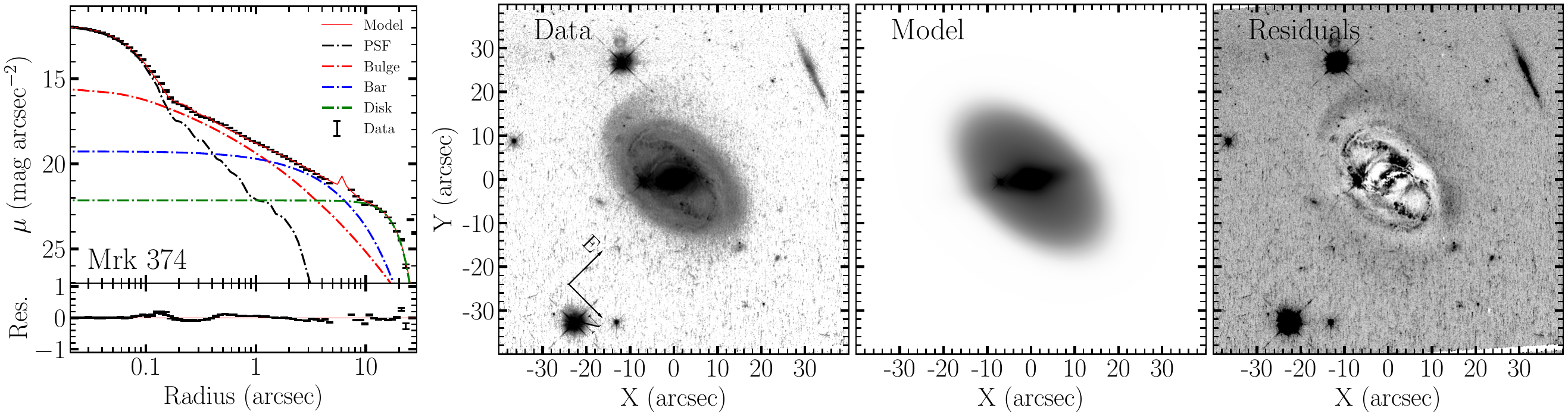}
	\includegraphics[width=0.9\textwidth]{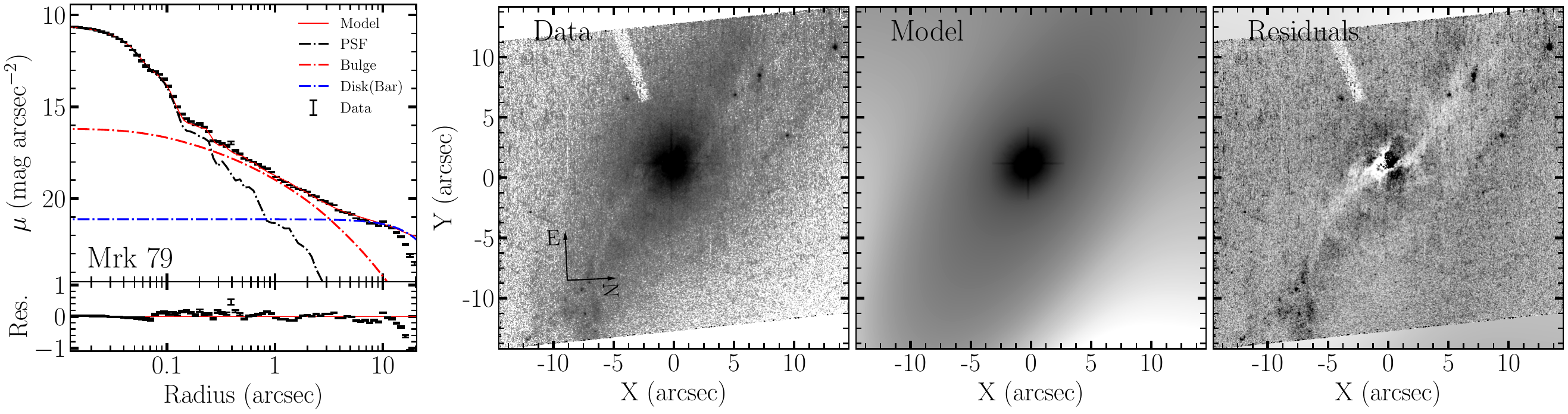}
	\includegraphics[width=0.9\textwidth]{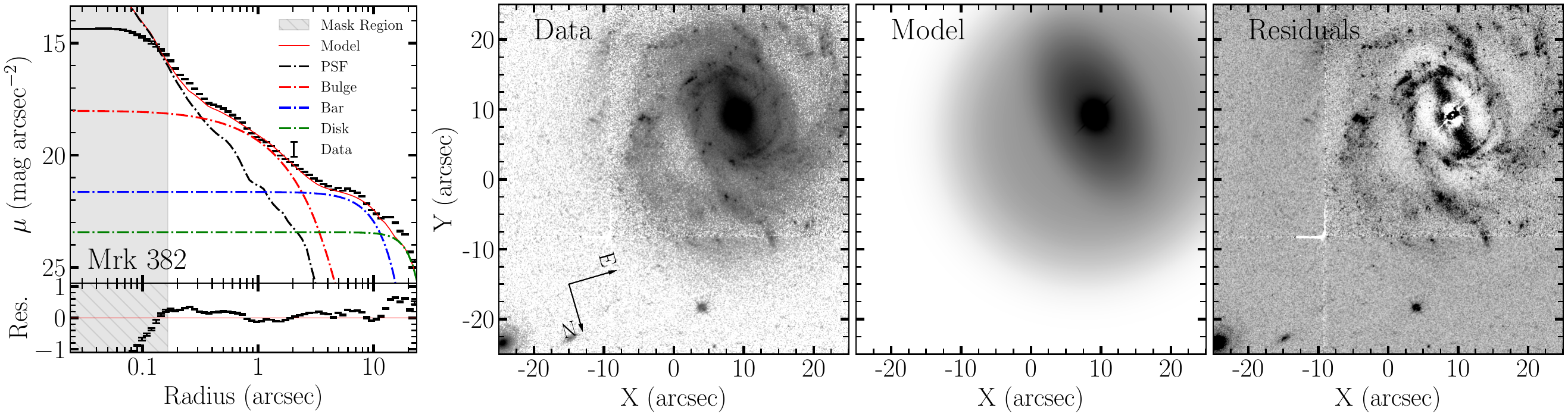}
	\figurenum{\ref{fig:galfit_1}}
	\caption{(Continued.)}
\end{figure}

\begin{figure}
	\centering
	\includegraphics[width=0.9\textwidth]{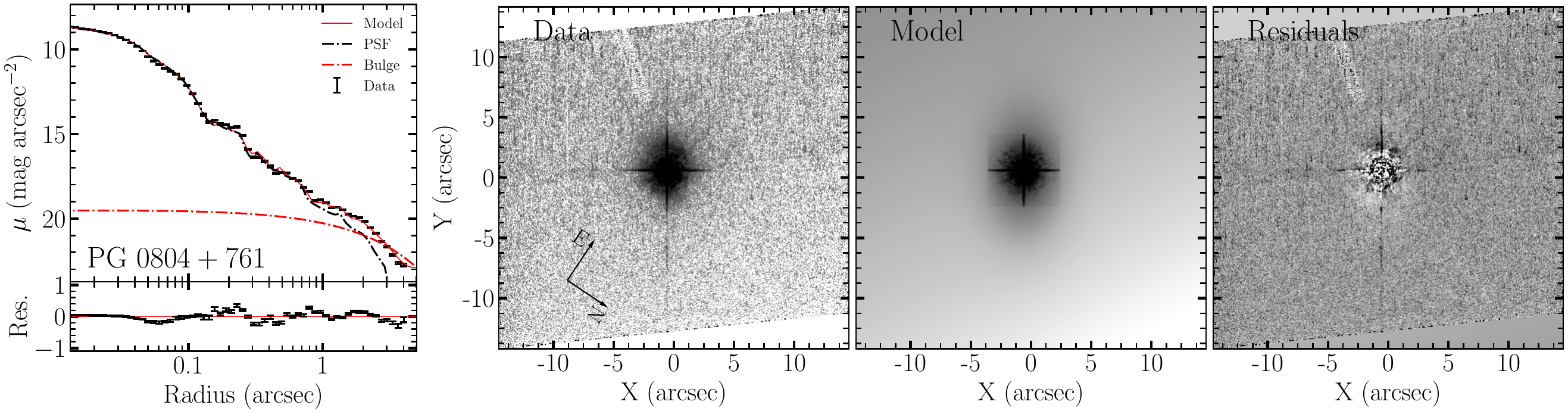}
	\includegraphics[width=0.9\textwidth]{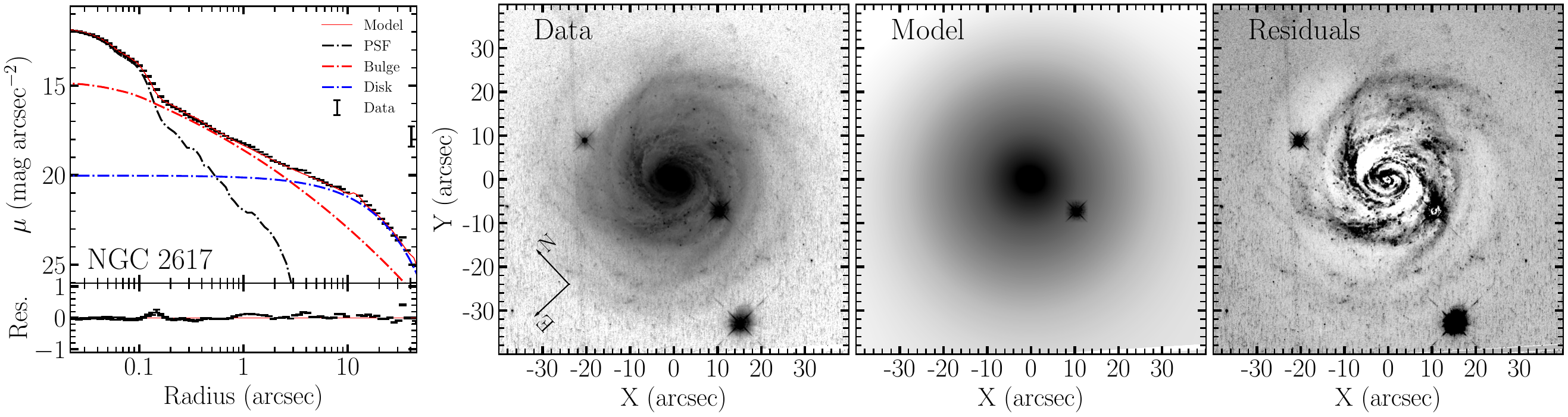}
	\includegraphics[width=0.9\textwidth]{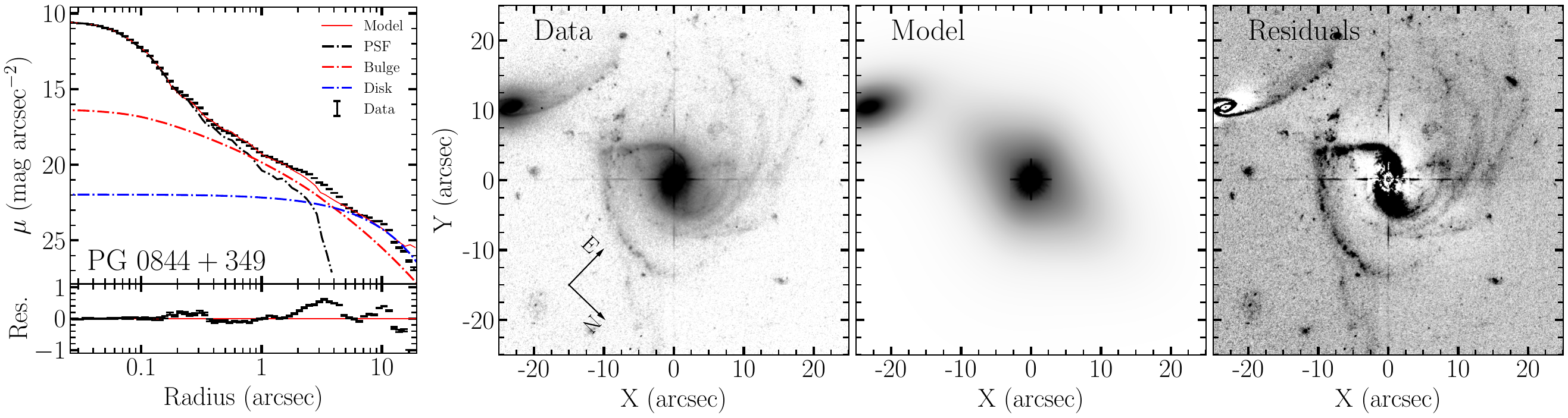}
	\includegraphics[width=0.9\textwidth]{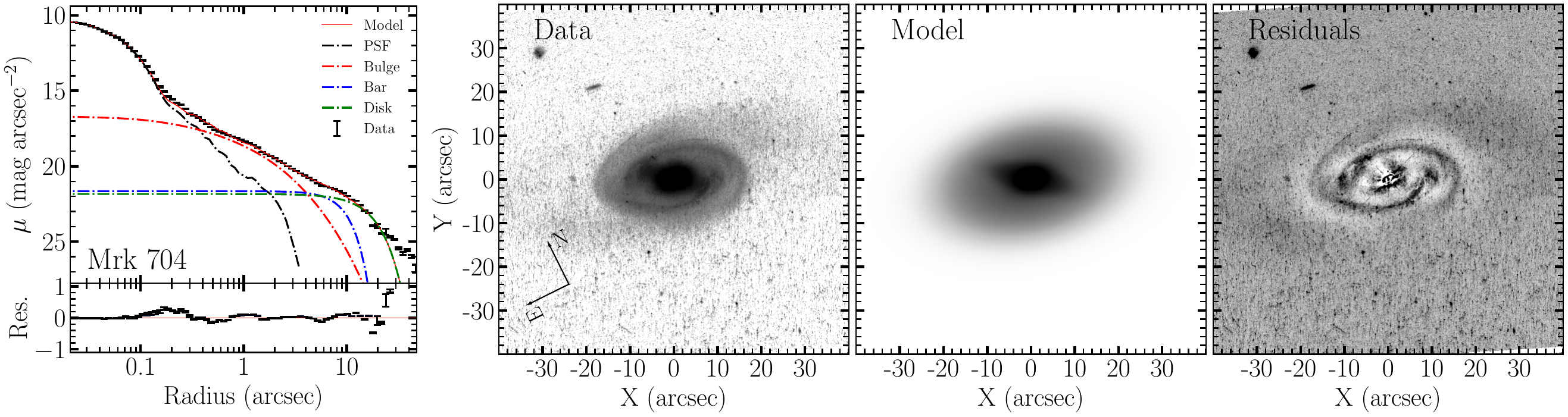}
	\includegraphics[width=0.9\textwidth]{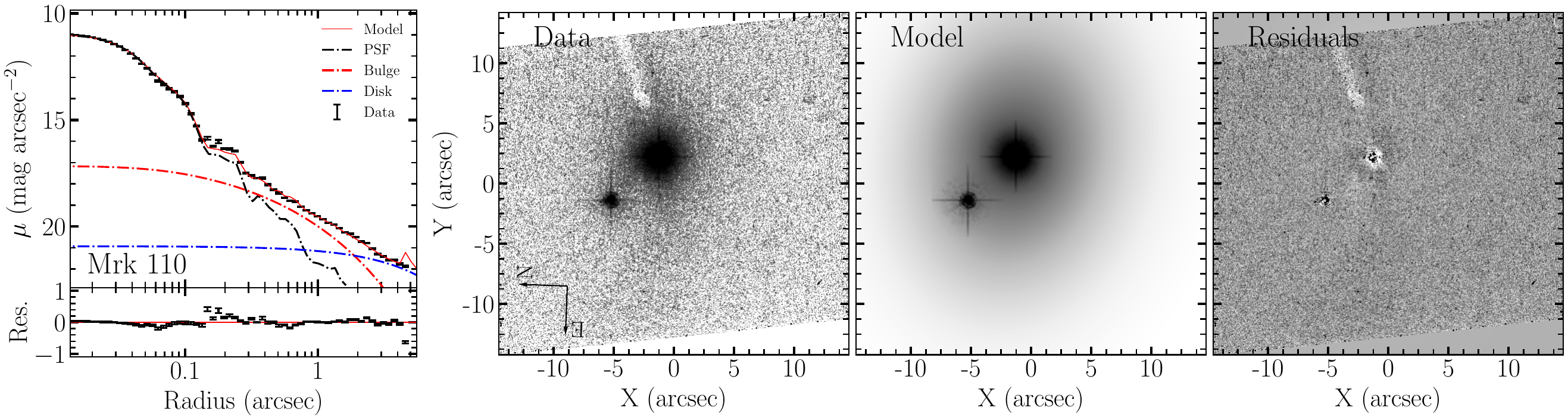}
	\figurenum{\ref{fig:galfit_1}}
	\caption{(Continued.)}
\end{figure}

\begin{figure}
	\centering
	\includegraphics[width=0.9\textwidth]{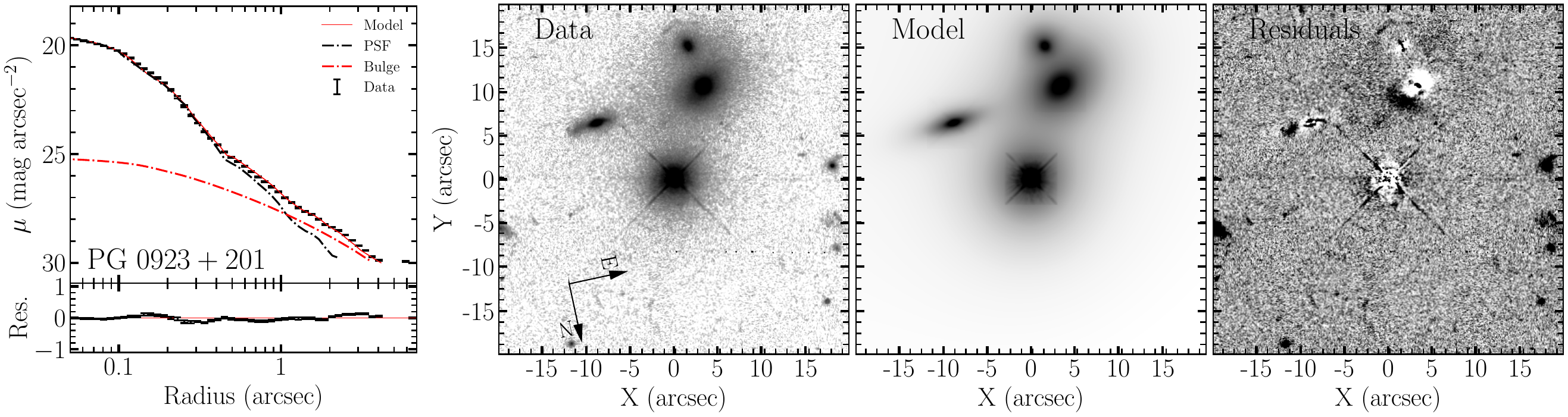}
	\includegraphics[width=0.9\textwidth]{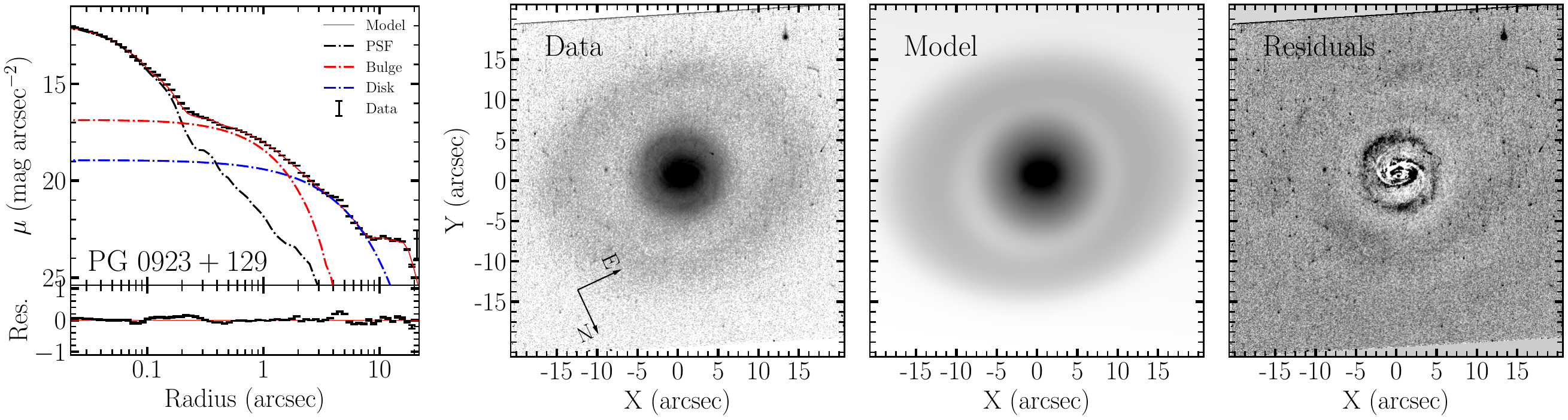}
	\includegraphics[width=0.9\textwidth]{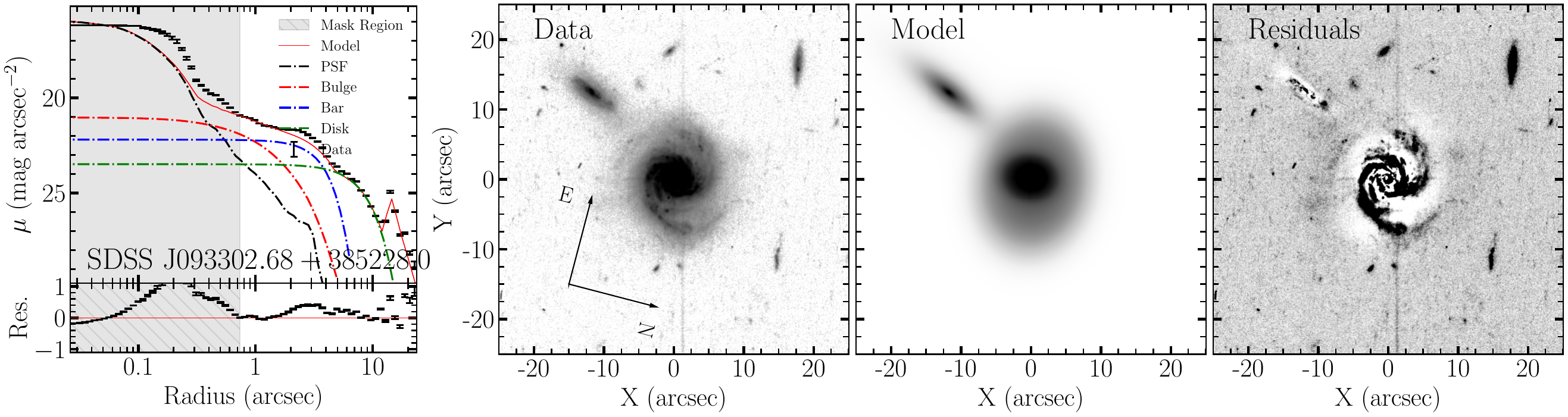}
	\includegraphics[width=0.9\textwidth]{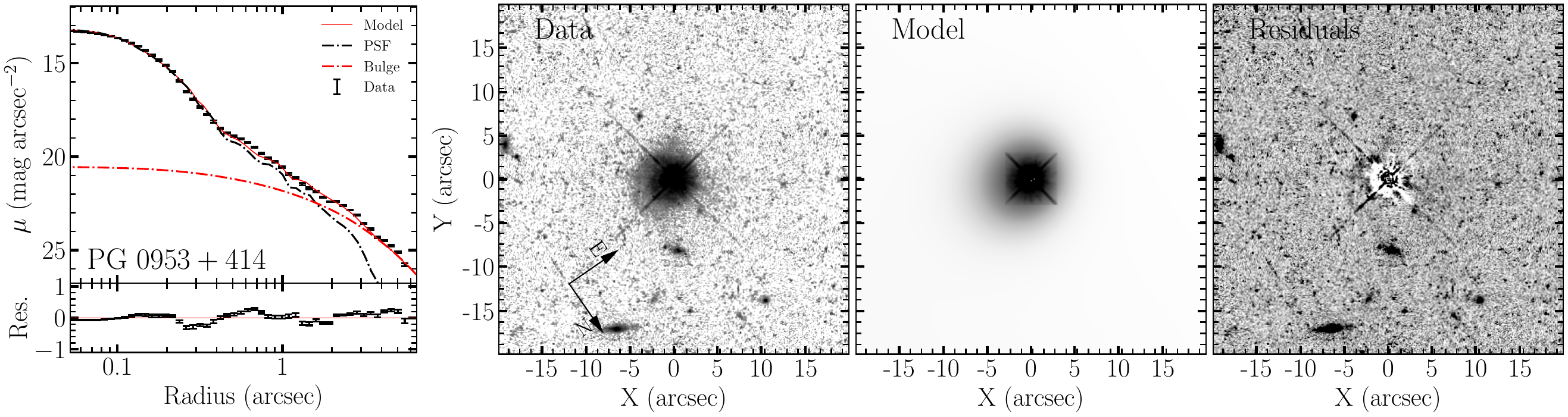}
	\includegraphics[width=0.9\textwidth]{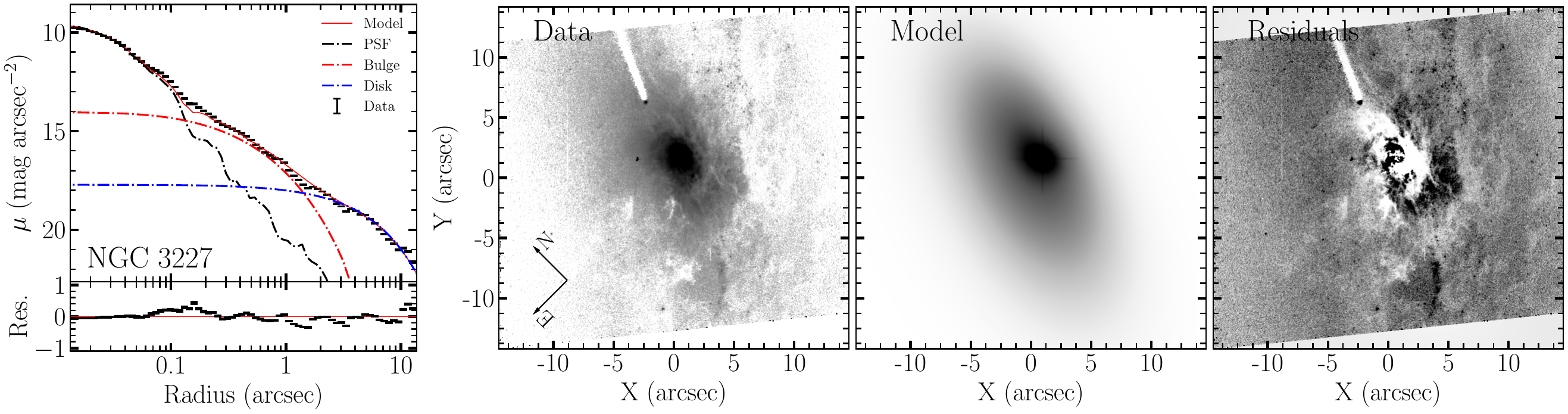}
	\figurenum{\ref{fig:galfit_1}}
	\caption{(Continued.)}
\end{figure}

\begin{figure}
	\centering
	\includegraphics[width=0.9\textwidth]{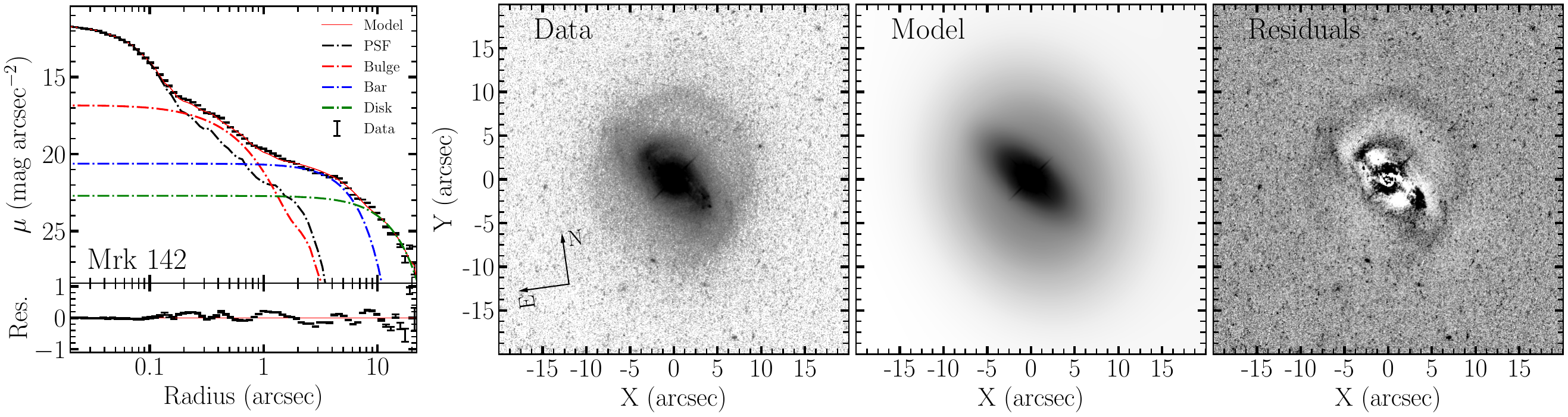}
	\includegraphics[width=0.9\textwidth]{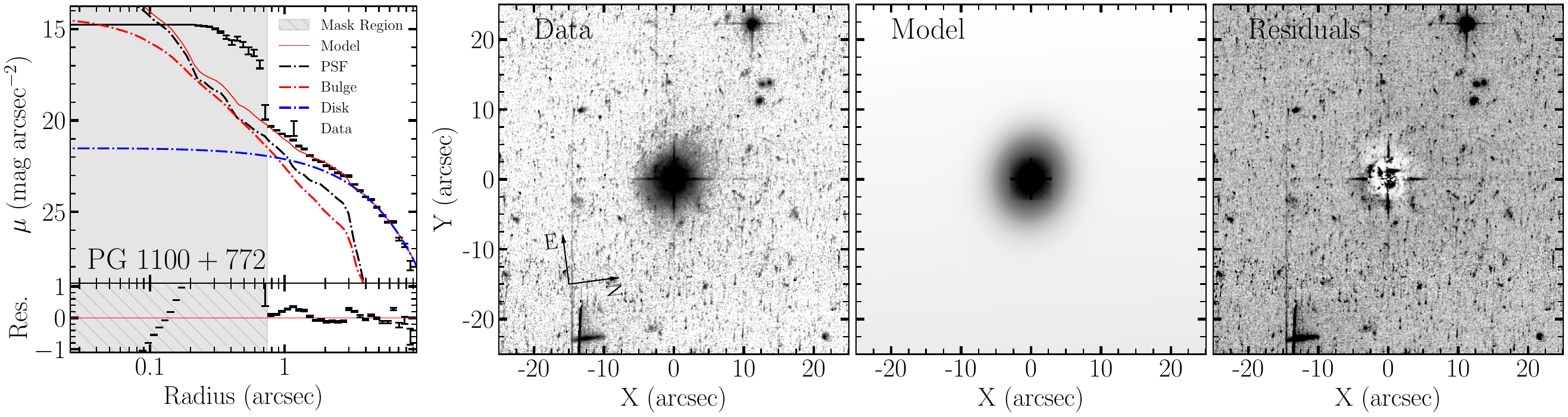}
	\includegraphics[width=0.9\textwidth]{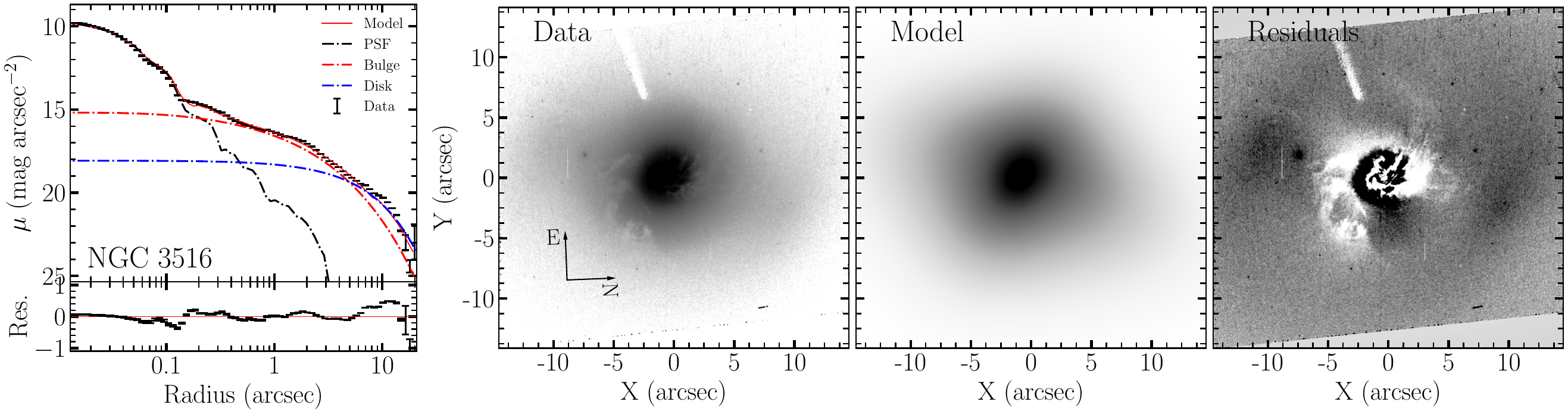}
	\includegraphics[width=0.9\textwidth]{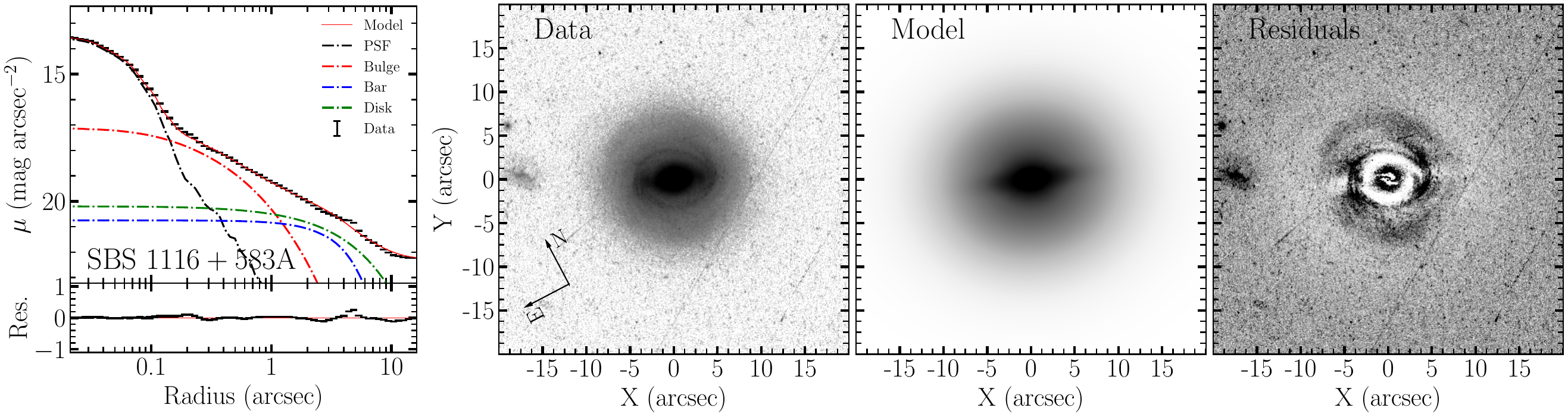}
	\includegraphics[width=0.9\textwidth]{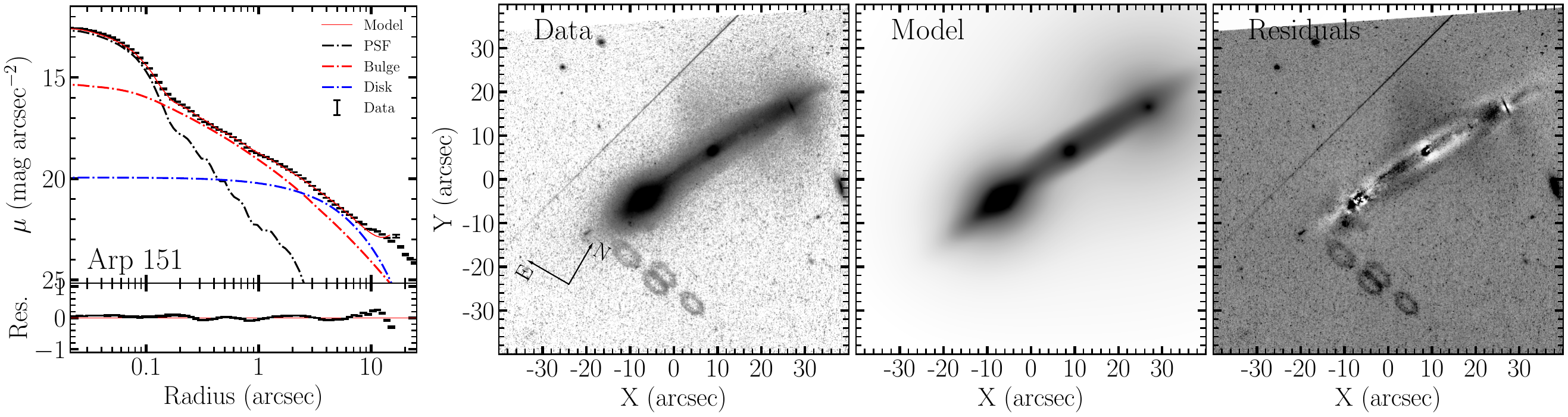}
	\figurenum{\ref{fig:galfit_1}}
	\caption{(Continued.)}
\end{figure}

\begin{figure}
	\centering
	\includegraphics[width=0.9\textwidth]{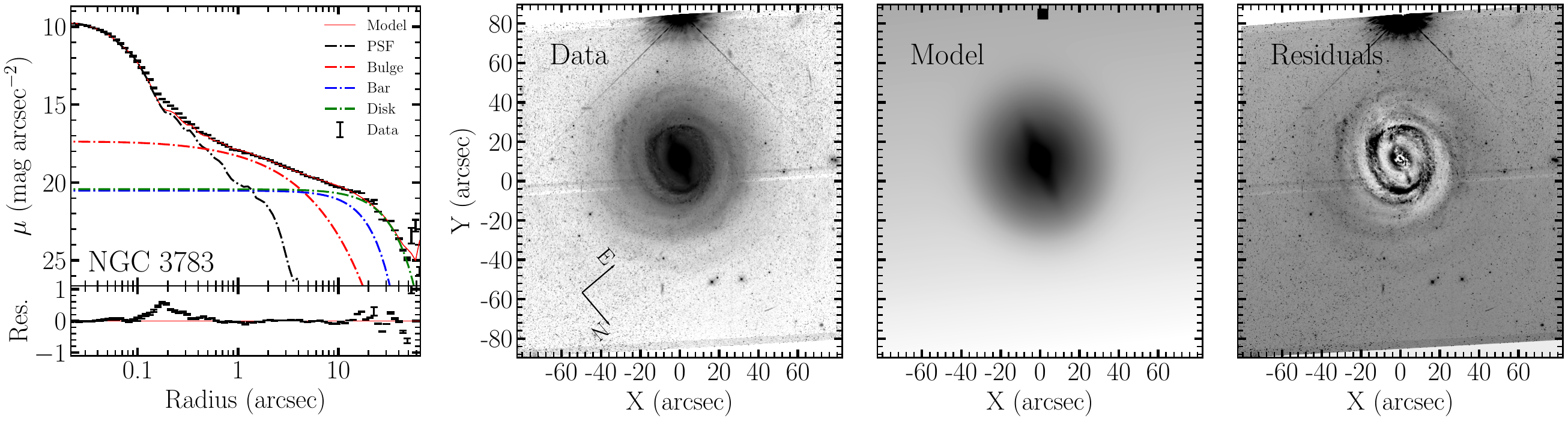}
	\includegraphics[width=0.9\textwidth]{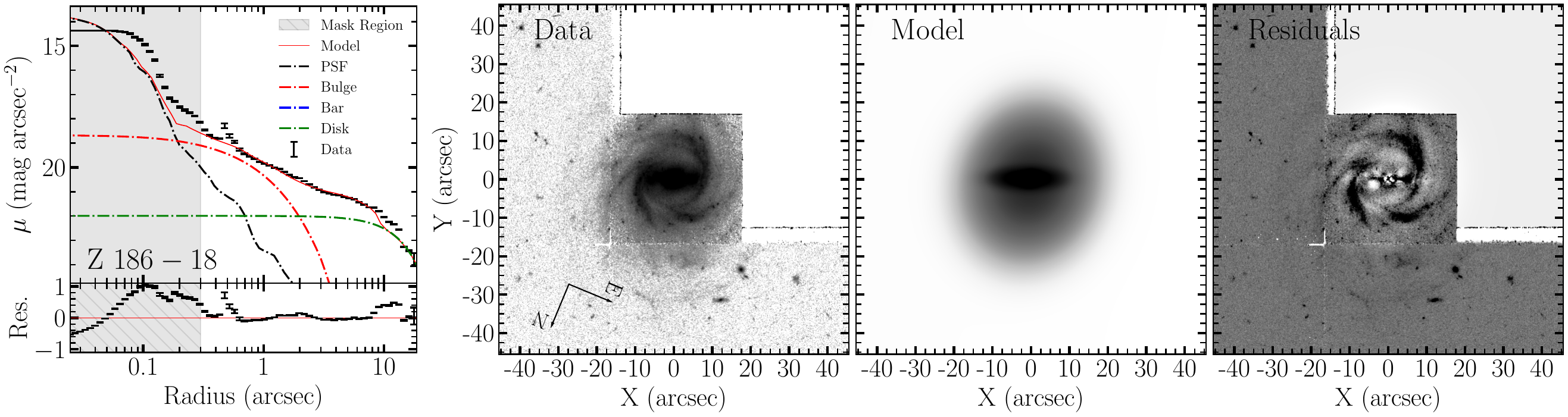}
	\includegraphics[width=0.9\textwidth]{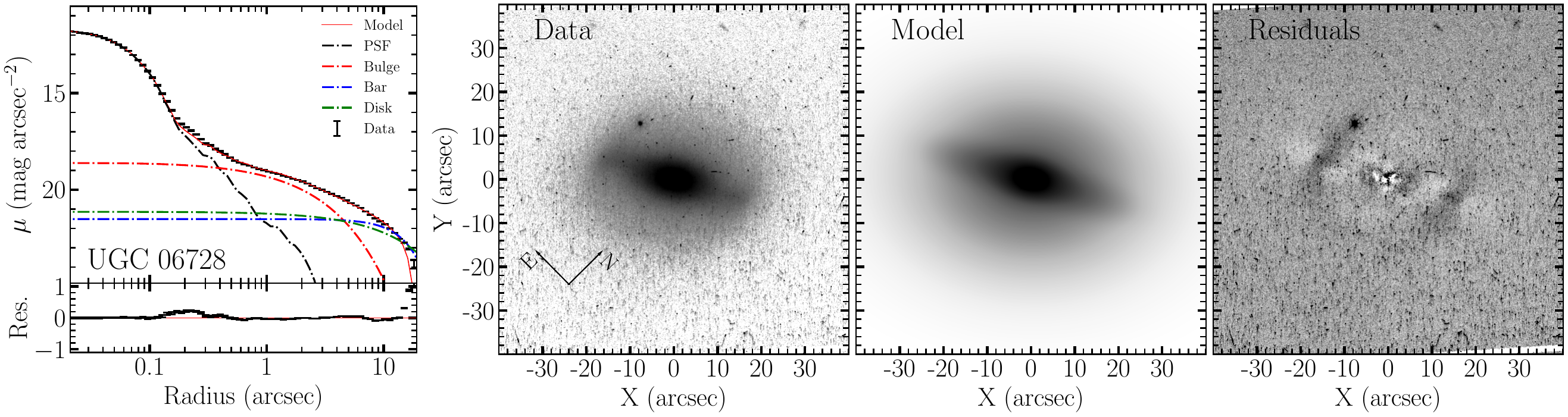}
	\includegraphics[width=0.9\textwidth]{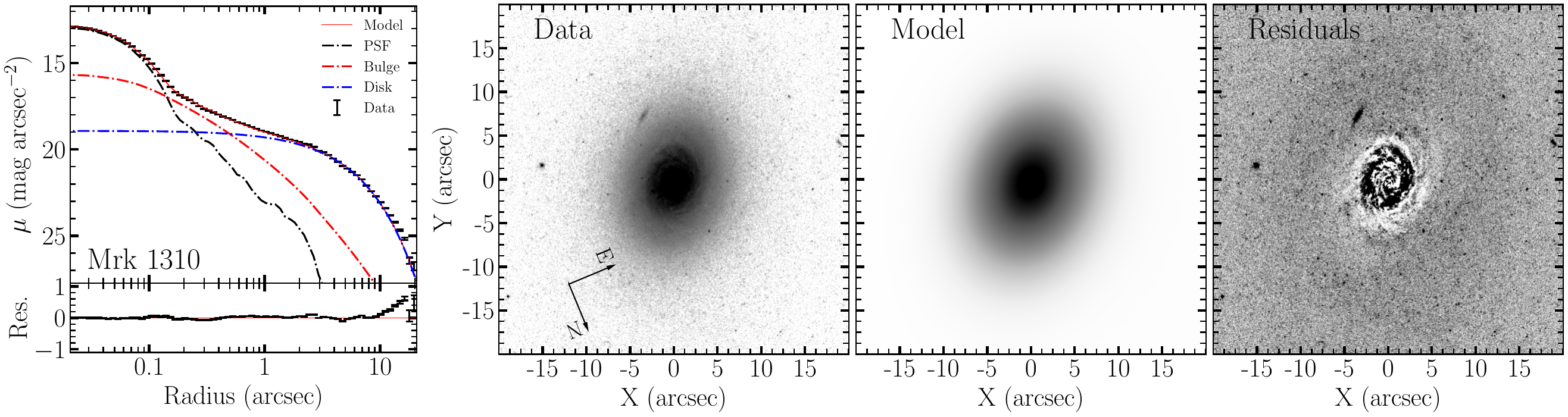}
	\includegraphics[width=0.9\textwidth]{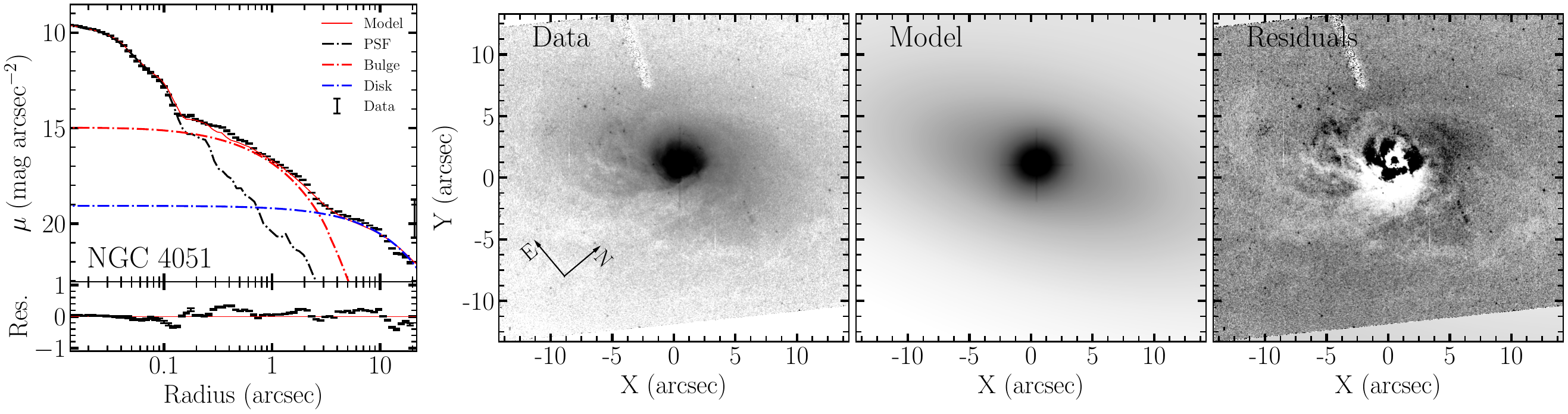}
	\figurenum{\ref{fig:galfit_1}}
	\caption{(Continued.)}
\end{figure}

\begin{figure}
	\centering
	\includegraphics[width=0.9\textwidth]{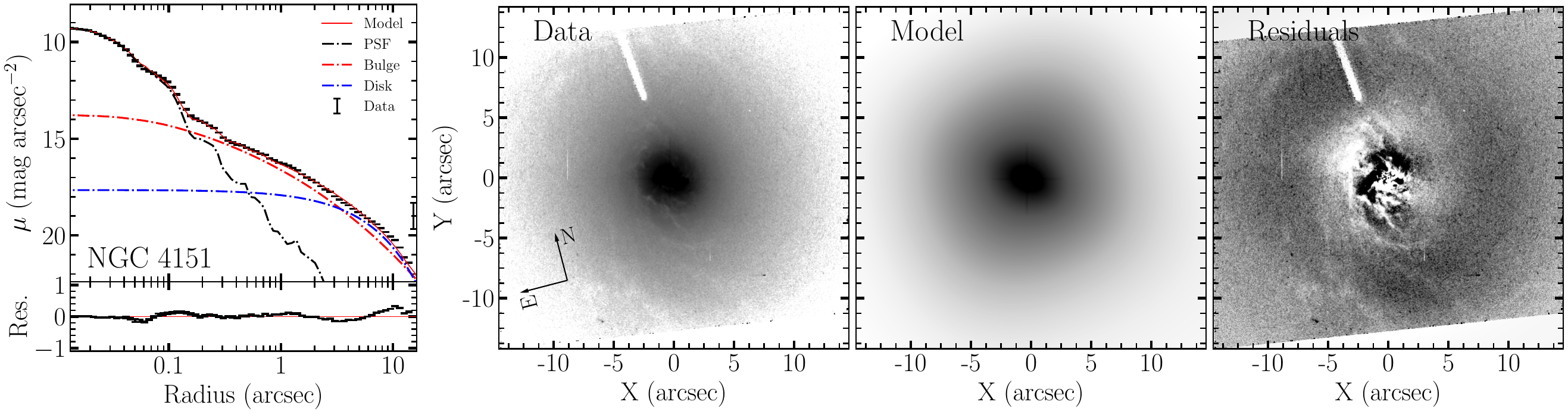}
	\includegraphics[width=0.9\textwidth]{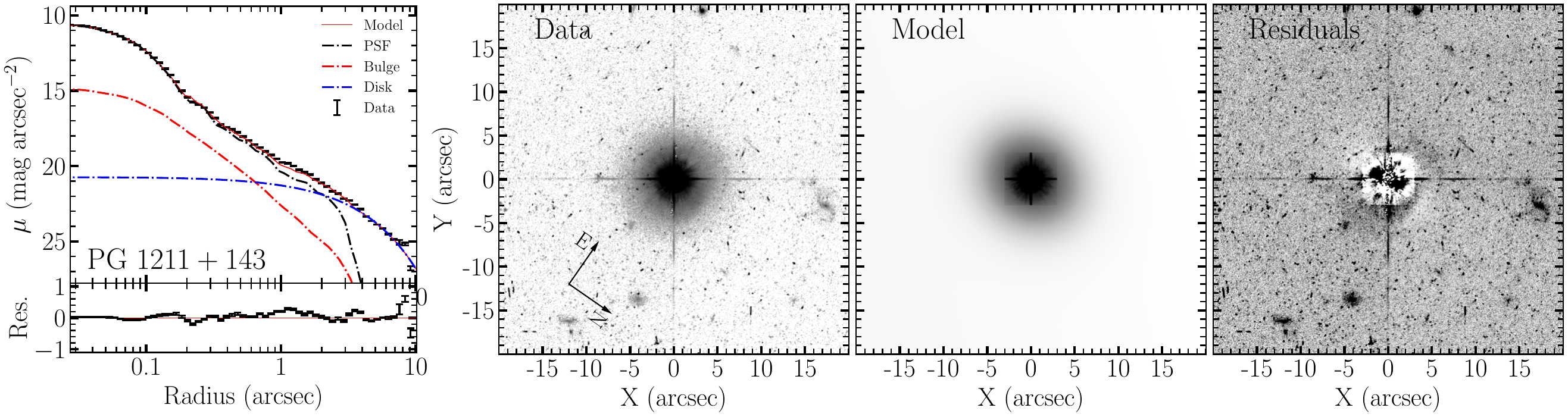}
	\includegraphics[width=0.9\textwidth]{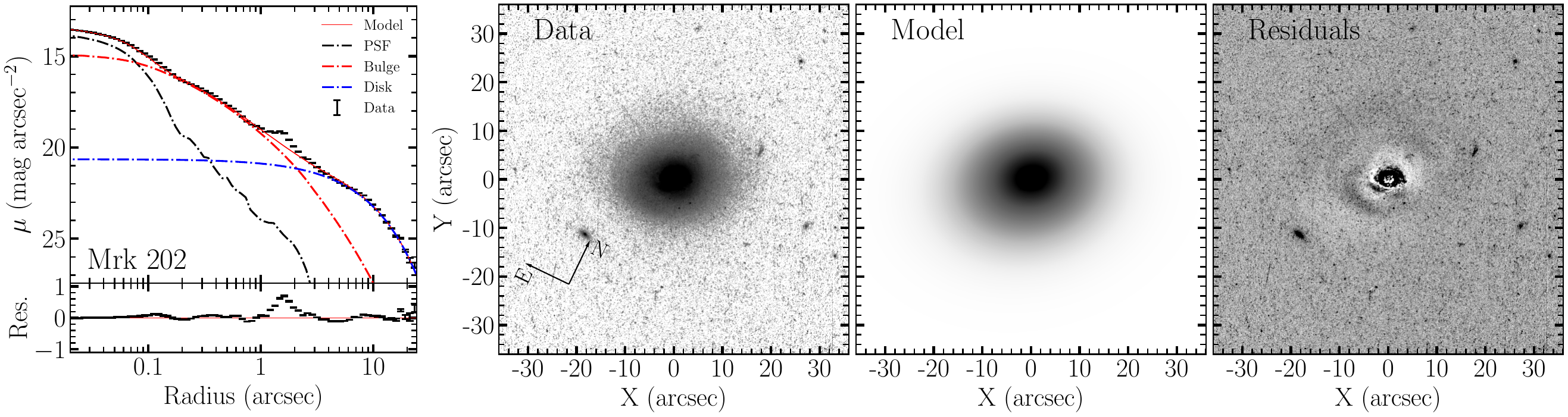}
	\includegraphics[width=0.9\textwidth]{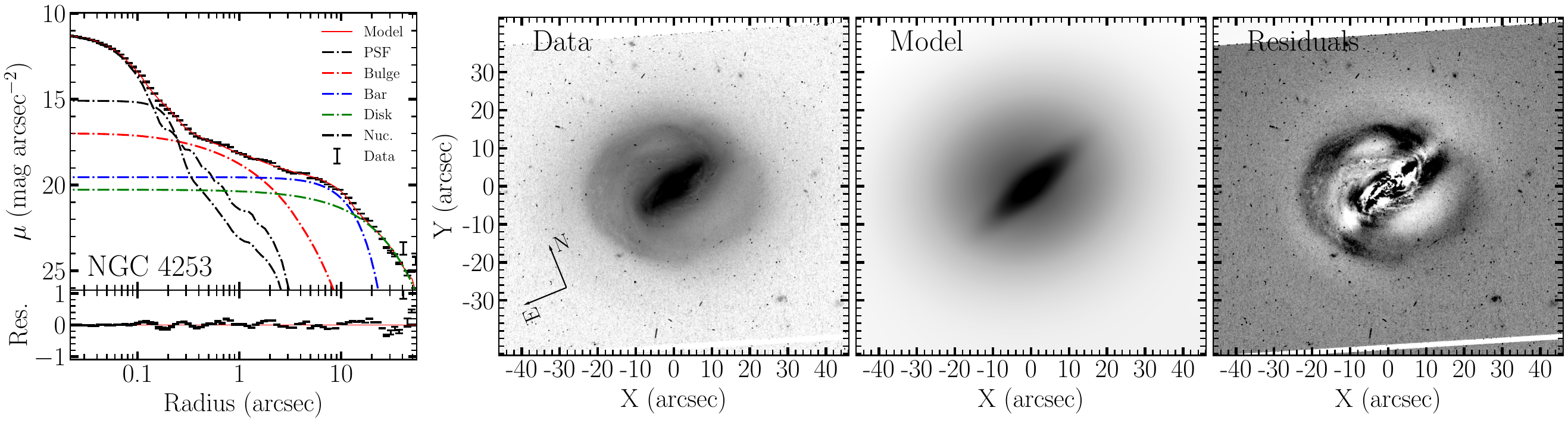}
	\includegraphics[width=0.9\textwidth]{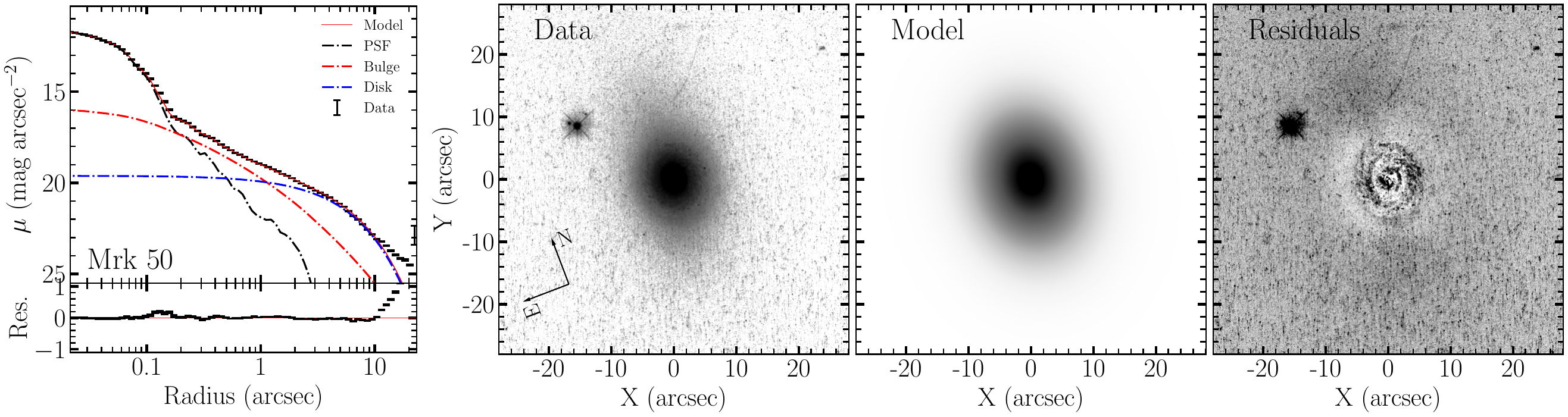}
	\figurenum{\ref{fig:galfit_1}}
	\caption{(Continued.)}
\end{figure}

\begin{figure}
	\centering
	\includegraphics[width=0.9\textwidth]{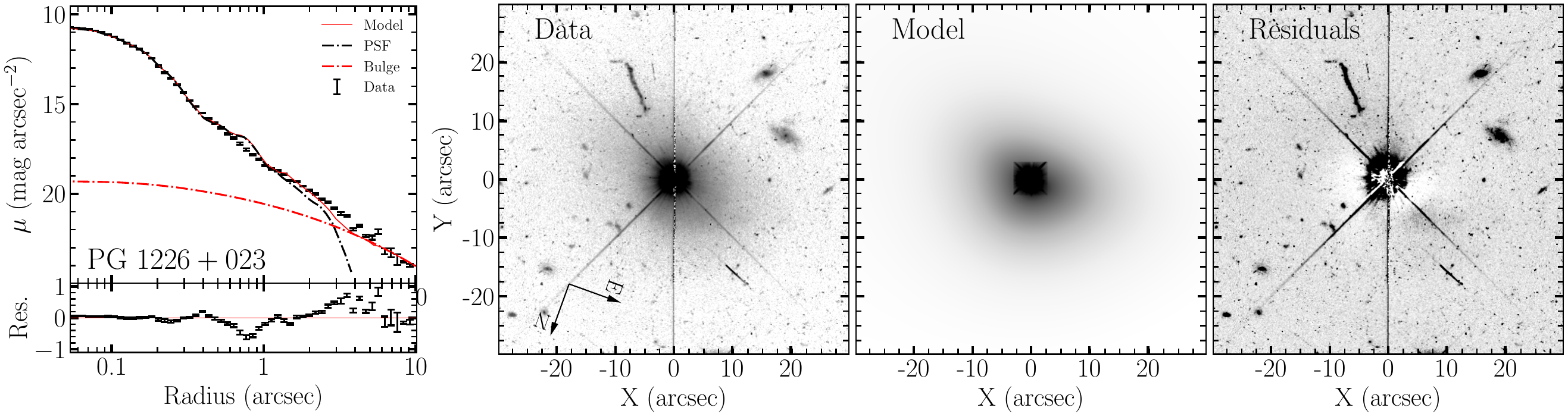}
	\includegraphics[width=0.9\textwidth]{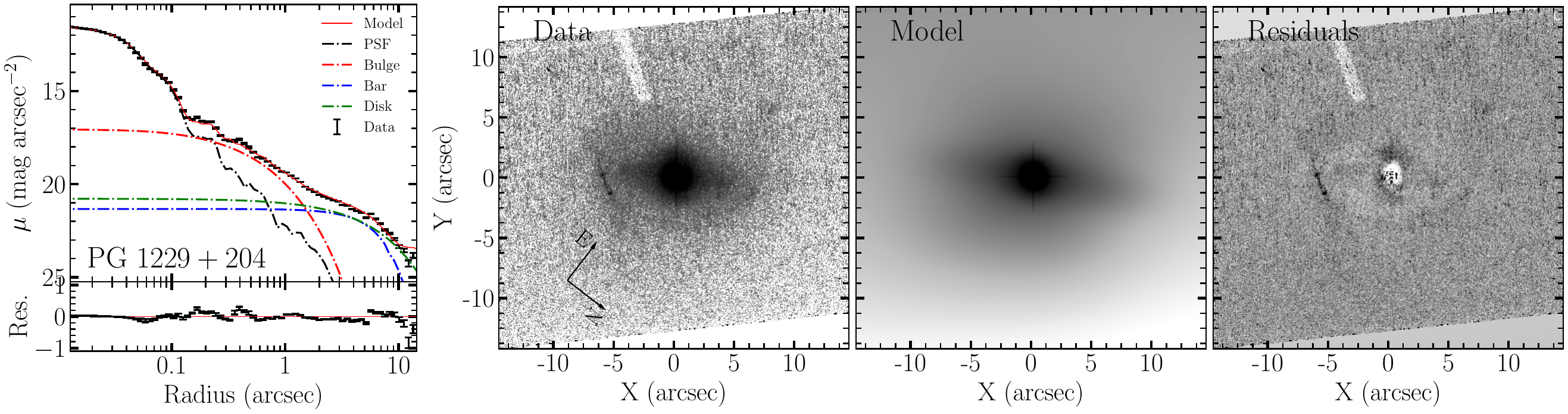}
	\includegraphics[width=0.9\textwidth]{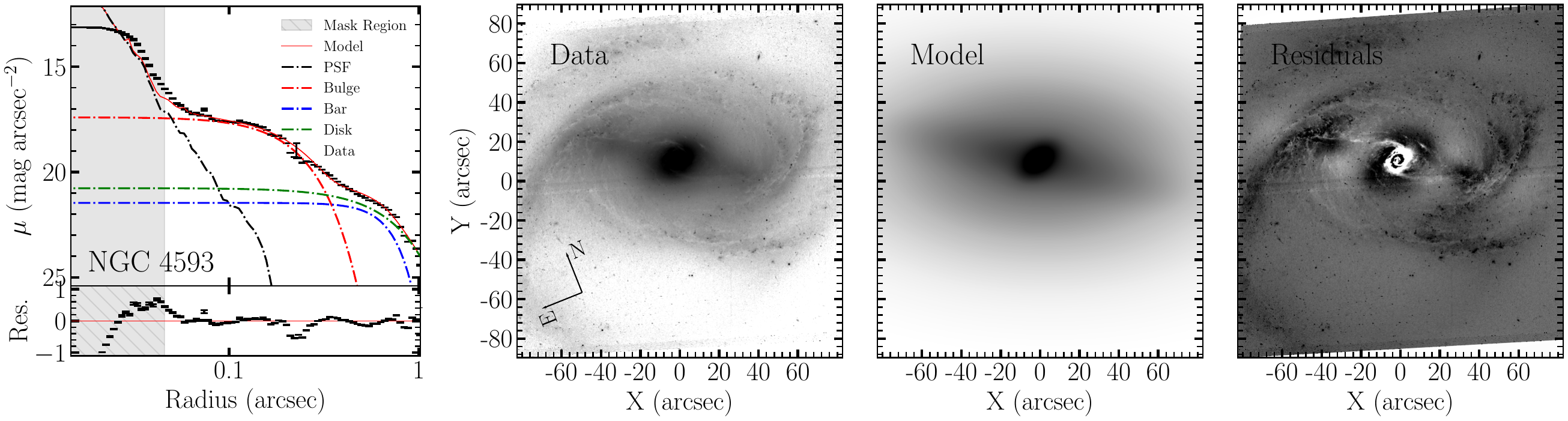}
	\includegraphics[width=0.9\textwidth]{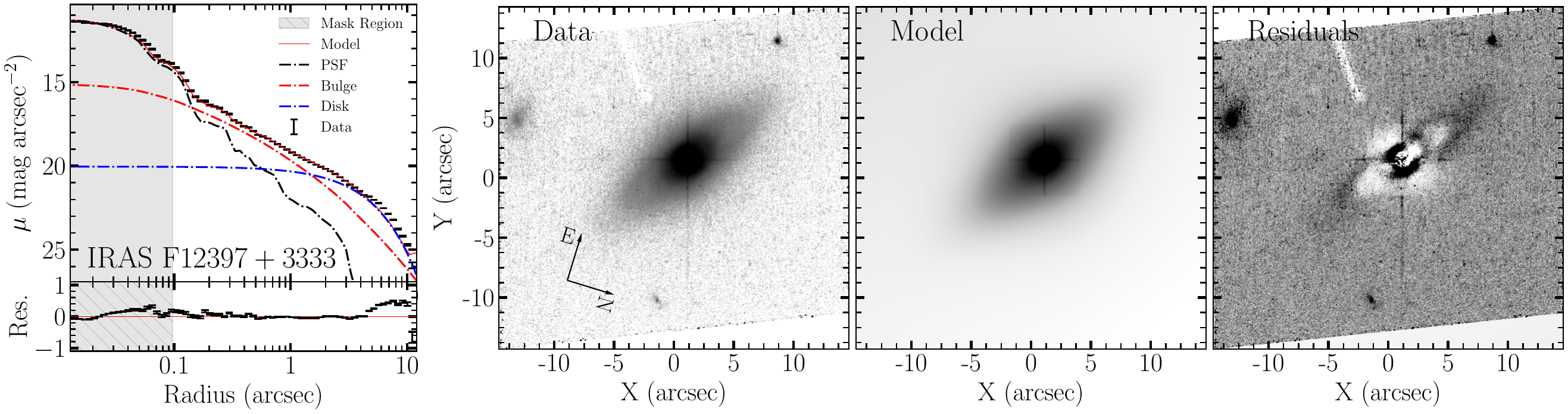}
	\includegraphics[width=0.9\textwidth]{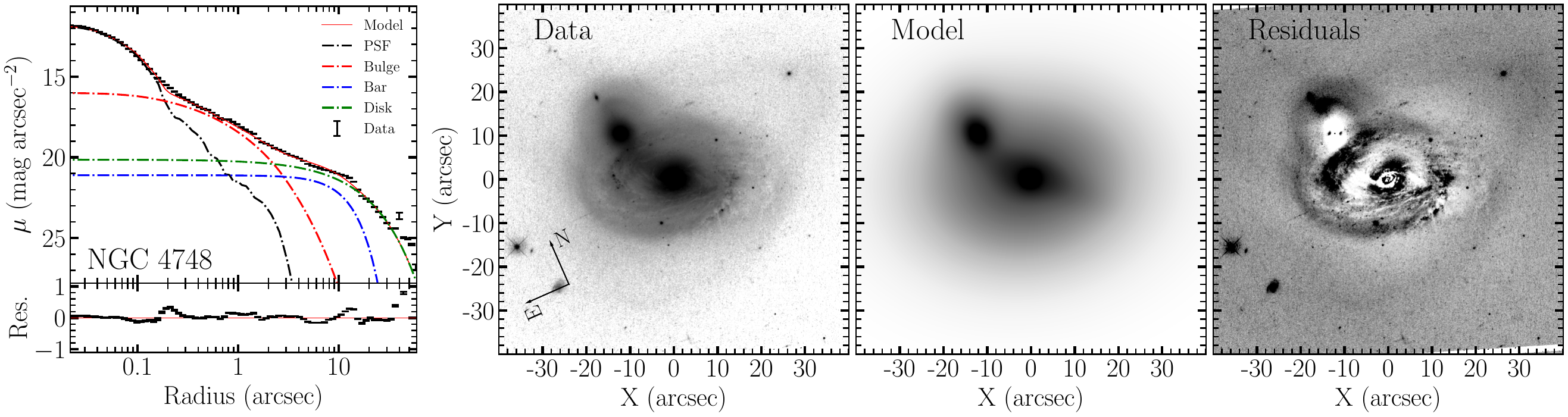}
	\figurenum{\ref{fig:galfit_1}}
	\caption{(Continued.)}
\end{figure}

\begin{figure}
	\centering
	\includegraphics[width=0.9\textwidth]{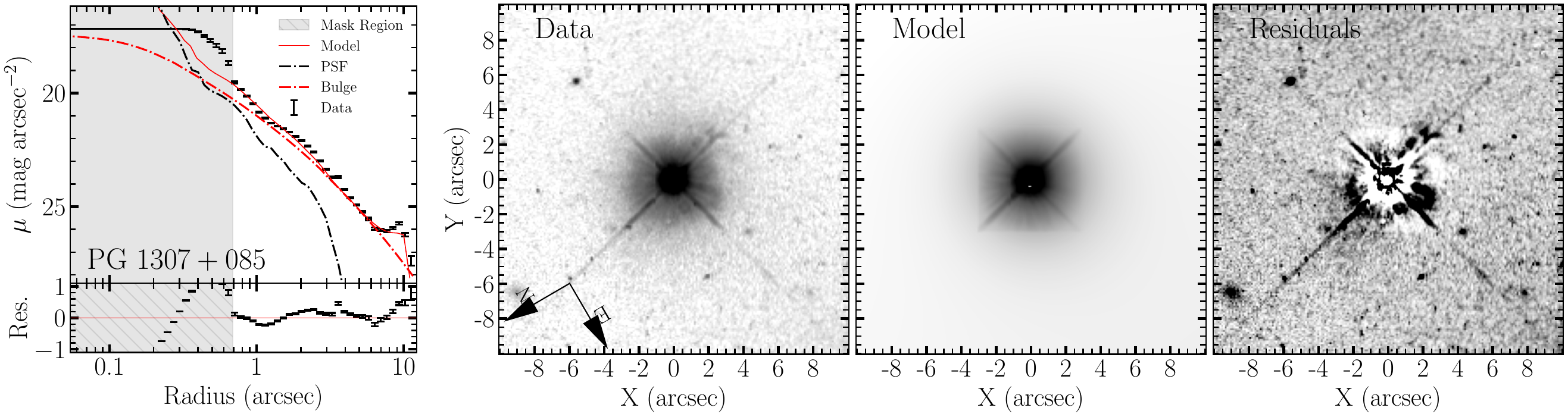}
	\includegraphics[width=0.9\textwidth]{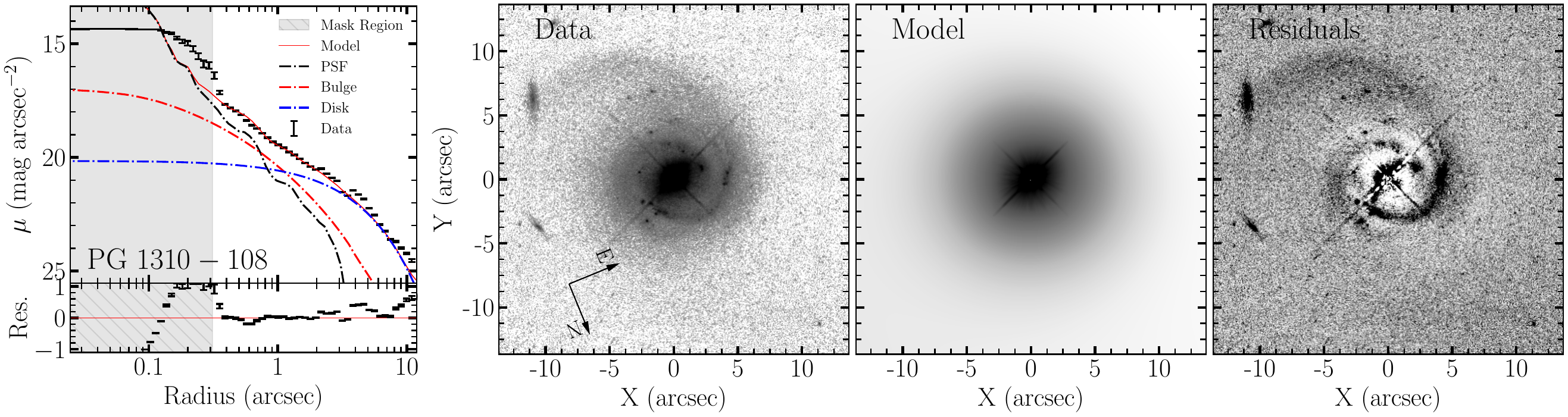}
	\includegraphics[width=0.9\textwidth]{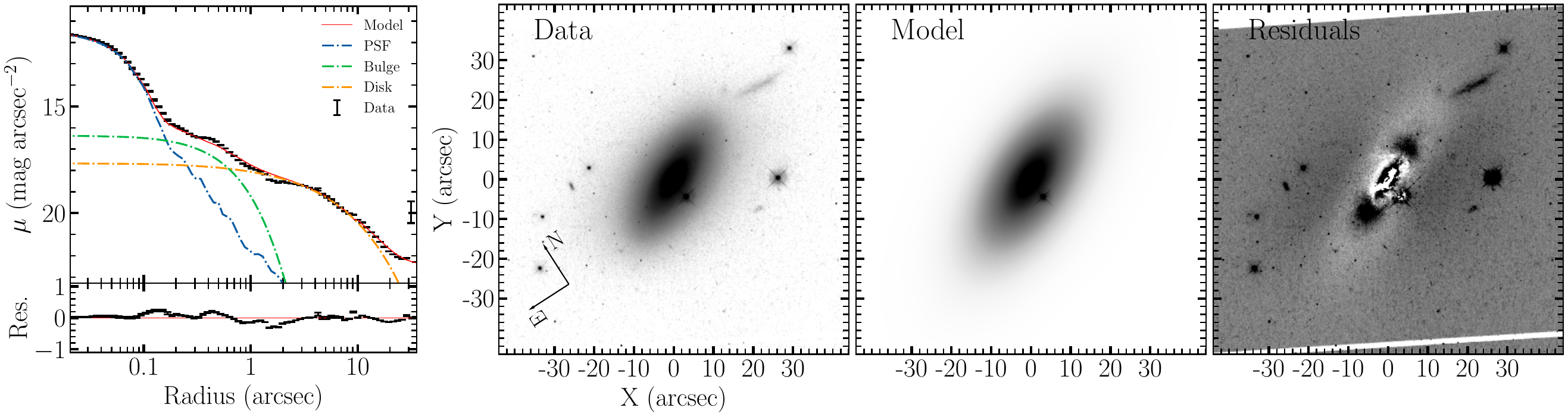}
	\includegraphics[width=0.9\textwidth]{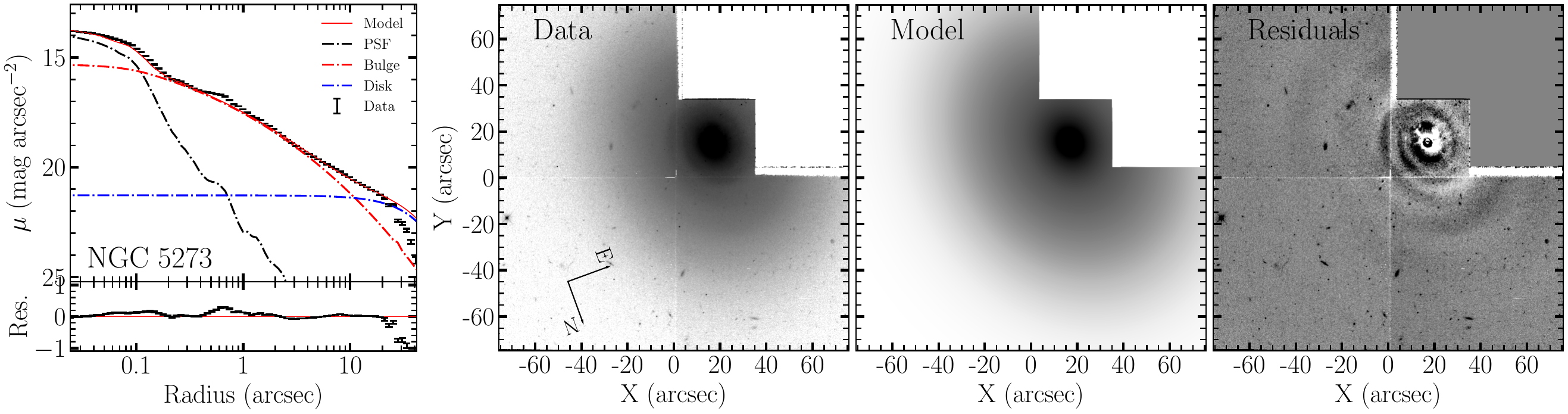}
	\includegraphics[width=0.9\textwidth]{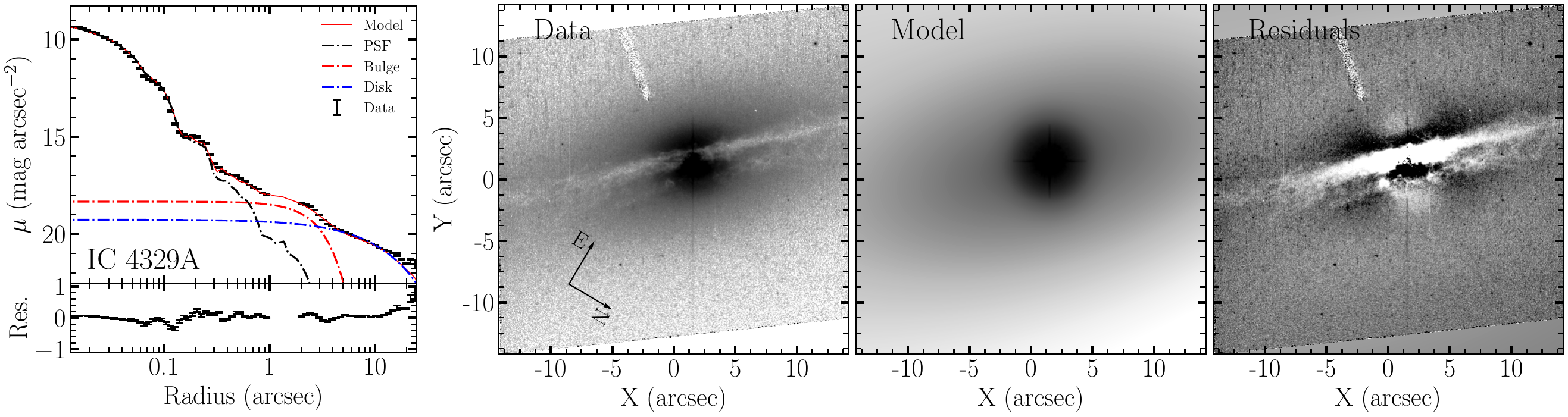}
	\figurenum{\ref{fig:galfit_1}}
	\caption{(Continued.)}
\end{figure}

\begin{figure}
	\centering
	\includegraphics[width=0.9\textwidth]{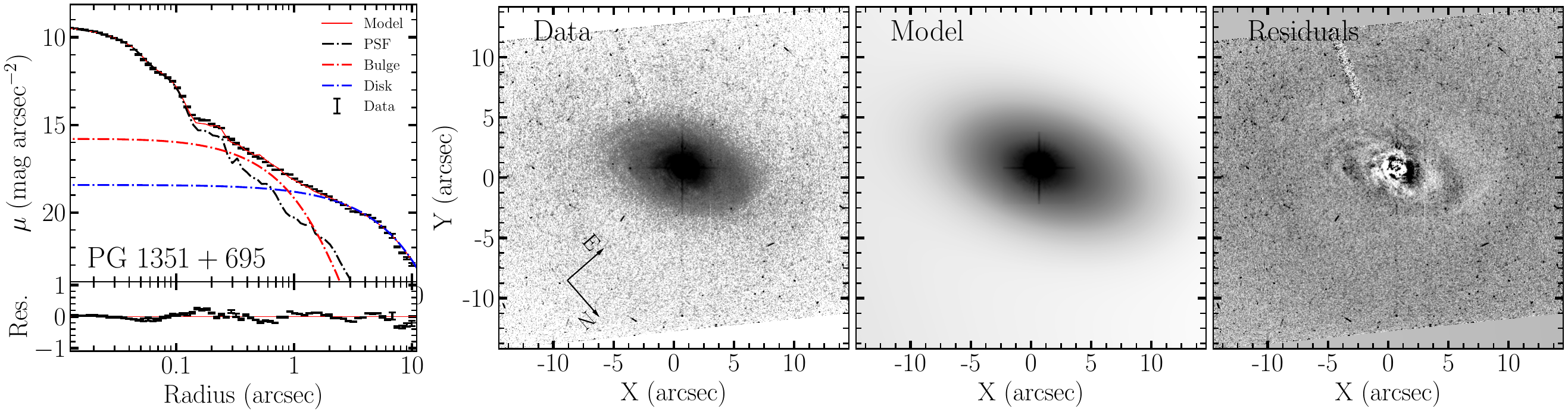}
	\includegraphics[width=0.9\textwidth]{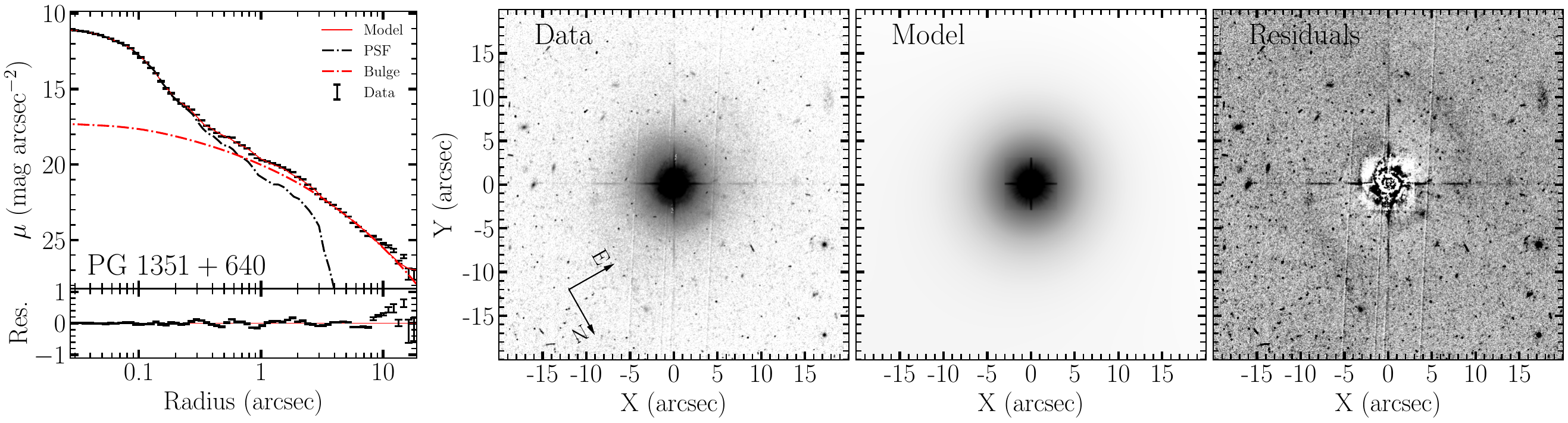}
	\includegraphics[width=0.9\textwidth]{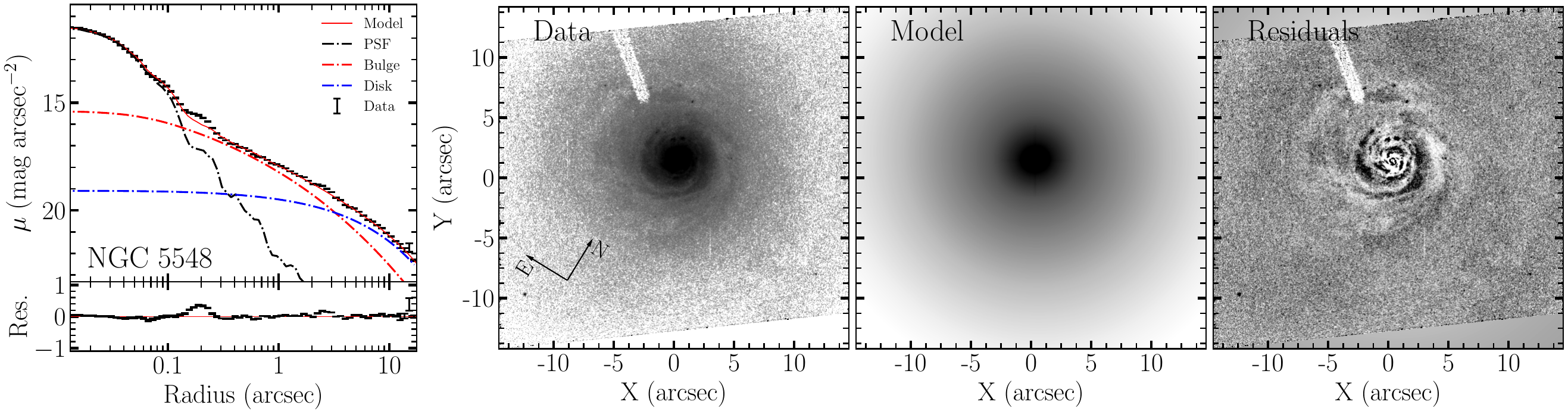}
	\includegraphics[width=0.9\textwidth]{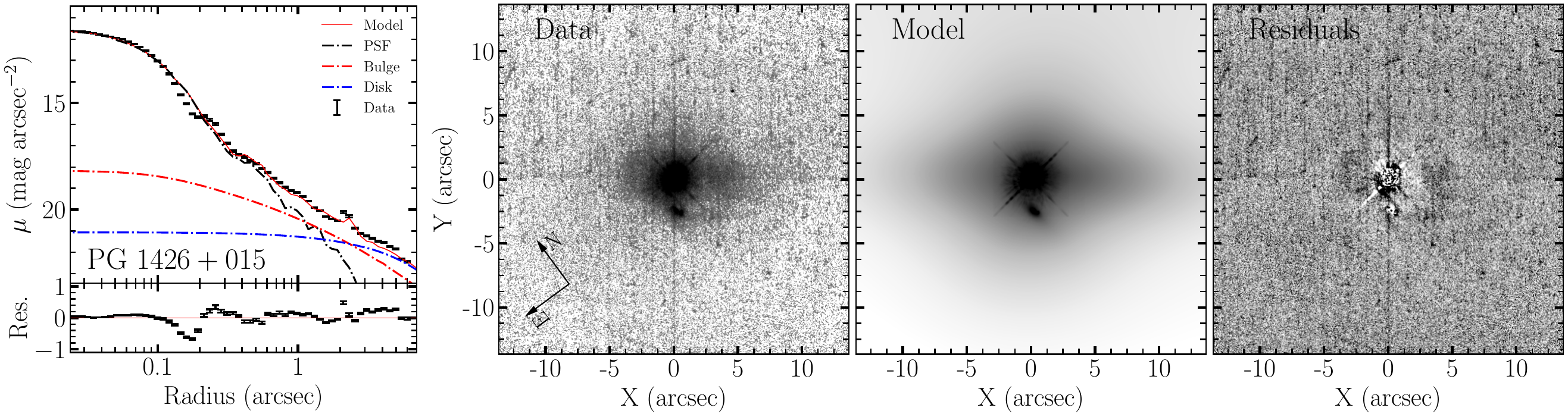}
	\includegraphics[width=0.9\textwidth]{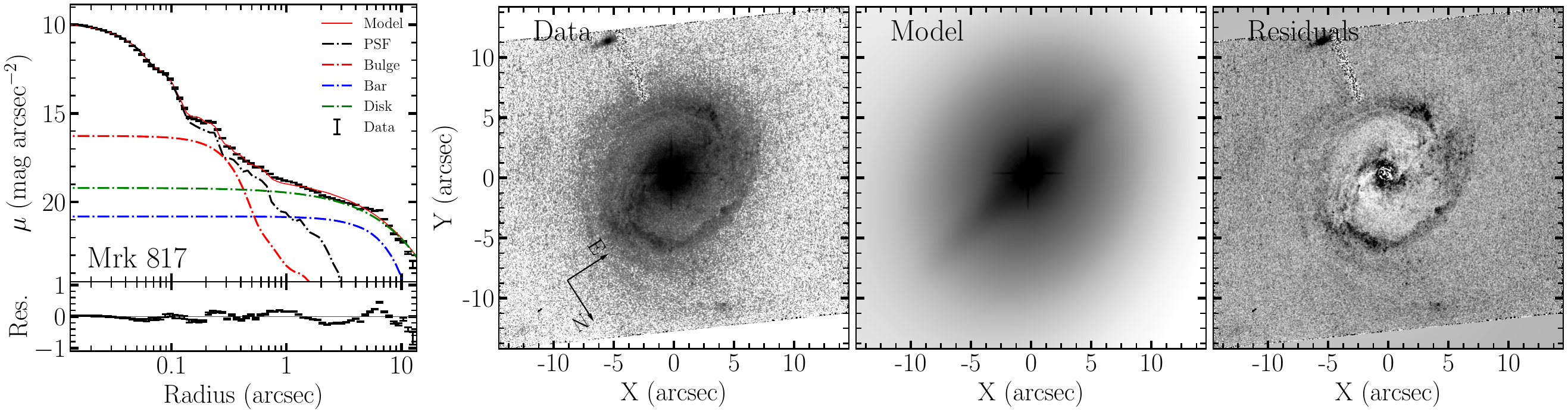}
	\figurenum{\ref{fig:galfit_1}}
	\caption{(Continued.)}
\end{figure}

\begin{figure}
	\centering
	\includegraphics[width=0.9\textwidth]{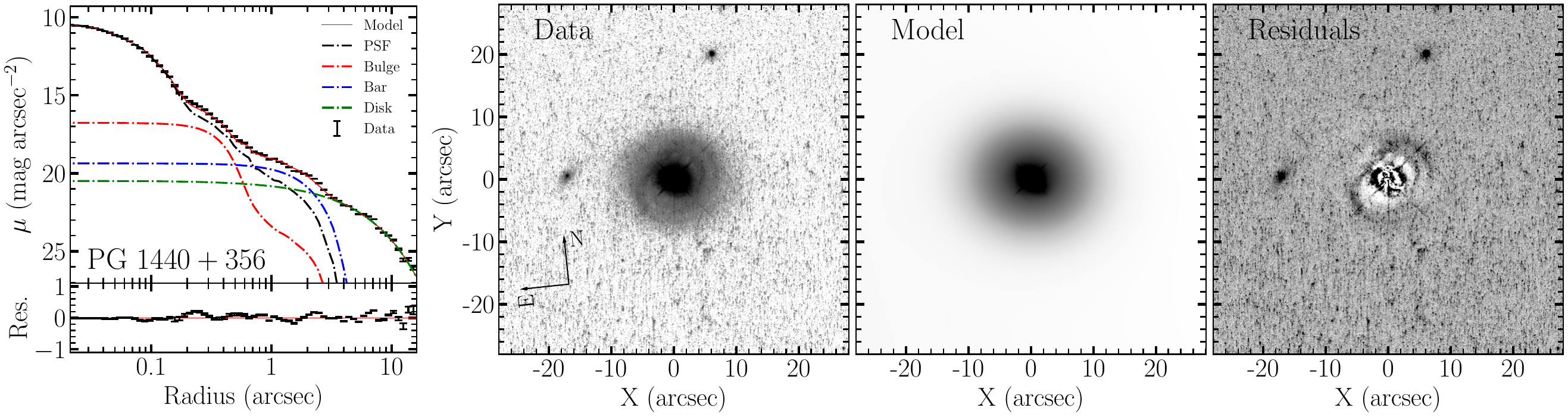}
	\includegraphics[width=0.9\textwidth]{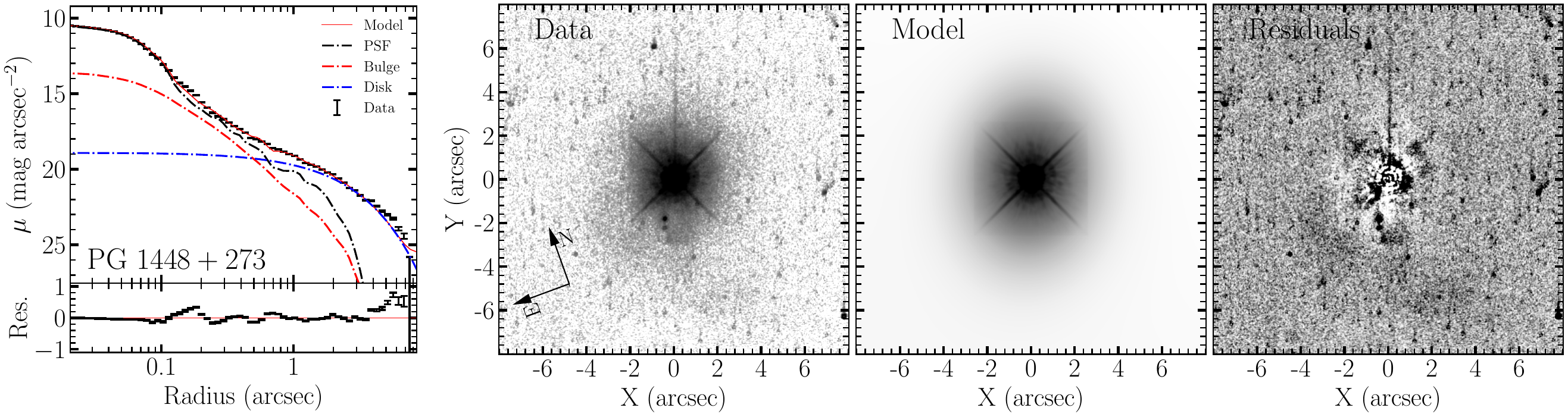}
	\includegraphics[width=0.9\textwidth]{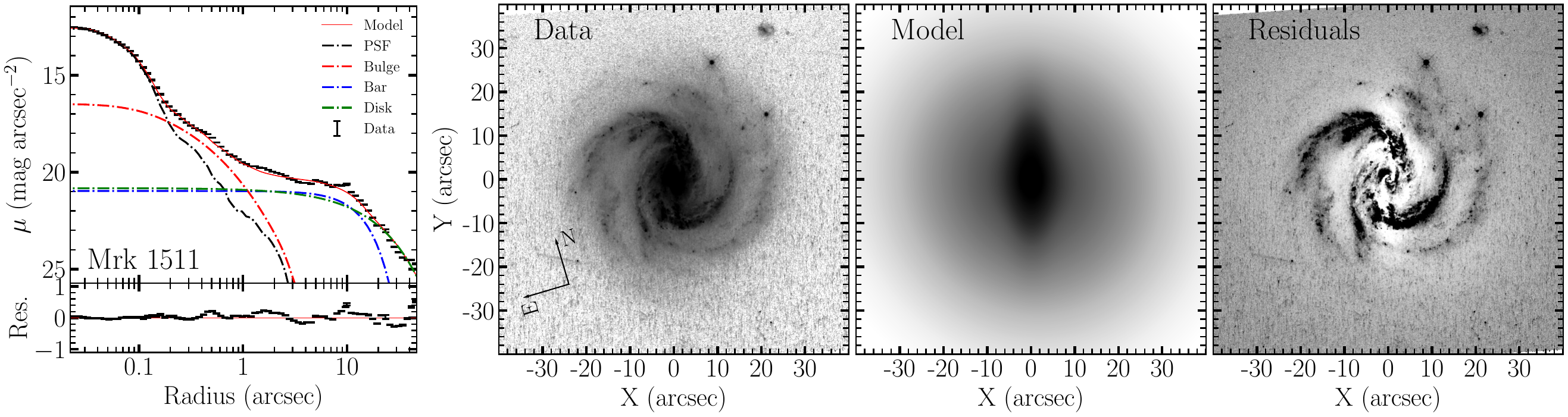}
	\includegraphics[width=0.9\textwidth]{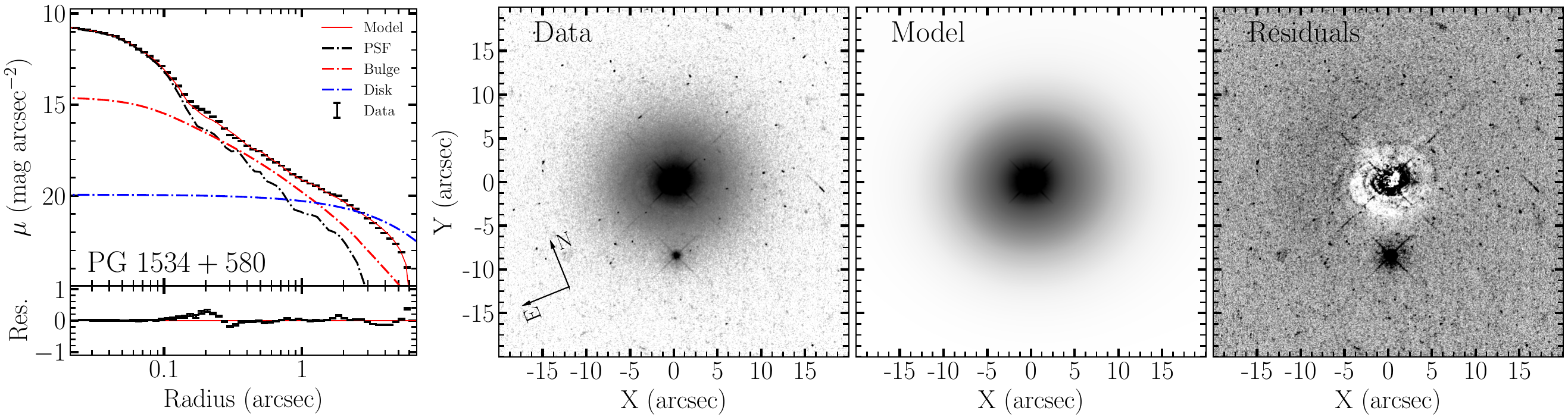}
	\includegraphics[width=0.9\textwidth]{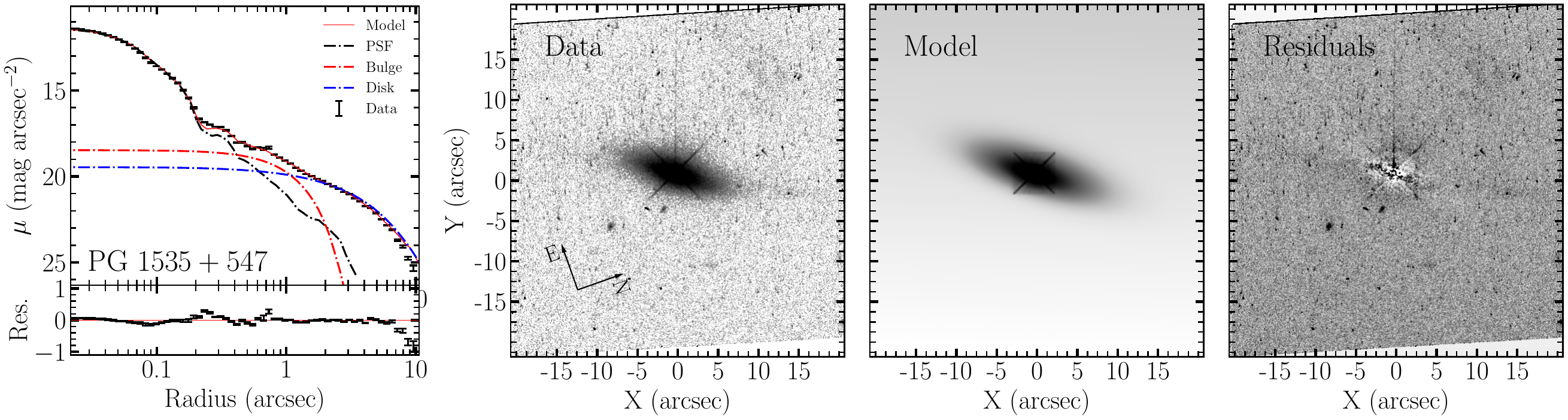}
	\figurenum{\ref{fig:galfit_1}}
	\caption{(Continued.)}
\end{figure}

\begin{figure}
	\centering
	\includegraphics[width=0.9\textwidth]{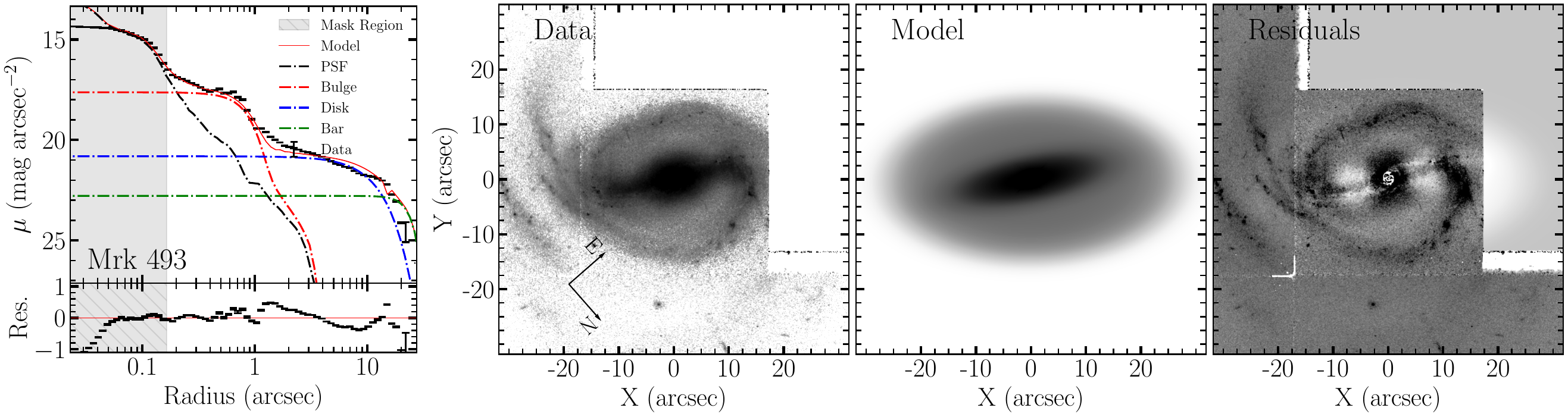}
	\includegraphics[width=0.9\textwidth]{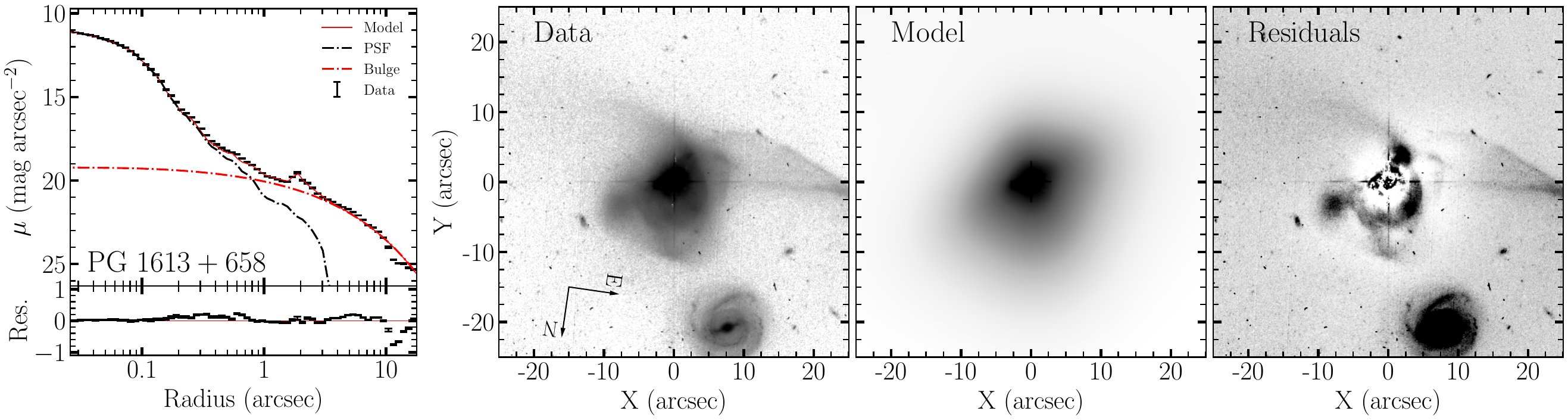}
	\includegraphics[width=0.9\textwidth]{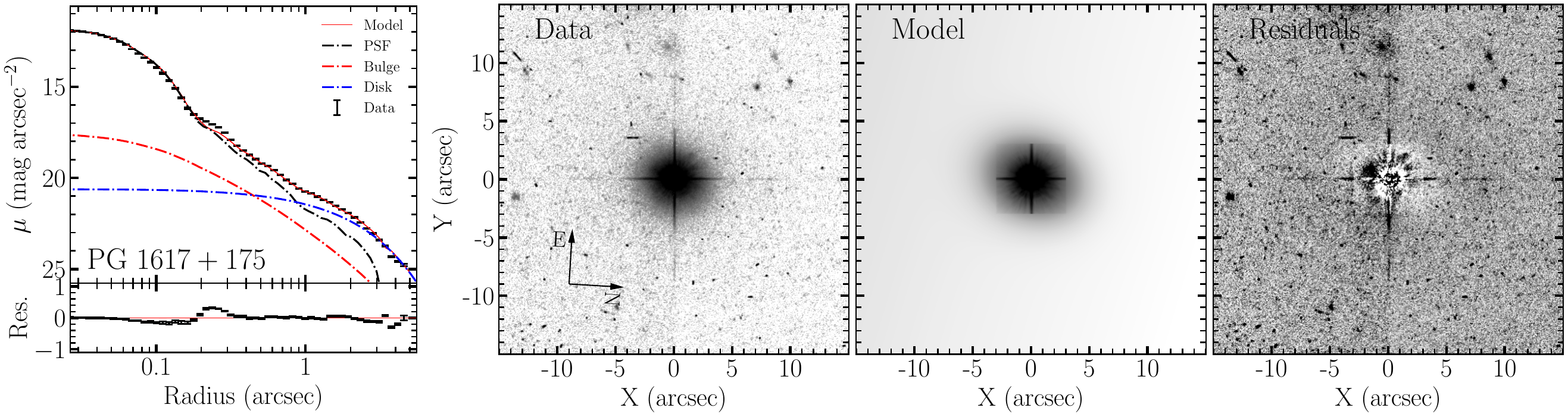}
	\includegraphics[width=0.9\textwidth]{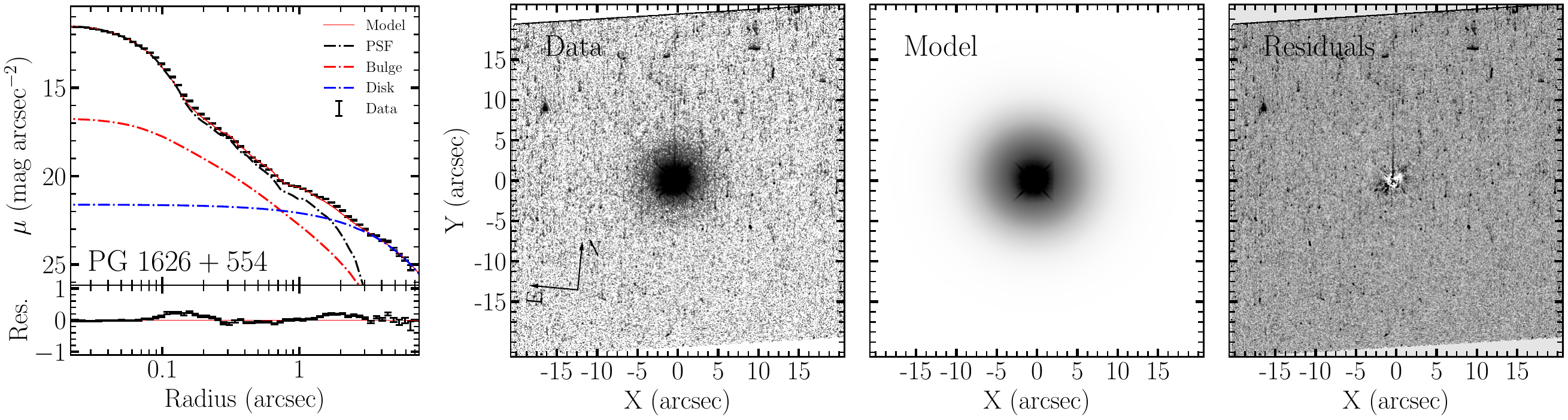}
	\includegraphics[width=0.9\textwidth]{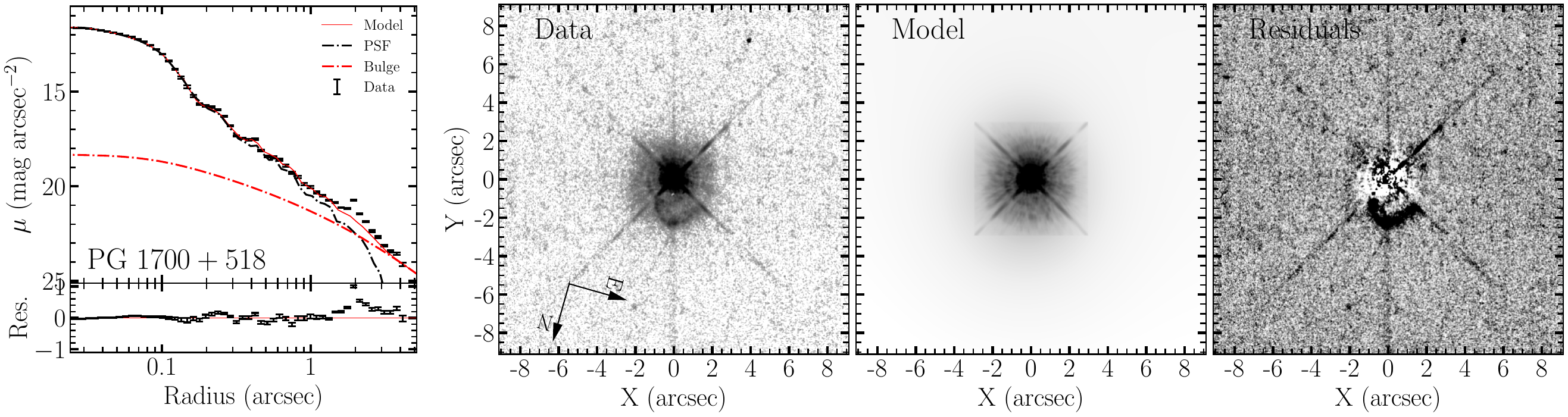}
	\figurenum{\ref{fig:galfit_1}}
	\caption{(Continued.)}
\end{figure}

\begin{figure}
	\centering
	\includegraphics[width=0.9\textwidth]{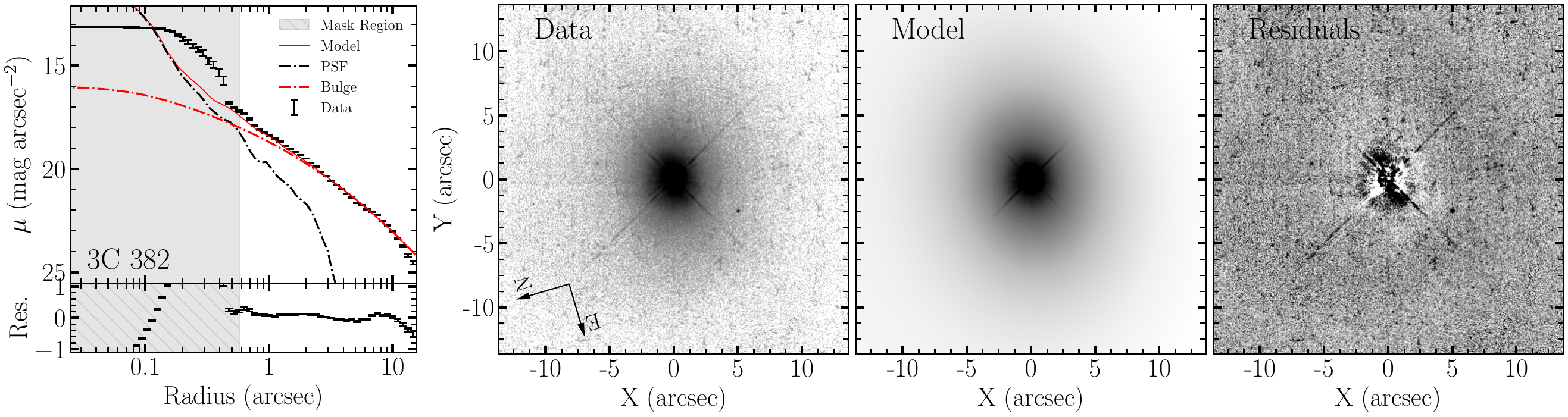}
	\includegraphics[width=0.9\textwidth]{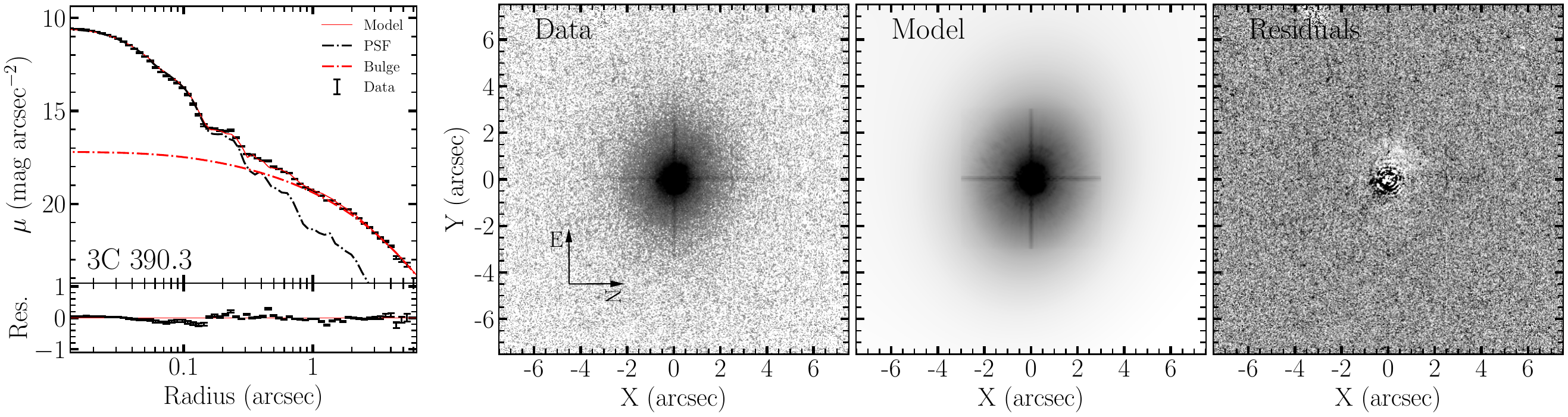}
	\includegraphics[width=0.9\textwidth]{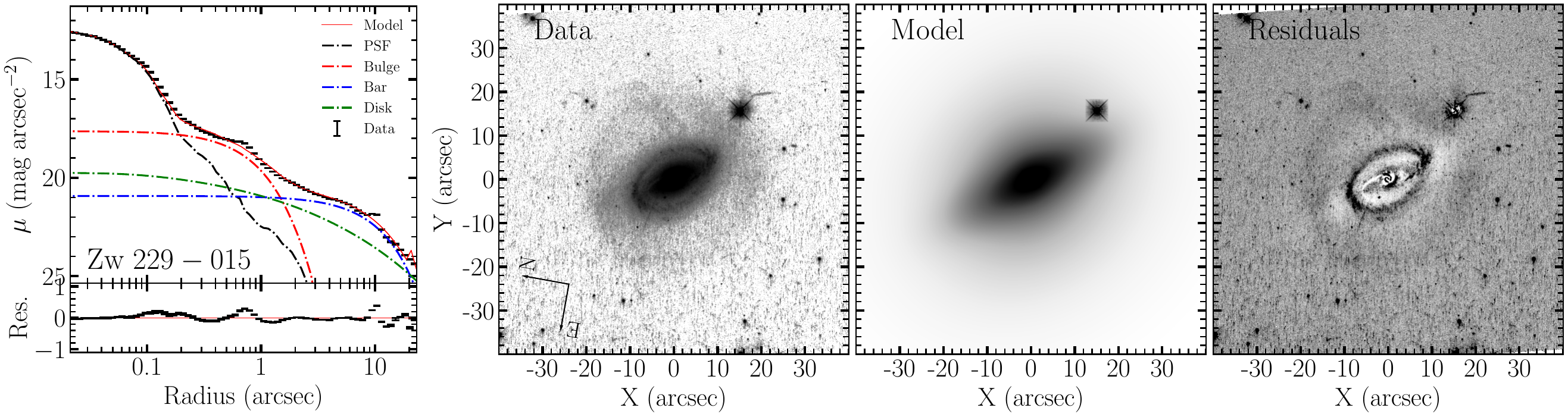}
	\includegraphics[width=0.9\textwidth]{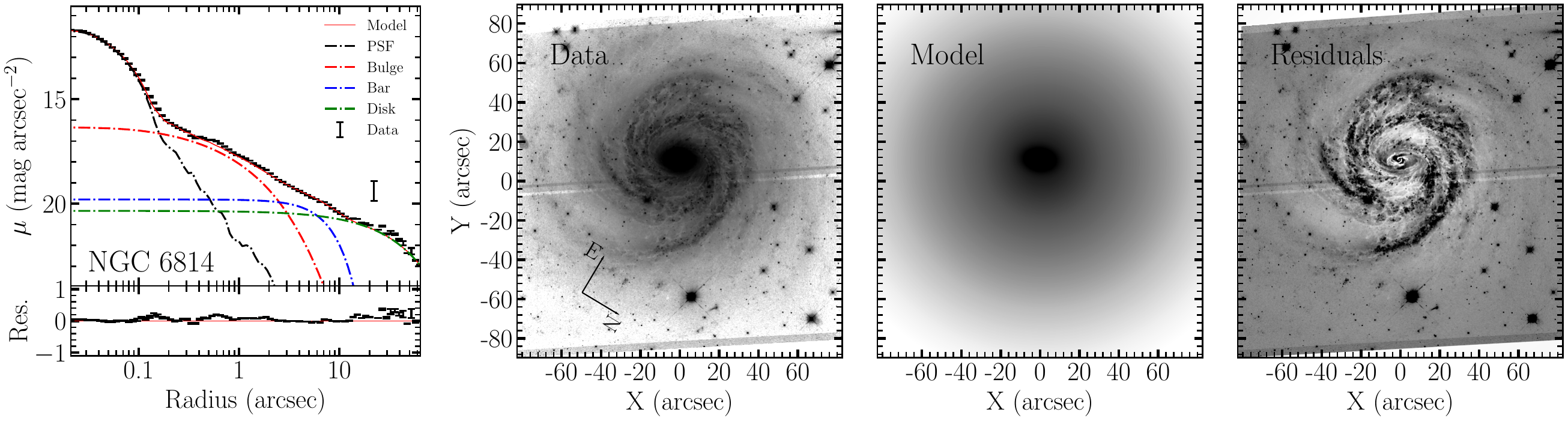}
	\includegraphics[width=0.9\textwidth]{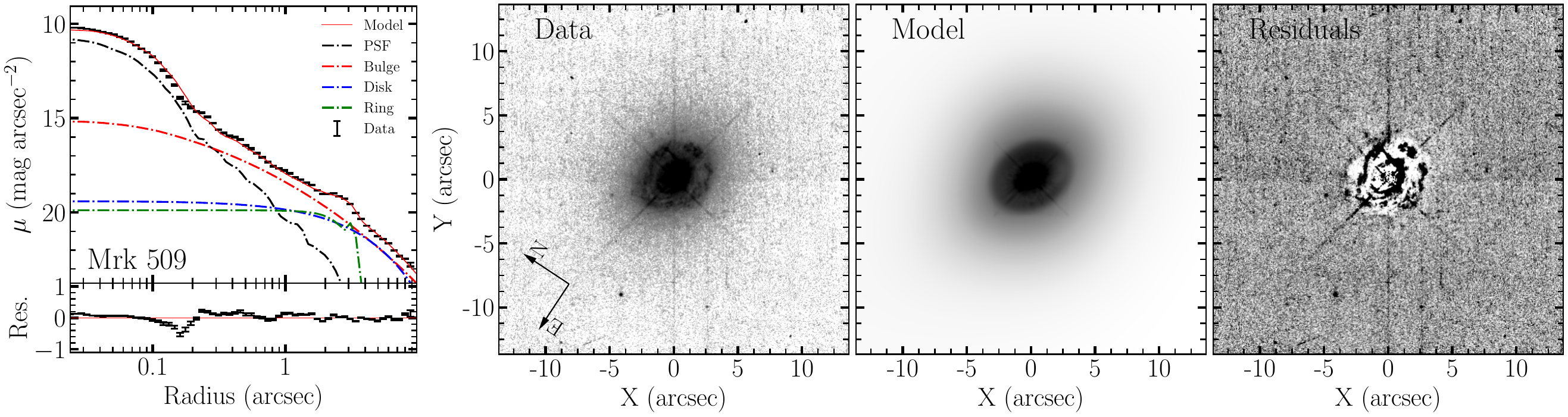}
	\figurenum{\ref{fig:galfit_1}}
	\caption{(Continued.)}
\end{figure}

\begin{figure}
	\centering
	\includegraphics[width=0.9\textwidth]{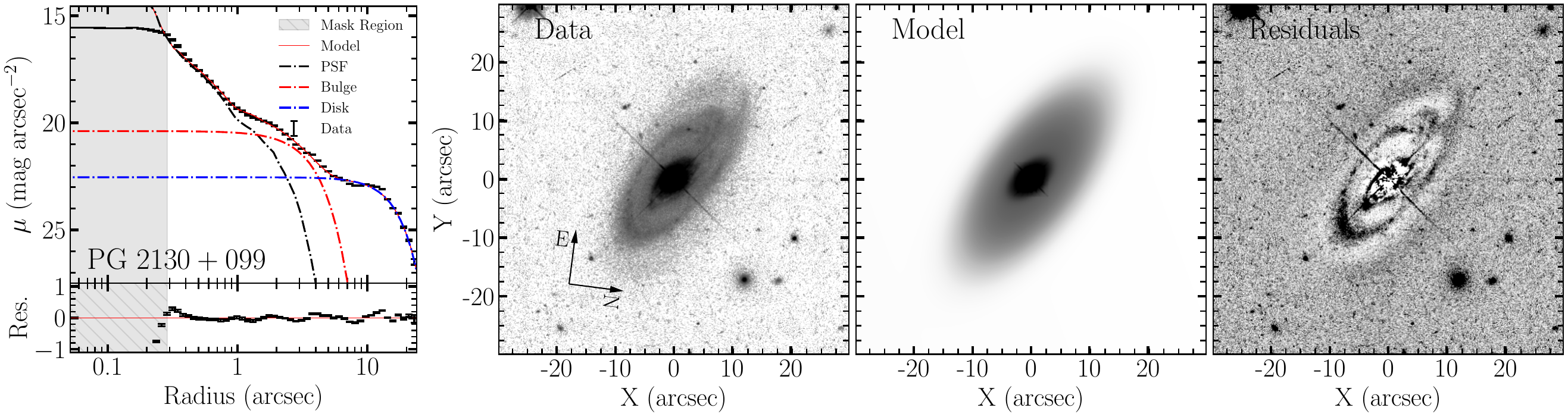}
	\includegraphics[width=0.9\textwidth]{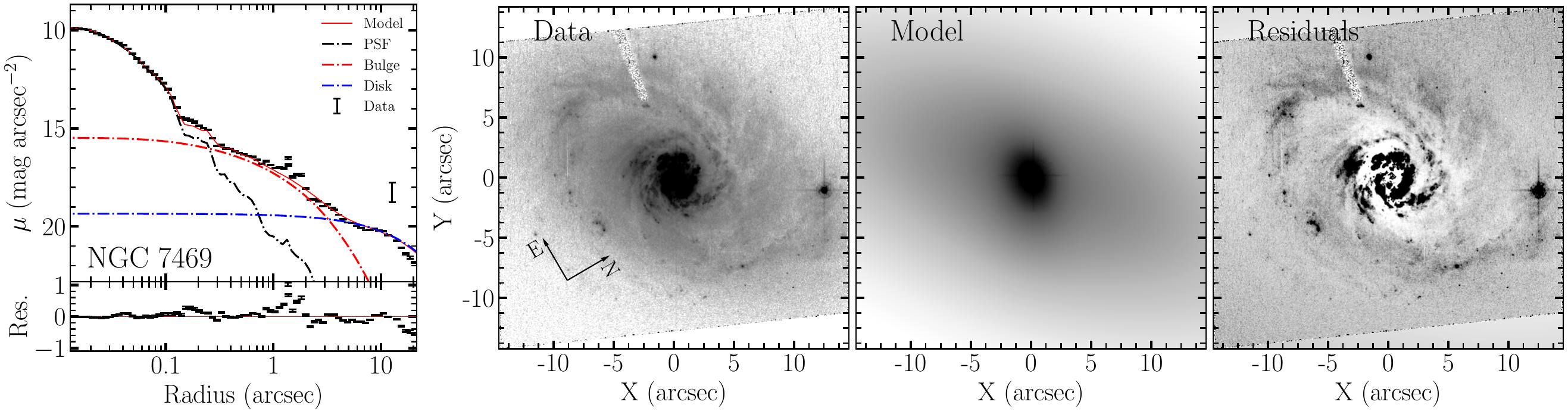}
	\figurenum{\ref{fig:galfit_1}}
	\caption{(Continued.)}
\end{figure}